\documentclass[10pt]{iopart}
\usepackage[normalem]{ulem}
\usepackage{multirow}
\usepackage[utf8]{inputenc}
\usepackage{iopams}
\usepackage{hyperref}
\usepackage{color}  
\expandafter\let\csname equation*\endcsname\relax
\expandafter\let\csname endequation*\endcsname\relax
\usepackage{mathtools}
\usepackage[]{graphicx}
\usepackage{multicol}
\usepackage{cooltooltips}
\newcommand{\iotab}{\lower3pt\hbox{$\mathchar'26$}\mkern-7mu\iota}

\usepackage[colorinlistoftodos]{todonotes}
\graphicspath{{./}
	{./figs/png/}
}

\begin{document}
%***************************************************************************************

\title[A quasi-isodynamic configuration with good confinement of fast ions at low plasma $\beta$]{A quasi-isodynamic configuration with good confinement of fast ions at low plasma $\beta$
}
\author{E. Sánchez, J. L. Velasco, I. Calvo and S. Mulas}

%\corresp{\email{edi.sanchez@ciemat.es}},
\address{
Laboratorio Nacional de Fusión, CIEMAT, 28040, Madrid, Spain
}

\ead{edi.sanchez@ciemat.es}
\vspace{10pt}
%\begin{indented}
%	\item[]? 2022
%\end{indented}

%***************************************************************************************
\begin{abstract}
A new quasi-isodynamic stellarator configuration optimized for the confinement of  energetic ions at low plasma $\beta$ is obtained. The numerical optimization is carried out  
using the STELLOPT suite of codes. New proxies to measure closeness to quasi-isodynamicity and quality of fast ion confinement have been included. The new configuration has poloidally closed contours of magnetic field strength, {low magnetic shear and a rotational transform profile allowing an island divertor}. It shows ideal and ballooning magnetohydrodynamic stability up to $\beta=5\%$, reduced effective ripple, with $\epsilon_{eff}<0.5\%$ in the plasma core. 
 Even at low $\beta$, the configuration approximately satisfies the maximum-$J$ property,  and the confinement of fast ions is good at $\beta\sim1.5\%$ and becomes excellent at reactor values, $\beta\sim 4\%$.
 An evaluation of the $D_{31}$ neoclassical mono-energetic coefficient supports the {expectation} of a reduced bootstrap current for plasmas confined in quasi-isodynamic configurations. 
A set of filamentary coils that preserve the good confinement of fast ions in the core is presented.

\end{abstract}
%% Uncomment for PACS numbers
%\pacs{52.25.Xz, 52.55.Hc, 52.65.−y, 52.55.−s}
%
% Uncomment for keywords
%\vspace{2pc}
%\noindent{\it Keywords}: stellarator, gyrokinetic simulations, flux tube,   full surface, global, ITG,  zonal flows, GENE,  stella, GENE-3D, EUTERPE

\ioptwocol

%
%***************************************************************************************
\section{Introduction}
%***************************************************************************************

The stellarator is an attractive concept, alternative to the tokamak, for a nuclear fusion reactor based on magnetic confinement. It offers some advantages with respect to the tokamak {derived from the fact that in stellarators most of the magnetic field is externally generated, % through magnetic coils, 
without requiring an inductive toroidal current to produce part of the confining magnetic field. The absence of inductive current makes  steady-state operation easier.  Besides, the smallness of plasma currents in stellarators makes them less prone to current instabilities or disruptions than tokamaks}.  
 However, while in a tokamak the axisymmetry ensures that all collisionless particles are confined, the  fully three-dimensional magnetic configuration in a stellarator has important and, in general, deleterious consequences for the confinement of particles and energy. In a generic stellarator, the transport of particles and energy due to the inhomogeneities of the magnetic field combined with inter-particle collisions (the so-called neoclassical transport) is significantly larger than that in a tokamak at  low collisionality. 
However, in principle, it is possible to design stellarator magnetic configurations that confine particles and energy as well as tokamaks; these are called omnigenous magnetic fields \cite{hall_three-dimensional_1975, cary_omnigenity_1997,LandremanM2012,parra_less_2015}. For this purpose, careful tailoring of the magnetic configuration is required, which is commonly known as stellarator optimization. 

A  magnetic field  is called omnigenous if the orbit-averaged radial magnetic drift vanishes for all particles. 
{In these magnetic fields the second adiabatic invariant, $J$, is constant on the flux surface for all trapped-particle trajectories. An omnigeneous magnetic field is said to satisfy the maximum-$J$ property if $J$ has negative radial derivative over the whole plasma radius. This property has been theoretically predicted to have beneficial effects on the stabilization of interchange \cite{rosenbluth_lowfrequency_1968} and Trapped Electron Mode (TEM) turbulence \cite{Helander2013}.}

Omnigeneity has several implications for the shape of the magnetic field. In omnigenous fields, for each flux surface, the contours of constant  magnetic field strength close on themselves, either poloidally, toroidally, or helically. The distance (along the field line) between two points with the same magnetic field  strength at both sides of a local minimum is independent of the field line. Furthermore, the contours of absolute maxima of field strength on a flux surface are straight lines, {and the values of the local maxima (and minima) of the magnetic field strength are also independent of the field line.}

Two particular classes  of omnigenous magnetic fields, quasi-symmetric (QS) and quasi-isodynamic (QI), have received particular interest \footnote{Note that the term quasi-symmetric (respectively, quasi-isodynamic) is often used in the literature to refer to magnetic configurations that exactly satisfy the property of quasi-symmetry (respectively, quasi-isodynamicity) as well as to configurations that satisfy the property only approximately. We will also follow this slightly loose terminology in this paper.}

 In QS fields, the magnetic field strength {has the form} $B=B(\psi, M \theta -N \phi )$ in Boozer coordinates \cite{Nuhrenberg1988,Boozer1995}, with $\psi$ the magnetic toroidal flux labeling the flux surface, $N$ and $M$ integer numbers and $\phi$ and $\theta$ Boozer angles.  Three different forms of quasi-symmetric configurations can be, in principle, obtained depending on the values of the numbers $N$ and $M$. We have quasi-axisymmetry (QA) for $N=0$ \cite{zarnstorff_physics_2001}, quasi-helical symmetry (QH) for $N \neq 0, M \neq 0$ \cite{anderson_helically_1995}, and quasi-poloidal symmetry (QP) for $M=0$ \cite{spong_physics_2001}
Significant progress has been made in the last  few years in the search for new quasi-symmetric configurations, and QA and QH configurations have been obtained recently with tiny deviations of the magnetic field from perfect symmetry \cite{landreman_magnetic_2022}.  In these configurations, the neoclassical transport, as predicted by the effective ripple \cite{nemov_evaluation_1999}, is very small, and the confinement of energetic ions is also  significantly improved with respect to previous configurations \cite{landreman_optimization_2022}.

In QI configurations, the magnetic field is omnigenous with magnetic field strength contours that close poloidally \cite{subbotin_integrated_2006}.
{Results for QI configurations are more scarce than for QS  ones, even though the QI class of {omnigenous} devices offers some advantages over the QS class}. In QI devices, 
the current generated by neoclassical processes (bootstrap current) can be kept very small. {For a perfectly omnigenous stellarator with poloidally closed $B$ contours the bootstrap current vanishes \cite{subbotin_integrated_2006,Helander2009, LandremanM2012}.} However, in QA or QH configurations the scaling of the bootstrap current with $\beta$ is not that favorable \cite{landreman_optimization_2022}. 
 
 {QI magnetic configurations generated with modular coils, exhibiting magnetohydrodynamic (MHD) stability at high plasma $\beta$, small Pfirsch-Schlüter currents, low magnetic shear and a rational surface at the plasma edge allowing an island divertor define the Helias \cite{Wobig1999} concept, a category to which Wendelstein 7-X (W7-X) \cite{Grieger1992} belongs}. 

 {W7-X is the largest stellarator whose magnetic configuration was optimized using numerical methods, and its good confinement of thermal species has been recently demonstrated} \cite{Beidler2021}. This has been a success for the stellarator line of research and the numerical optimization effort in particular. The effective ripple figure of merit has reached in the W7-X standard configuration a level that is considered enough for a reactor \cite{Beidler2021,Alonso2022}. However, there is still room for improvement beyond W7-X. The bootstrap current in the {so-called} standard configuration is too large for a reactor. It is small enough in {the so-called high mirror configuration; however, the effective ripple in this configuration} is too large, and there is no configuration {accessible with the W7-X coils that has at the same time} small enough values of both bootstrap current and effective ripple. 
 
 Furthermore, it is known that in Helias configurations the confinement of fast ions improves with $\beta$  \cite{faustin_fast_2016,velasco_model_2021} and good fast ion confinement is expected to be achievable in W7-X configurations only at high $\beta$, which is still to be confirmed experimentally. However, the fast ion confinement expected for W7-X is probably not sufficiently good for a reactor. A good confinement of energetic particles, particularly that of 3.5 MeV fusion-generated alpha particles, is crucial for a fusion reactor because  prompt alpha losses can deposit into the first wall an intolerable amount of energy.  On the other hand, it is required that the alphas remain in the reactor for a sufficiently long time, so that they heat the D-T fuel, thus allowing a self-sustained chain reaction with a minimum need for external auxiliary energy input once ignition is reached.   {Furthermore, good confinement at low values of $\beta$ is highly desirable for two reasons. First, it would facilitate the design of high-field reactors, which could operate at a smaller $\beta$. Additionally, 
 	in a reactor with standard magnetic field values, alpha particles will be generated in the reactor before it reaches its high-$\beta$ physics design point, and they need to be confined.} {However, the best confinement properties of W7-X, including the maximum-$J$ property, are expected to appear at high $\beta$}. {In recent years, several QI configurations have been obtained  \cite{goodman_constructing_2022,jorge_single-field-period_2022,mata_direct_2022}  that exhibit good confinement of  energetic ions. However, the configuration that we will present in this work is, to our knowledge, the first one to achieve this in a satisfactory range of $\beta$ while maintaining at the same time good MHD stability properties.}
 
{Even though the neoclassical transport of energy and thermal particles can be reduced to acceptable levels through the optimization of the configuration, turbulence is another major factor limiting the performance of a stellarator in terms of confinement, as demonstrated in W7-X. The performance obtained so far is in general below the expectations based on neoclassical calculations  \cite{bozhenkov_high-performance_2020} and it was only  improved over the empirical ISS04 scaling \cite{Yamada2005} in particular scenarios with turbulence largely suppressed \cite{bozhenkov_high-performance_2020}. 
The confinement of energetic ions and the turbulent transport are two key elements presently under consideration for the design of a new generation of stellarators. Other engineering-based  criteria should also be taken into account in the numerical optimization to ensure that a configuration can actually be constructed \cite{warmer_w7-x_2016,lion_deterministic_2022,lion_general_2021}. The success in the simultaneous optimization according to all these criteria  could define the stellarator as a preferred option over the tokamak as a fusion reactor. 
} {We leave the assessment of turbulent transport of this configuration for a separate publication, but we will show that it is close to fulfilling the maximum-$J$ property even at low $\beta$, which is expected to be beneficial for TEM turbulence.}
      
In this work, we present a new four-field period QI configuration with moderate aspect ratio, $A \sim 10$ \footnote{For comparison, the aspect ratios for other QI configurations are: $A \sim 11$ for W7-X \cite{Grieger1992}, $A \sim 12$ for QIPC \cite{subbotin_integrated_2006}, $A \sim 9$ for HSR4/18, and $A \sim 12$ for HSR5/22 \cite{beidler_helias_2001}.},
 optimized for MHD stability and reduced neoclassical transport of both thermal and energetic ions.
  It has low magnetic shear and a rotational transform profile with $\iotab<1$, allowing a 4/4 island divertor. 
The configuration properties are analyzed in detail and compared with configurations of W7-X. It has an effective ripple smaller than 0.5\% in the plasma core, is predicted to be ideal and ballooning MHD stable up to $\beta=5\%$, shows  very good confinement of fast ions at low $\beta$ (1.5\%), and exceptionally good confinement at $\beta=4-5\%$, which supposes a significant improvement with respect to W7-X\footnote{We will study configurations at different values of volume-averaged $ \beta$ that we will simply name as $\beta$.}. Besides, the bootstrap current is estimated to be small, as expected for a QI configuration. 
We also present a first feasibility study of simplified (filamentary) magnetic coils capable of generating this configuration keeping the good fast ion confinement. 

 The rest of the paper is organized as follows. The tools and the strategy used  for the optimization are introduced in section \ref{secOptimStrat}. In section \ref{secConfigProps}, we present the optimization results, and the properties of the optimized configuration are compared to several reference cases.
 In section \ref{secCoils} a set of modular coils  that can generate the optimized configuration is presented. Finally,  the results are summarized and some conclusions are drawn in section \ref{secSumandConc}.

%***************************************************************************************
\section{Numerical tools and optimization strategy}\label{secOptimStrat}
%***************************************************************************************
%
In this work, we use the STELLOPT suite of codes \cite{STELLOPT_doecode_12551} as the basic tool for optimization. It provides a general framework for varying the parameters defining a magnetic configuration, evaluating  the configuration properties and optimizing it according to a cost function that depends on these properties. 
STELLOPT includes different  global,  such as the Genetic Algorithm with Differential Evolution (GADE), and gradient-based optimization algorithms, such as  the Levenberg-Marquardt (LMDIFF) algorithm, which is the one used here.

The suite includes many specialized codes for evaluating the magnetic configuration properties. We use the code NEO for the assessment of the thermal particle transport, through the calculation of the effective ripple metric, $\epsilon_{eff}$ \cite{nemov_b_2005}. $\Gamma_{\alpha}$  \cite{velasco_model_2021}, $\Gamma_c$ \cite{nemov_poloidal_2008} and several other proxies related with the confinement of energetic ions have been included in STELLOPT \footnote{This has been done through the neoclassical code KNOSOS \cite{velasco_knosos_2020}, that computes the necessary orbit-averated quantities}
 {In addition to these codes that are integrated in the suite and executed within the optimization loop, we  use  the codes COBRA \cite{sanchez_cobra_2000}, MAKEGRID \cite{VMECFBTutorial}, and BNORM \cite{merkel_integral_1986} in a standalone manner for evaluating the ballooning stability and generating coil-dependent data for free-boundary equilibrium VMEC calculations, respectively (see section \ref{secCoils}).  We use the code ASCOT \cite{hirvijoki_ascot_2014} for the evaluation of the fast ion confinement in selected configurations obtained through the optimization loop. 

The basic parameters defining a magnetic configuration that are varied in the optimization process are those determining the MHD equilibrium that is computed with the VMEC code \cite{Hirshman1983}. For vacuum configurations, these are the Fourier coefficients describing the plasma boundary, either the current or rotational transform profile, and the magnetic axis,  whose initial guess is required by VMEC. 
 For the optimization at finite $\beta$, we also allow variations of the pressure profile through the PRESCALE input and the toroidal flux at the plasma edge (PHIEDGE in VMEC input).

The cost function to be optimized (minimized) in STELLOPT is defined  as:
%*******************************
\begin{equation}
\chi^2 = \sum _i {\frac{(f_i^{target}-f_i^{equil})^2}{\sigma _i}},
\label{eqCostFunction}
\end{equation}
%*******************************
where $f_i^{target}$ represents the value of {property $f_i$} established as a goal in the optimization and $f_i^{equil}$ represents the value of this property calculated for the configuration under evaluation within the optimization loop. The value $\sigma_i$ represents the tolerance associated to the target value $f_i^{target}$. Note that $\sigma _i$ plays several roles. On the one hand, its value defines the tolerance for an specific target property, $f_i$. On the other hand, the relative size of the different $\sigma_i$ allows us to modify the  importance of a specific property, $f_i$, as compared to others, and to modify the shape of the cost function to minimize in the optimization process. The optimal values of these $\sigma _i$ are not known a priori and have to be determined empirically, in general.

%*******************************
%*******************************
\begin{table*}
	\centering
	\caption{Target values used in the optimization for different properties. The targets with * are defined in a set of radial positions. Note that not all the targets are used in all phases of the optimization process (see section \ref{secOptimStrat} for details).}
	\begin{tabular}{|c|c|c|c|}
		\hline
		{ }& \bfseries{Property} & \bfseries{Meaning/Purpose} & \bfseries{Target value} \\
		\hline
		\multirow{6}{*}{\rotatebox{90}{{standard metrics }}}   & A 	& aspect ratio	&	9.5 \\
		& $\beta$	&	average $\beta$ value & 	1.5 \\
		& $I_t$	& toroidal current &	0	\\
		& $\epsilon_{eff}$	&	effective ripple \cite{nemov_b_2005}& 0* \\
		& $W$	&	magnetic well (as defined in \cite{greene_brief_1998})& $\sim 0.1$* \\
		& $\iotab$	&	rotational transform & $< 1$* \\
		\\
		\multirow{6}{*}{\rotatebox{90}{{new proxies }}}   &
		$\Gamma_c$ & 	 proxy for fast ion confinement \cite{nemov_poloidal_2008} & 0* \\
		& $\Gamma_{\alpha}$ & 	 proxy for fast ion confinement \cite{velasco_model_2021}& 0* \\
		& $VBT$ & 	alignment of B maxima (see text) & 0* \\
		& $VBB$ & 	alignment of B minima (see text) & 0* \\
		& $WBW$ & 	width of the central magnetic valley (see text) & 0* \\
		& $RDB00$ & radial derivative of $B_{00}$ (see text) & $\sim 0.2$* \\
		\hline
	\end{tabular}
	\label{tabTargets}
\end{table*}
%*******************************
%*******************************		
		
In the cost function  \ref{eqCostFunction}, we include the targets  listed in Table \ref{tabTargets}. First, we use several general targets, such as  
aspect ratio, $A$, average $\beta$, and the toroidal current, $I_t$, which require a unique evaluation for each configuration. 
We also use a set of targets evaluated at several radial positions for each configuration: the magnetic well, $W$ (as defined in \cite{greene_brief_1998}), the rotational transform, $\iotab$, and the effective ripple, $\epsilon_{eff}$. 
In addition to these ``standard" targets we have defined a set of proxies specifically related to  the confinement of energetic ions and the quasi-isodynamicity of the magnetic field, which we present next.

		%*******************************
		%*******************************
		\subsection{New proxies for fast ion confinement}\label{secNewProxies}
		%*******************************
		%*******************************

The well-known $\Gamma_c$ proxy \cite{nemov_poloidal_2008} is defined as 
%*******************************
\begin{equation}
\Gamma_c(s) = \frac{\pi}{4\sqrt{2}}\left< \int _{B_{min}^{-1}}^{B^{-1}} d\lambda \frac{B}{\sqrt{1-\lambda B}}(\gamma_c^*)^2\right>,
\end{equation}
%*******************************
with 
%*******************************
\begin{equation}
\gamma_c^*(s) = \frac{2}{\pi}\arctan \frac{\partial_{\alpha}J}{\partial_s J}=\frac{2}{\pi}\frac{\overline{ \mathbf{v_M} \cdot \nabla s}}{|\overline{ \mathbf{v_M} \cdot \nabla \alpha}|},
\end{equation}
%*******************************
where $s$ is the VMEC radial coordinate (normalized toroidal flux), $\alpha$ is the field line label, $\alpha=\theta-\iotab \zeta$, with $\theta$ and $\zeta$ the generalized poloidal and toroidal angle magnetic coordinates, respectively; $\lambda = {v_{\perp}^2}/{Bv^2}$ is the pitch angle, with $v_{\perp}$ the component of the velocity $v$ perpendicular to the magnetic field, and ${B}$ the magnetic field modulus; $J$ is the second adiabatic invariant, $\mathbf{v_M}$ is the magnetic drift velocity, the overline $\overline{\dots}$ means orbit average and $\left<\dots\right>$ means flux surface average. The proxy $\Gamma_c$ has already been used in previous works for the optimization of fast ion confinement \cite{bader_advancing_2020}.
 In addition, the newly defined $\Gamma_{\alpha}$ proxy \cite{velasco_model_2021} 
%*******************************
\begin{multline}
%\begin{equation}
%	\nonumber
	\Gamma_{\alpha}(s) = \frac{1}{2} \left< \int _{B_{min}^{-1}}^{B^{-1}} d\lambda\frac{B}{\sqrt{1-\lambda B}} \right. \\ \left.H((\alpha_{out}-\alpha)\overline{v_M\cdot\nabla\alpha})
	H((\alpha-\alpha_{in})\overline{v_M\cdot\nabla\alpha}) 	\right>
	\label{GammaAlphaFormula}
%\end{equation}
\end{multline}
%*******************************
 is also used in this work. In Eq. (\ref{GammaAlphaFormula}), $H$ is the Heavyside function. 
Note that this new proxy is a refinement of the well-known proxy $\Gamma_c$, as demonstrated in \cite{velasco_model_2021} for W7-X configurations. $\Gamma_{\alpha}$ estimates  the fraction of ions born on a flux surface that are lost by interpreting the distribution in phase space of the quantity $\gamma_c^*$, instead of taking its average. We refer the reader to \cite{velasco_model_2021} for the details of the derivation of  $\Gamma_{\alpha}$  and the demonstration of its performance.
$\Gamma_{\alpha}$ has been used in this work together with $\Gamma_c$ for a fast evaluation of the energetic ion confinement, which is then confirmed with ASCOT guiding-center calculations.
We have implemented the proxies $\Gamma_{c}$ and $\Gamma_{\alpha}$ in STELLOPT \cite{ciematBranch}. 

In addition to these proxies for fast ion confinement, a set of proxies explicitly related to the deviation from quasi-isodynamicity has been implemented.
 First, two proxies related to the alignment of magnetic field strength extrema: VBT (Variance of $B$ at the Top) and VBB (Variance of $B$ at the Bottom) are defined as the variance of the magnetic field maxima and minima, respectively, on a flux surface. For  VBT, two slightly different versions are considered. The first one is defined as
%*******************************
\begin{equation}
 	VBM(s) = \frac{\sum _{i} (B_{max}(s, \theta_i) - {B_{M}(s)})^2}{B_{00}(s)^2N_{\theta}},
\end{equation}
%*******************************
where 
$B_{max} (s, \theta_i)$ is the maximum value of $B$ at the flux surface $s$ for poloidal angle $\theta_i$, ${B_{M}}(s)$ is the mean value of $B_{max} (s, \theta_i)$ maxima, $B_{00}(s)=B(s)$ is the $m=0$, $n=0$ component of the magnetic field (in Boozer coordinates) for the flux surface at s, and $N_{\theta}$ is the resolution in the $\theta$ angle coordinate. 
Based on the property that for a perfectly quasi-isodynmic configuration the $B_{max} (s, \theta_i)$ maxima are aligned at $\zeta=0$ \cite{cary_omnigenity_1997}, an alternative form of the proxy VBT, called VB0, has been defined as
%*******************************
\begin{equation}
VB0(s) = \frac{\sum _{i} (B (s, \zeta=0, \theta_i) - {B_{M0}}(s))^2}{B_{00}(s)^2N_{\theta}},
\end{equation}
%*******************************
which is similar to VBM, but assumes that the maximum values of $B$ are aligned along the line  $\zeta=0$. Here ${B_{M0}}(s)$ is the mean  value of $B (s, \zeta=0, \theta_i)$. We have used both VBM and VB0 without a noticeable difference in results, and we refer generically to VBT without specifying which version is used in each case.
 
The proxy VBB is defined as
%*******************************
\begin{equation}
	VBB(s) =  \frac{\sum _{i} (B_{min}(s, \theta_i) - {B_{m}}(s))^2}{B_{00}(s)^2N_{\theta}},
\end{equation}
%*******************************
with $B_{min} (s, \theta_i)$  the minimum value of $B$ at the flux surface $s$ for poloidal angle $\theta_i$ and ${B_{m}}(s)$ the mean value of $B_{min} (s, \theta_i)$ minima.
These minima  are not assumed to be located along a specific curve known a priori, but the minimum value along each constant-$\theta$ line, for any $\zeta$, are considered in the calculation of the variance, as for the VBM proxy. The target value for VBT and VBB proxies is set to zero  to reduce the variance of both extrema. 
Proxies similar to VBT and VBB have already been used in ROSE \cite{drevlak_optimisation_2019}.

A fourth proxy, WBW, related to the width of the magnetic field central valley (in Boozer coordinates) is defined as 
%*******************************
\begin{equation}
	WBW(s) =  \frac{\sum _{ij} B(s, \zeta_i, \theta_j) \zeta_i^ 2}{4\pi^2 B_{00}(s)N_{\theta}N_{\zeta}},
\end{equation}
%*******************************
with $N_{\zeta}$ the resolution in the $\zeta$ angle. 
The target for this proxy is set to zero to increase the width of the central valley of the magnetic field strength on a flux surface (reduce the width of the region of maxima of B at $\zeta=0$). The beneficial effect of a wide valley in the magnetic field strength at the center of the period was already discussed  in \cite{drevlak_optimisation_2019}.

In addition to these proxies, a function has been defined in STELLOPT for the calculation of the radial derivative of the $B_{00}$ component (in Boozer coordinates) of the magnetic field, which accounts for the diamagnetic effect, particularly relevant in finite-$\beta$ equilibria. Namely,
%***************************************
\begin{equation}
	\centering
		RDB00 = \frac{1}{B_0}\frac{dB_{00}}{ds},
		\label{eq:DefnRdB00Proxie}
\end{equation}
%***************************************
with $B_0$ the average magnetic field at the magnetic axis.
The optimization goal   is to increase the value of this radial derivative as much as possible, in general. Then, the target value for this proxy is set positive and large ($\sim 0.2$).

All the proxies described in this section are radially local quantities evaluated at several radial positions. 
The target values used  for the properties optimized in this work are listed in Table \ref{tabTargets}. Note that not all the targets were used in all the stages of the optimization process.  The values included in the table are those used in the final phase of the optimization.

	%***************************************************************************************
	\subsection{Optimization strategy}
	%***************************************************************************************
	{Our goal is to find a QI configuration   with moderate aspect ratio, around $A \sim 10$,   and good confinement of both thermal and energetic ions. We seek a configuration with a rotational transform profile with low (positive) magnetic shear, with an island at the edge that allows an island divertor, and avoiding low order rationals in the confined region. 
With all the proxies previously described included in the code and their target values set, the optimization loop is followed trying to obtain a configuration with reduced deviation from quasi-isodynamicity, which we could expect to be translated into reduced transport of energetic ions, also enforced by the specific proxies $\Gamma_c$ and $\Gamma_{\alpha}$.

We start with two magnetic configurations from W7-X, the standard configuration (STD) and a high mirror configuration (HM), scale \footnote{The scaling is done manually by scaling the boundary coefficients RBCs and ZBSs in the VMEC input consistently with the PHIEDGE input parameter.} them to the target aspect ratio  and modify the number of periods  in the VMEC input.  
After scaling and setting the number of periods of the configurations, some basic properties, such as the rotational transform profile, the magnetic well, and the effective ripple change in the resulting VMEC equilibria. In particular, the effective ripple was significantly worse (larger) than that of the starting configuration.

Starting with these configurations, we optimize for neoclassical (NC) transport and MHD stability with STELLOPT including targets for  $\epsilon_{eff}$,  
 defined at several (3 to 7) radial positions. We fix the rotational transform profile  in the VMEC input. 
 We  work with vacuum configurations in this process and allow the variation of all the RBC and ZBS boundary coefficients.
After this initial optimization, we select a configuration with four periods, the same rotational transform profile as that of the standard W7-X configuration, and larger magnetic well than the initial configurations (STD and HM). The effective ripple, $\epsilon_{eff}$,  was also significantly improved  with respect to the initial configurations.

With this four-period optimized configuration, we start the optimization of fast ion confinement. We work with finite $\beta$ equilibria, which facilitates the optimization, by setting a quadratic pressure profile in the VMEC input (AM input) and defining a target for $\beta=1.5\%$. We allow the variation of $\iotab$ and define targets for it, {such that it is is close to (and below) 1 at the edge and avoid low order rationals in the plasma}.  
$W$  and $\epsilon_{eff}$, and also for the proxies related to the fast ion confinement and the quasi-isodynamicity, 
 
 $\Gamma_c$, $\Gamma_{\alpha}$, VB0, VBM, VBB, WBW, and RDB00,   as listed in table  \ref{tabTargets}. The properties and proxies with radial dependence are evaluated in several (3 to 7) radial positions.

  Including all these ingredients, finding a global minimum of the cost function can be difficult. 
 If a large number of modes describing the plasma boundary vary in the optimization process the function to minimize becomes  very complicated, and the optimizer can get stuck in local minima. The problem would be difficult even for a global optimization algorithm such as GADE. To overcome these difficulties, we follow a multi-stage optimization strategy, similar to that in \cite{landreman_calculation_2021}. We  start with a small number of Fourier modes of the plasma boundary  (RBCs and ZBSs coefficients in VMEC input) that are allowed to vary in the optimization, and increase this number progressively. In a first stage, only modes of the boundary with $ |n| \leq 2$ and $0 \leq m \leq 2$ are allowed to vary. Note that, although the modes of the boundary that are allowed to vary are limited, the boundary definition, coming from previous stages, contains modes with larger mode numbers and also the magnetic field still shows a spectrum containing modes with $|n|$  and $m$ larger than 2. Note that the description of the boundary used in VMEC is not optimal \cite{henneberg_representing_2021}.
 
 Once the variables and targets are fixed, the cost function is fully defined when the tolerances of the targets, $\sigma_i$  in equation (\ref{eqCostFunction}), are set.  We follow an iterative process in which, starting from the same initial configuration, we define different sets of tolerances for the targets, and run an optimization loop with each set. %The target values are refined in the process, approaching those in Table \ref{tabTargets}. 
 Among the new configurations obtained, we select the best one according to its closeness to the target values. With this configuration, we repeat he process until no further improvement can be obtained for the number of modes selected\footnote{The final results in an optimization process using LMDIFF depends strongly on the starting point and the optimization function, which depends on the tolerances set for the targets. Then, strictly speaking, we cannot say that this configuration is the best possible.}. 
  Then, we increase this number to
  allow the variation of modes with $|n| \leq 3$ and $0 \leq m \leq 3$, and repeat the process. We continue the loop until $|n| \leq 7$ and $0 \leq m \leq 7$, which does not produce significant changes with respect to the previous step.

With this strategy, we have obtained a configuration that is very close to omnigeneity, as we will show, and exhibits also good fast ion confinement, while keeping MHD stability and reduced $\epsilon_{eff}$. 
 In the next section, we describe the magnetic configuration obtained following this procedure and  compare it with the standard and high mirror configurations of W7-X.

 %
 %
 %***************************************************************************************
 %***************************************************************************************
 \section{Optimization results}\label{secConfigProps}
 %***************************************************************************************
 %***************************************************************************************

In this section, we discuss in detail the properties of the final optimized configuration, with particular emphasis on the ones considered in the optimization. {We start analyzing the magnetic field structure in section \ref{secGenProps}. Then, we  analyze the MHD stability and neoclassical transport in section \ref{secStandardMetrics}, and the values of the new proxies defined for fast ion confinement in \ref{FIMetrics}. The goodness of the fast ion confinement predicted by the values of $\Gamma_c$ and $\Gamma_{\alpha}$ proxies is confirmed by guiding-center calculations with ASCOT in section \ref{secASCOT}. The bootstrap current, which is not considered in the optimization loop, is discussed separately in section \ref{secBootstrap}.}

			%***************************************************************************************
		%***************************************************************************************
		\subsection{General properties of the optimized configuration}\label{secGenProps}
		%***************************************************************************************
		%***************************************************************************************
%*************************************	

The optimized configuration obtained has four periods, is relatively compact, with $A=9.94$, and has a magnetic field close to QI. 	
The magnetic field strength for this configuration, with  
$ \beta=1.5  \%$, at the flux surface with $s=1$ is shown in figure~\ref{ModBLastCFS1714}-left.
	%*************************************
	\begin{figure*}
		\centering
		\includegraphics[draft=false,  trim=20 10 20 10, clip, height=5.25cm]{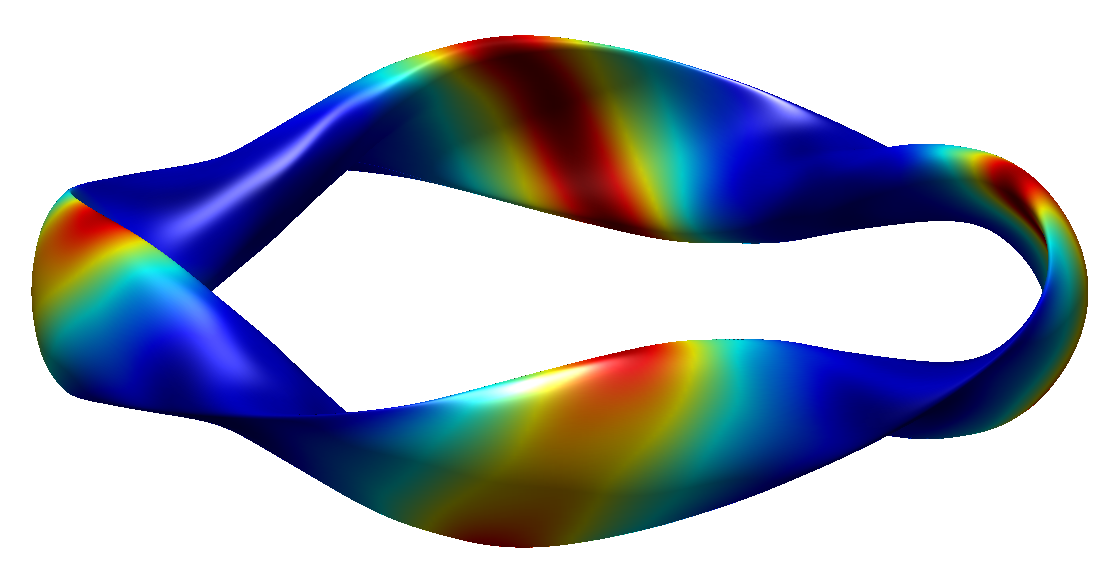}
		\raisebox{-0.15\height}{
			\includegraphics[draft=false,  trim=20 0 60 0, clip, width=5cm]{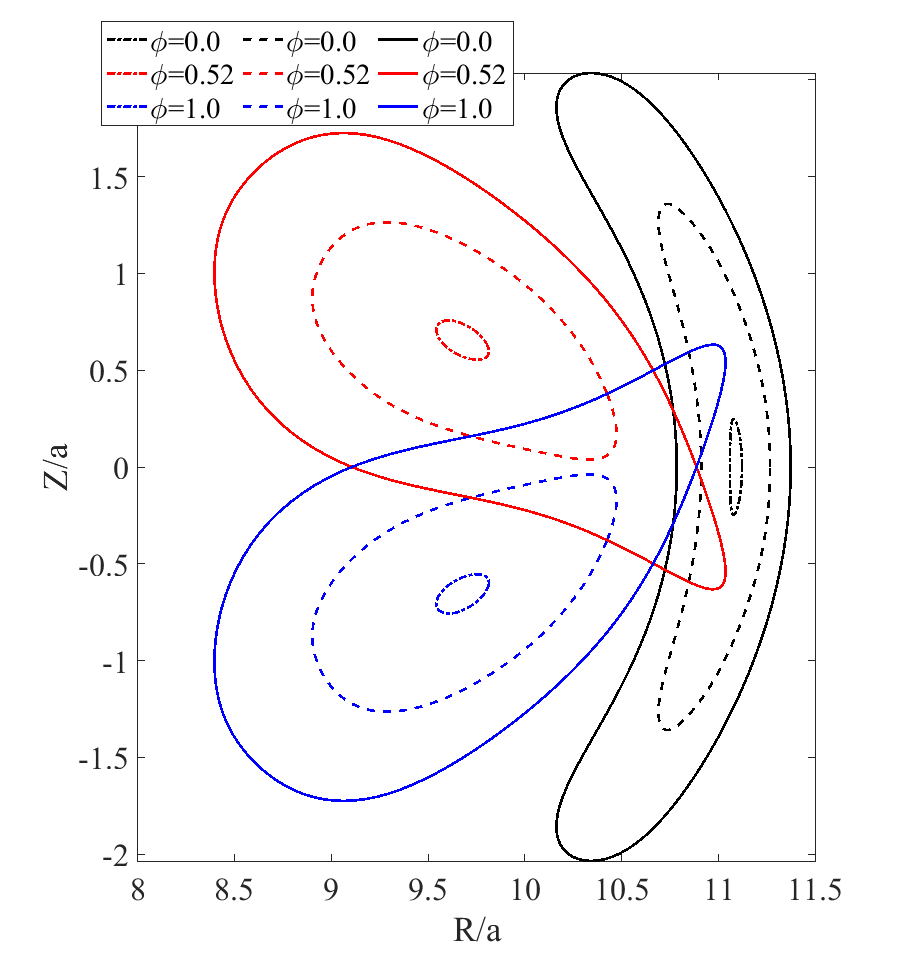}
		}
		\caption{Left: Magnetic field strength at the last closed flux surface for the optimized configuration, with  %$A=9.94$ and 
			{$ \beta =1.5\%$}. Right: Poincaré plots at toroidal angles $\phi=0,0.52,1$ for the optimized configuration  at $s=0.1,0.5,1$. $R$ and $Z$ are normalized with the minor radius, $a$.} 
	\label{ModBLastCFS1714}
	\end{figure*}
	%*************************************
	%*************************************
	\begin{figure*}
		\centering

		\includegraphics[draft=false,  trim=60 126 60 55, clip, width=6.15cm] {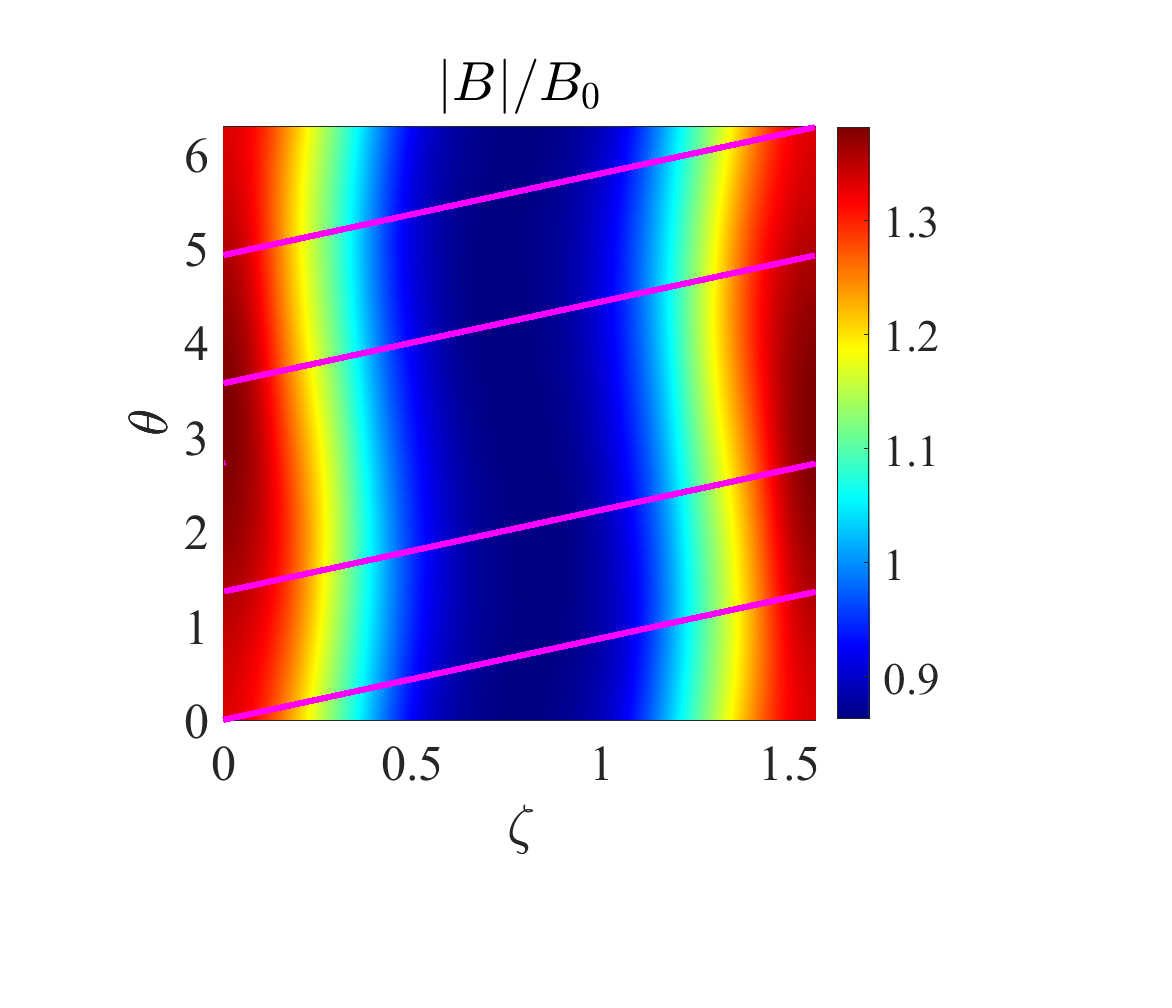} 
		\includegraphics[draft=false,  trim=10 73 10 10, clip, width=9.cm] {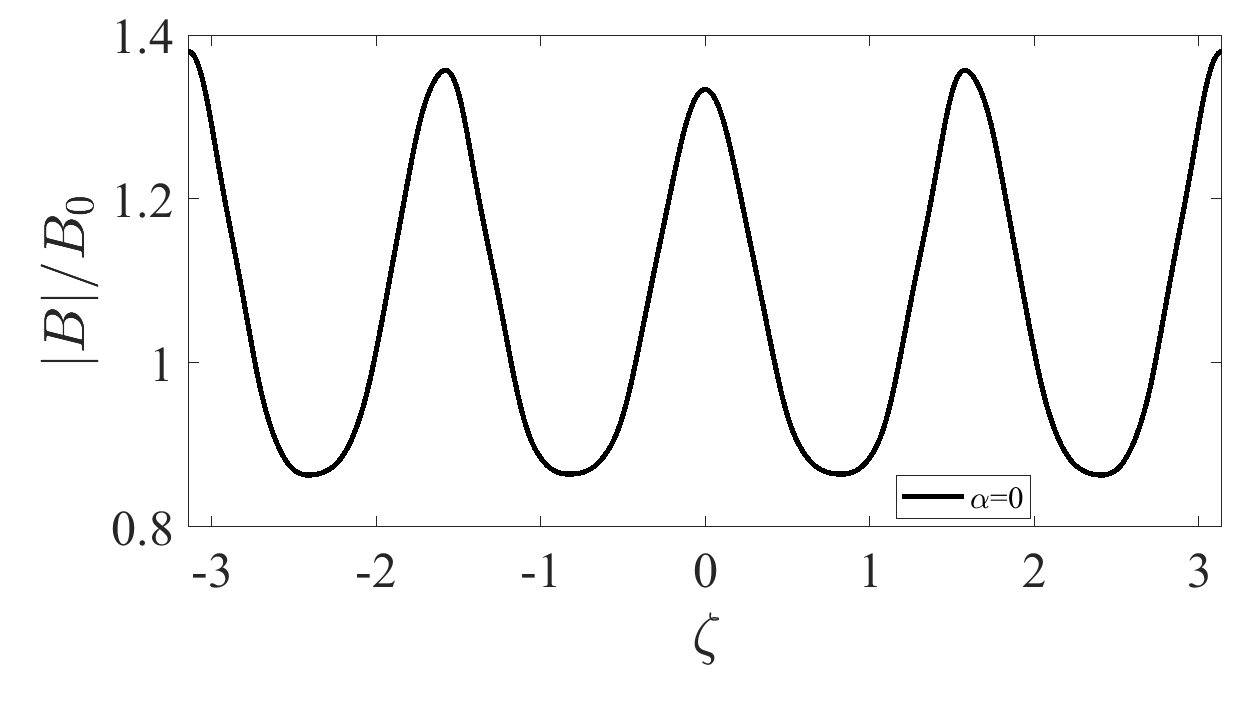} \\
		\includegraphics[draft=false,  trim=60 126 60 55, clip, width=6.15cm] {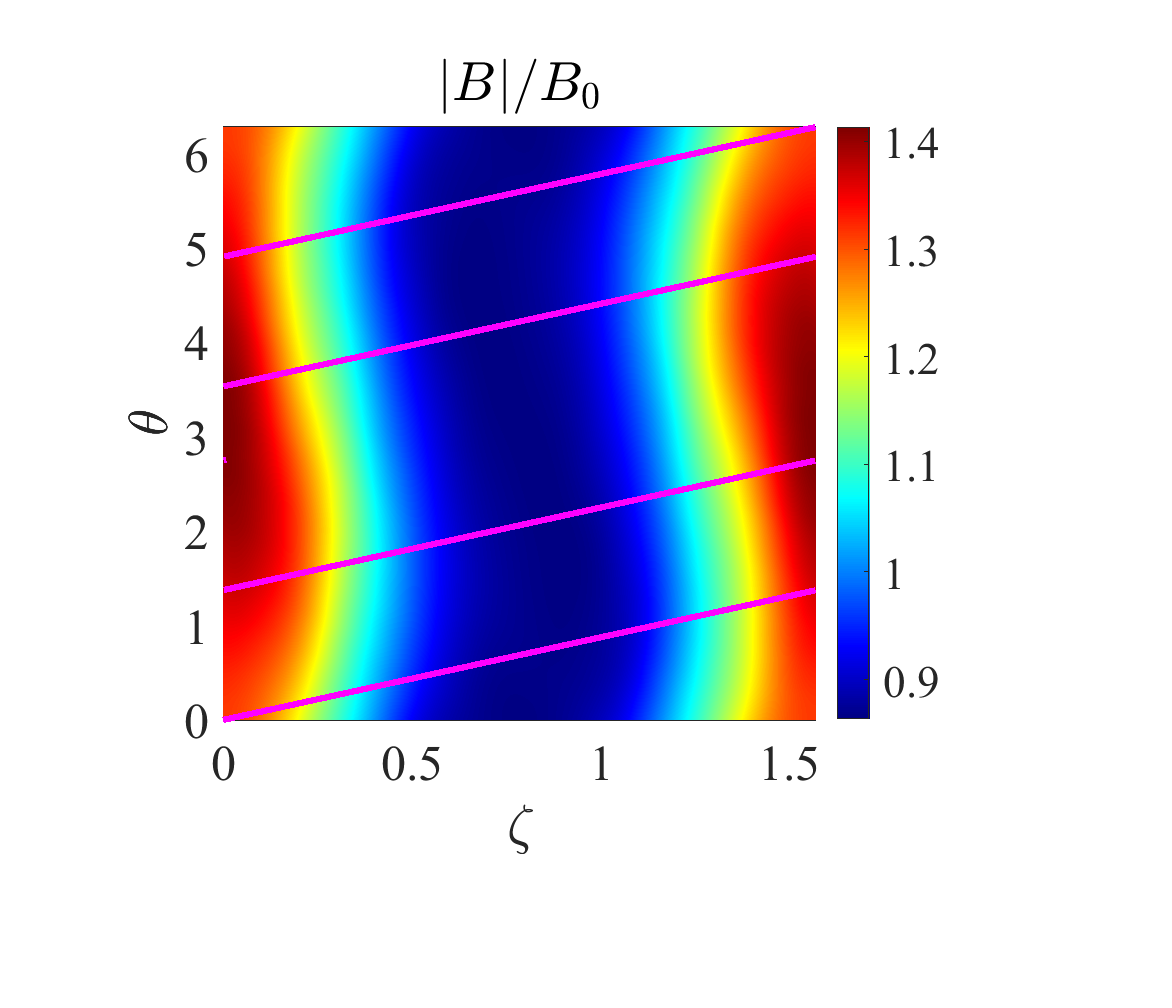} 
		\includegraphics[draft=false,  trim=10 73 10 10, clip, width=9.cm] {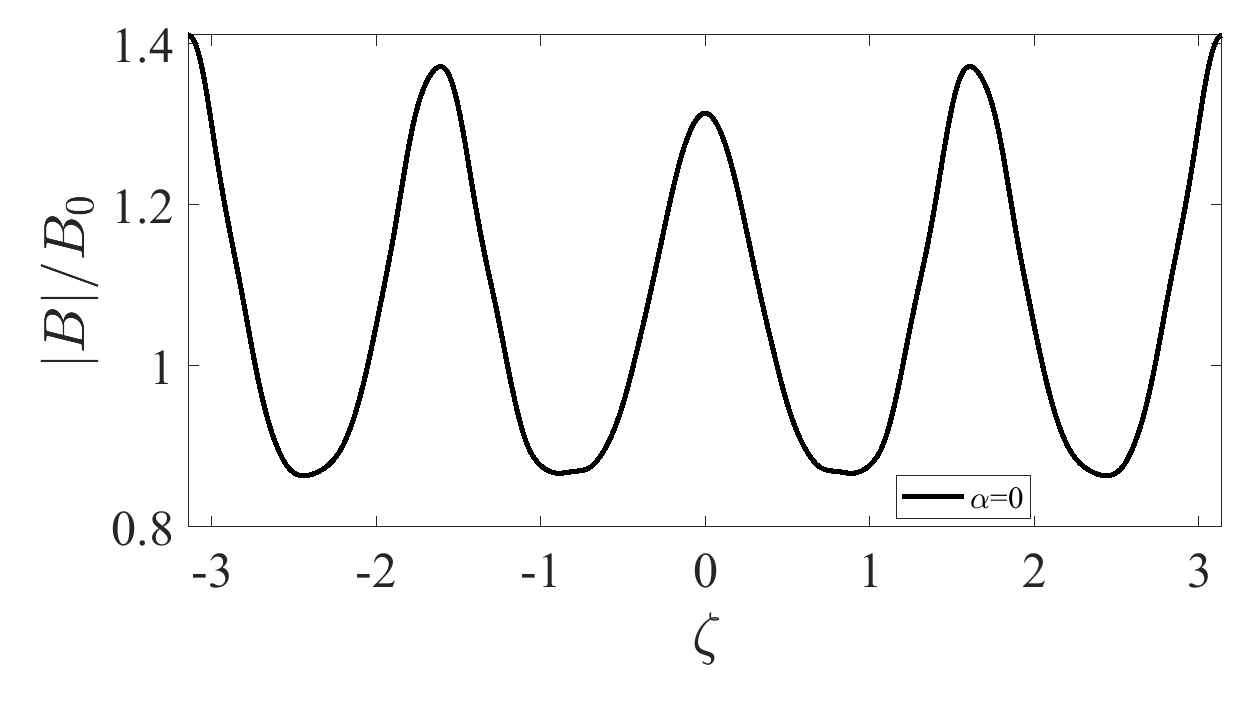} \\
		\includegraphics[draft=false,  trim=60 70 60 55, clip, width=6.15cm] {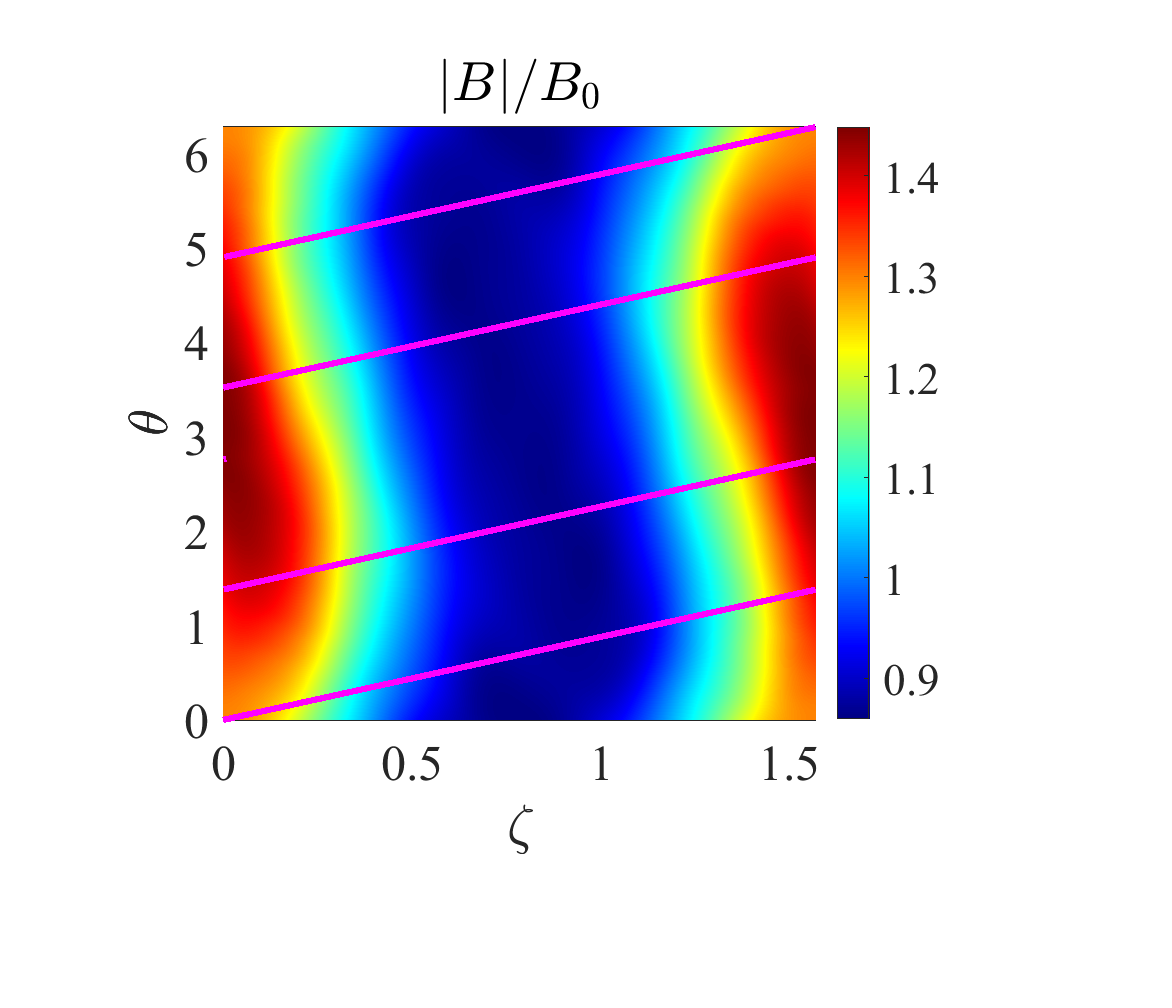} 
		\includegraphics[draft=false,  trim=10 18 10 10, clip,  width=9.cm] {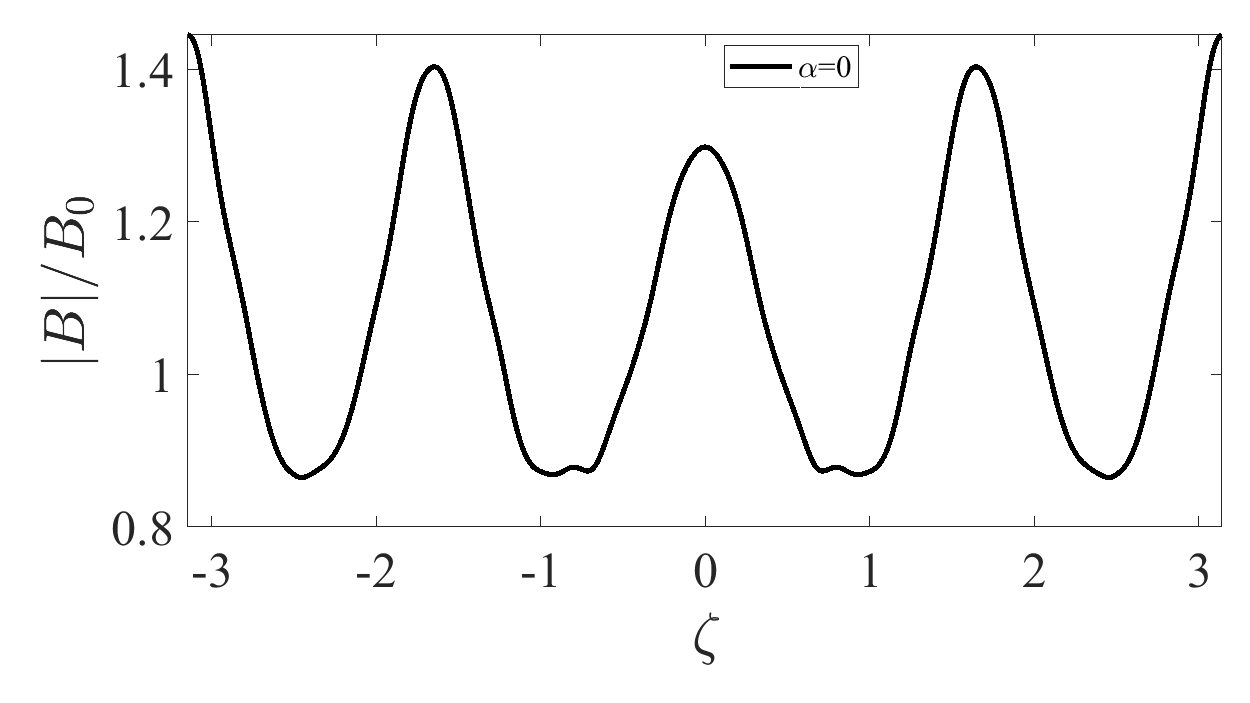} \\		
		\caption{Magnetic field strength (normalized with  $B_0$) versus Boozer angles coordinates $\zeta$ and $\theta$ at three radial positions, $r/a=0.25$ (top), $0.5$ (middle), and $0.75$ (bottom) for the optimized configuration, with $\beta=1.5\%$ (left) and magnetic field strength along the field line starting at $(\theta, \zeta)=(0,0)$ point for the same radial positions (right). The field line with $\alpha=0$ is shown in pink. Note that we map onto the first period segments of the field line
			lying on other periods.}
		\label{fig:BfieldAlingnement}
	\end{figure*}
	%*************************************

The QI structure (with magnetic field strength contours closing poloidally) can be readily recognized in this plot even at $s=1$. Note that in QI configurations, such as those of W7-X, the magnetic field strength is usually closer to have poloidally-closed contours near the magnetic axis than near the plasma edge. 
	
Figure \ref{ModBLastCFS1714}-right shows Poincaré plots at several toroidal angles for the optimized configuration. {Cuts for flux surfaces at $s=0.1,0.5,1$} are shown for toroidal angles $\phi=0,0.52,1$. The flux surface has a highly elongated section at toroidal angle $\phi=0$, 
which has been associated to drawbacks, such as a reduction of the plasma volume \cite{mata_direct_2022} or an increase of the coil complexity \cite{hudson_differentiating_2018}, 
in QI configurations. However, as we will see in section \ref{secCoils}, relatively simple coils can be found for this configuration.

We show some more details about the magnetic field structure in figure  \ref{fig:BfieldAlingnement}. The left panels in this figure show the (normalized to $B_0$) magnetic field strength versus Boozer angle coordinates $\zeta$ and $\theta$ for three magnetic flux surfaces at $r/a=0.25, 0.5,0.75$. 
The quasi-isodynamic structure of the magnetic field, with contours of a constant magnetic field closing poloidally, is readily seen in this figure. The magnetic field is closer to quasi-isodynamic near the magnetic axis (top panel), while near the outer boundary the deviation from quasi-isodynamicity increases.

The field line with $\alpha=0$  
is shown in pink. Note that we map onto the first period segments of the field line
lying on other periods. 
In the right panels of this figure, the magnetic field strength along the field line  $\alpha=0$  is shown. At $r/a=0.25$, the alignment of the magnetic field minima and maxima, as expected for a magnetic configuration close to omnigenous, is clear. The alignment is also significantly better in the inner radial positions. For $r/a=0.75$ the alignment of maxima and minima is significantly worse than for inner positions, indicating a larger deviation from omnigeneity. Getting good alignment in all radial positions is very difficult; then, tighter tolerances were put at the inner ones, where most of the fast particles are generated in a reactor. 
It can also be seen in the figure that the alignment of minima is better than that of the maxima, which is more relevant for fast ion confinement, as we will discuss later. In LHD, the alignment of the magnetic field minima produces a significant reduction in the effective ripple and in the fast ion losses \cite{murakami_neoclassical_2002}. 
Another noticeable feature in figure \ref{fig:BfieldAlingnement} is that the magnetic  valleys  are wider than the regions of maxima of B.

	%*************************************
	\begin{figure}
		\centering
		\includegraphics[draft=false,  trim=20 24 60 0, clip, width=8.25cm]{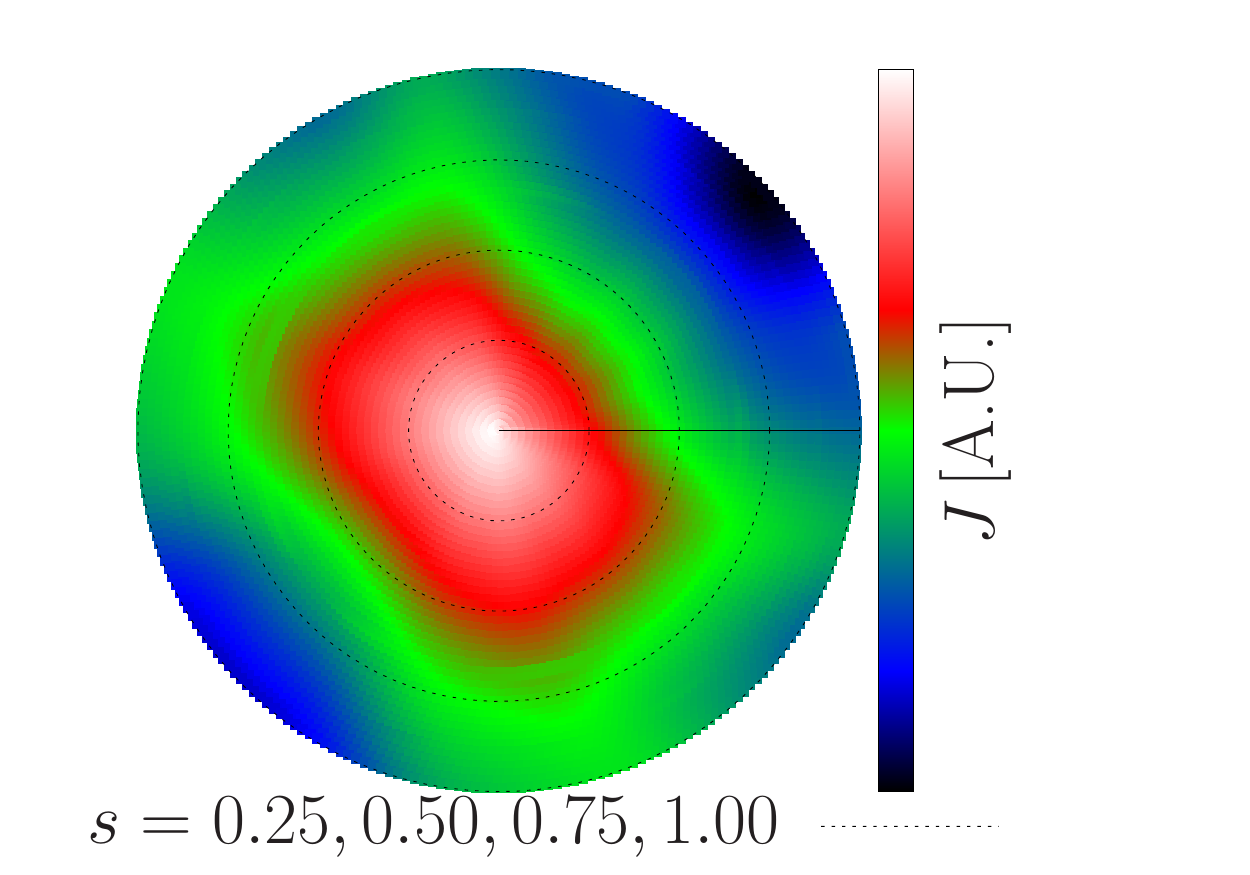}\\
		\includegraphics[draft=false,  trim=20 24 60 0, clip, width=8.25cm]{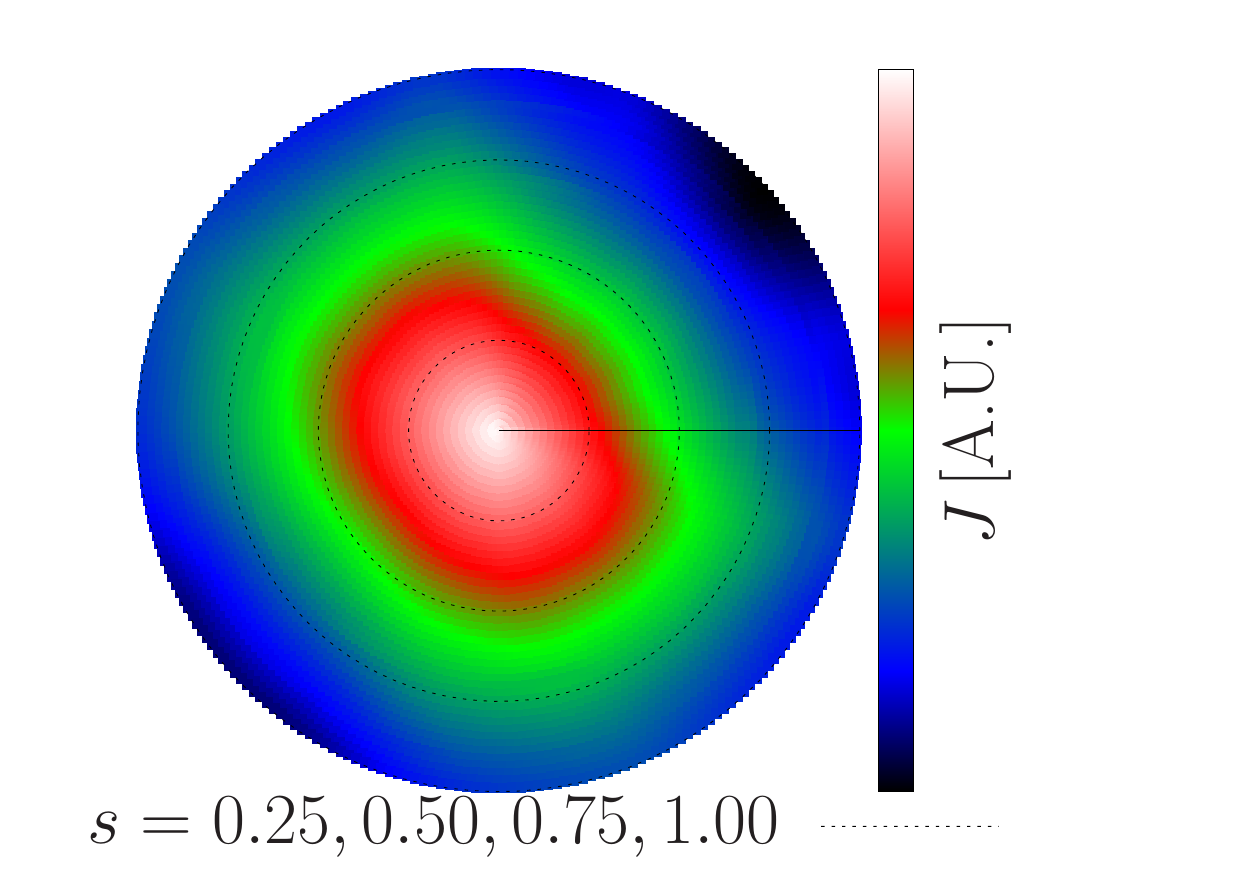}\\
		\includegraphics[draft=false,  trim=20 24 60 0, clip, width=8.25cm]{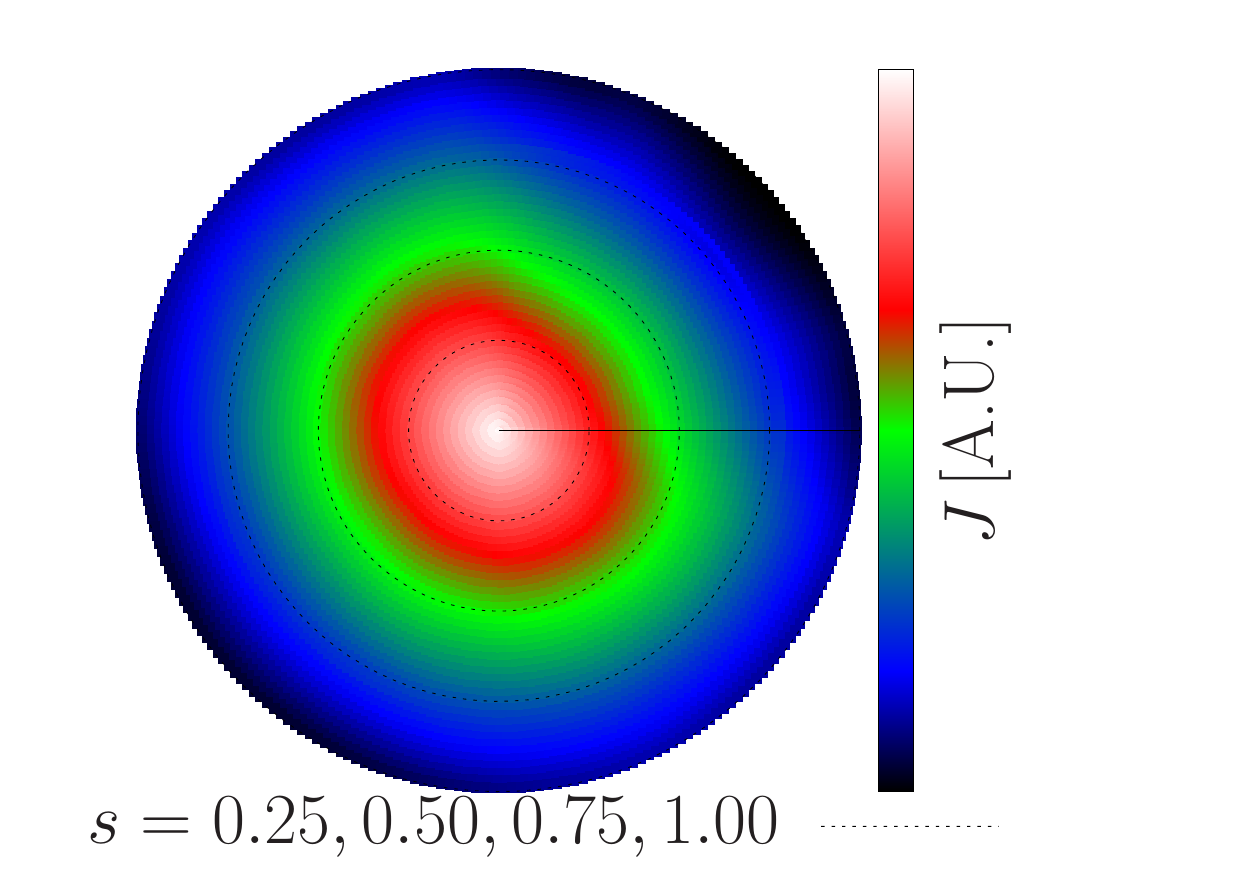}\\
		\caption{Maps of the second adiabatic invariant, $J$, in polar coordinates for the optimized configuration for $\beta=0.5\%$ (top), $\beta=1.5\%$ (middle) and  $\beta= 4.0\%$ (bottom). The radial and angular coordinates are the normalized toroidal flux, $s$, and the field line label $\alpha$, respectively. Constant radius  contours are shown for several radial positions $s=0.25, 0.5, 0.75, 1$ with dashed lines. The $J$ contours correspond to $\lambda \sim 1/B_{00}$.
		}
		\label{fig:JContours}
	\end{figure}
	%*************************************
	
{To complete the analysis of the magnetic field structure, we now look at the second adiabatic invariant, $J$,  whose contours,  for a particular value of} {the pitch-angle velocity $\lambda\approx 1/B_{00}$} {and several values of {$\beta$}, {$0.5\%,1.5\%,$ and $4.0\%$}, are shown in figure \ref{fig:JContours}. We can readily observe that {the contours of $J$ (especially those closer to the axis) are well aligned with the flux-surfaces. Furthermore,} $J$ is maximum at the center, $s=0$, and decreases toward the edge (maximum-$J$ property). As $\beta$ increases, the constant-$J$ contours approach concentric circles, such that $|\partial_{\alpha} J|$ decreases as well as $\partial_{s} J$ becomes more negative  in a wide range of $\alpha$ values, which positively impacts the confinement of trapped particles, as we will see next. {We note that circular contours of $J$ is a sufficient but not necessary condition for good fast ion confinement; closed contours of $J$ (i.e. contours that do not intersect $s=1$) is sufficient}. We can put the relevance of this result in perspective by comparing these plots with the equivalent ones for other devices. In other QI devices, 
the contours of J are not very close to concentric circles for $\beta \sim4\%$, and open contours are typically found for $\beta < 3\%$, see e.g. \cite{faustin_fast_2016,velasco_model_2021,subbotin_integrated_2006}.  }

	%*************************************
	\begin{figure}
		\centering
		\includegraphics[draft=false,  trim=15 67 30 10, clip, width=8.5cm]{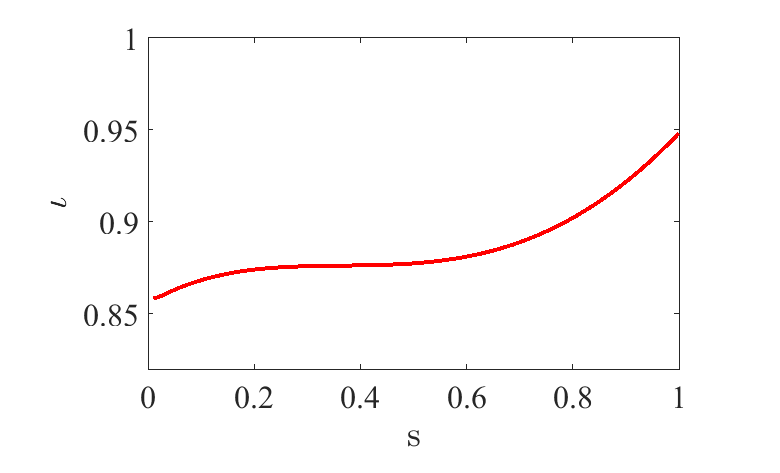}
		\caption{ Rotational transform  profile   versus radial position  for the optimized configuration, with   $\beta=1.5\%$.
		}
		\label{fig:iotaProfile}
	\end{figure}
	%*************************************
Figure \ref{fig:iotaProfile} shows the radial profile of the rotational transform for the optimized configuration with $\beta=1.5\%$. The global magnetic shear is small but positive and $0.85 < \iotab < 0.95$,  thus avoiding low order rationals and allowing a 4/4 island at the edge.
%***************************************************************************************
%***************************************************************************************
%\subsection{Standard metrics}\label{secStandardMetrics}
\subsection{MHD and neoclassical transport}\label{secStandardMetrics}
%***************************************************************************************
%***************************************************************************************

Now we analyze the MHD stability and neoclassical transport in this configuration through the standard metrics $W$ and $\epsilon_{eff}$.
Figure \ref{fig:mwYrippleProfile}-top shows the radial profile of magnetic well (including the finite $\beta$ plasma effect \cite{Freidberg1987, greene_brief_1998})  for the optimized configuration at $\beta=1.5\%$ compared with those  for the W7-X reference configurations at  $\beta=1.5\%$ and $\beta=0$. An increase of $W$ with $\beta$ (even above the target value) is observed in the W7-X configurations, which is
also obtained (not shown) in the optimized one. 
The large values of the magnetic well in all the radial domain ensure ideal MHD stability. Ballooning stability has also been studied with the code COBRA, up to a value of $ \beta =5\%$ using parabolic pressure profiles, showing that the configuration is ballooning stable for these pressure values.
	%*************************************
	\begin{figure}
		\centering
		\includegraphics[draft=false,  trim=15 67 30 10, clip, width=8.5cm]{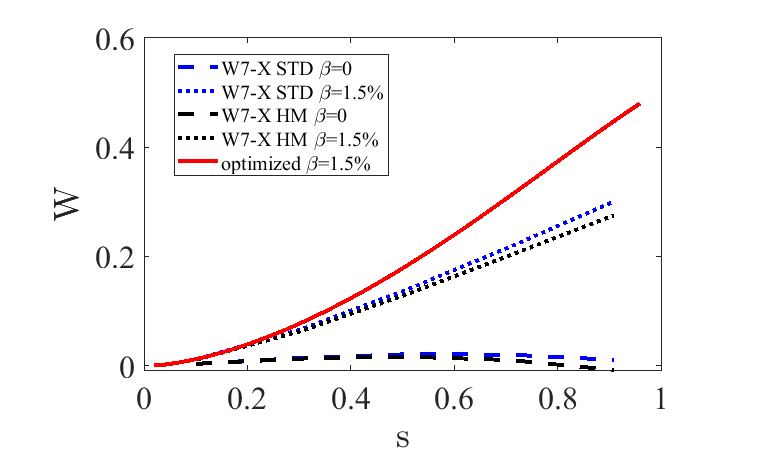}\\
		\includegraphics[draft=false,  trim=15 15 30 10, clip, width=8.5cm]{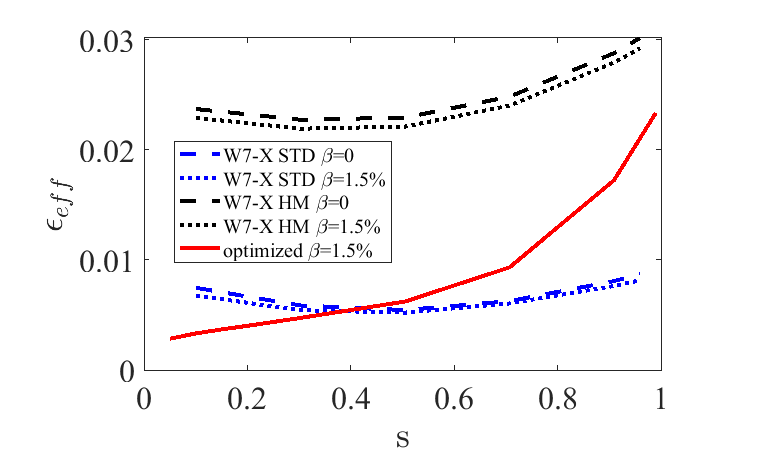}
		\caption{ Magnetic well, as defined in STELLOPT\cite{Freidberg1987, greene_brief_1998} (top), and effective ripple (bottom) profiles   versus radial position  for the optimized configuration, with   $\beta=1.5\%$ and the W7-X reference configurations at $\beta=0, 1.5\%$.}
		\label{fig:mwYrippleProfile}
	\end{figure}
	%*************************************

In figure \ref{fig:mwYrippleProfile}-bottom we show the effective ripple, $\epsilon_{eff}$\cite{nemov_b_2005} versus radial position for the optimized configuration and compare it with those for the W7-X configurations for $\beta=0$ and $\beta=1.5\%$. 
In the optimized configuration, the effective ripple is %smaller than for the high mirror configuration of W7-X. It is
 below $\epsilon_{eff} = 2\%$ in almost all the plasma radius and $\epsilon _{eff} < 0.5 \%$ in the plasma core $r/a < 0.5$, which is considered sufficient for a stellarator reactor \cite{Beidler2021,Alonso2022}. At the  plasma edge $\epsilon_{eff}$ increases significantly,  which is not very important, however, because the high collisionality at the edge makes the neoclassical transport less relevant in this region \cite{Dinklage2013}.

%***************************************************************************************
%***************************************************************************************
\subsection{{QI and fast ion confinement metrics}}\label{FIMetrics}
%***************************************************************************************
%***************************************************************************************
	%*************************************
	\begin{figure}
		\centering
		\includegraphics[draft=false,  trim=15 15 30 10, clip, width=8.5cm]{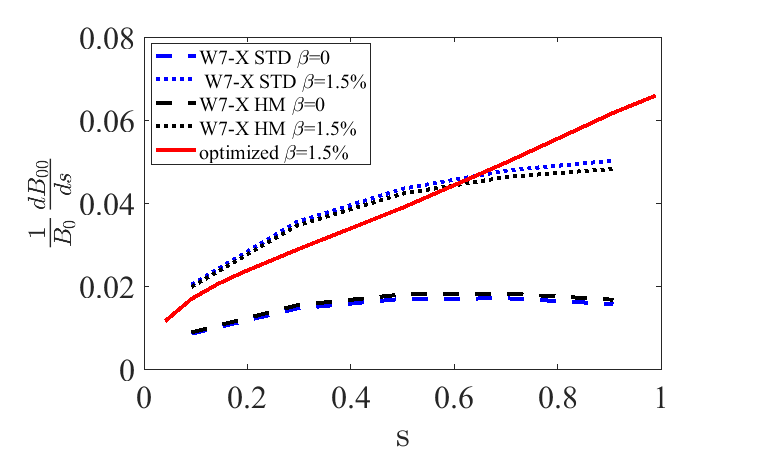}
		\caption{Values of the the radial derivative of $B_{00}$ (bottom) versus radial position for the optimized configuration, with  $\beta=1.5\%$. W7-X reference configurations are also shown for comparison.}%$B_0=5~\rm{T}$, $R=1.9~\rm{m}$,
		\label{fig:rdB00}
	\end{figure}
	%*************************************

Now, we analyze  the values of the proxies {most closely} related to quasi-isodynamicity and fast ion confinement for the optimized configuration. The quasi-isodynamicity character of the optimized configuration was already discussed in section {\ref{secGenProps}. Here we analyze the metrics defined to enforce QI in the optimization{, which contribute to improve the confinement of energetic particles, particularly the proxies $\Gamma_{\alpha}$ and $\Gamma_c$. The latter has already been used for the optimization of fast-ion confinement \cite{bader_advancing_2020}. In section \ref{secASCOT} we evaluate the confinement of energetic ions}.} 
	
Figure \ref{fig:rdB00} shows the profile of the proxy RDB00  versus the radial coordinate for the optimized configuration and the reference cases at $\beta=0 ~\rm{and}~ 1.5\%$. 
The  increase of RDB00 with $\beta$ for all the configurations is clear and reaches similar values in the optimized and reference cases at $\beta=1.5\%$.

Figure \ref{fig:VarBMinMax} shows the values of the proxies VBT and VBB versus the radial coordinate for the optimized configuration with $\beta=1.5\%$. The proxy values for the reference configurations at both $\beta=0$ and $\beta=1.5\%$  are  also shown  for comparison. 
The reduction of these metrics in the optimized configuration with respect to the reference configuration values is evident, which is consistent with the good alignment of the magnetic field extrema shown in figure \ref{fig:BfieldAlingnement}. The radial dependence is also very clear in this figure, showing quantitatively that the alignment of extrema is better in the inner region than in the outer one. Note that the values obtained for the proxy VBB are significantly smaller than those for VBT, which again is consistent with the better alignment of minima as compared to maxima in figure \ref{fig:BfieldAlingnement}.  

	%*************************************
	\begin{figure}
		\centering
		\includegraphics[draft=false,  trim=35 65 30 0, clip, width=8.5cm]{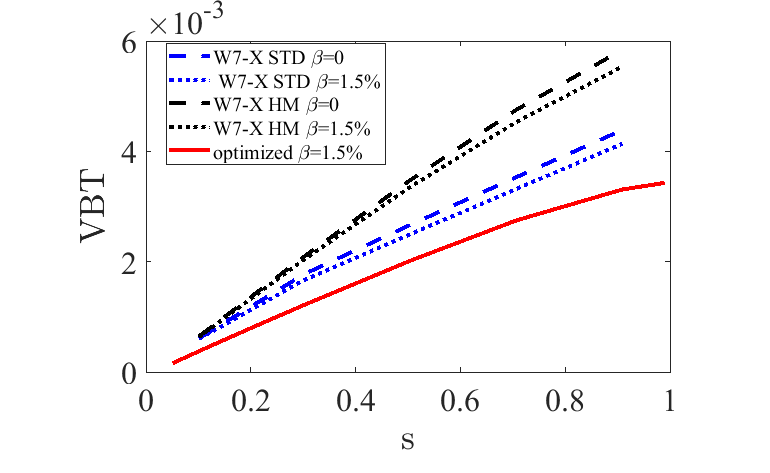}
		\includegraphics[draft=false,  trim=35 15 30 0, clip, width=8.5cm]{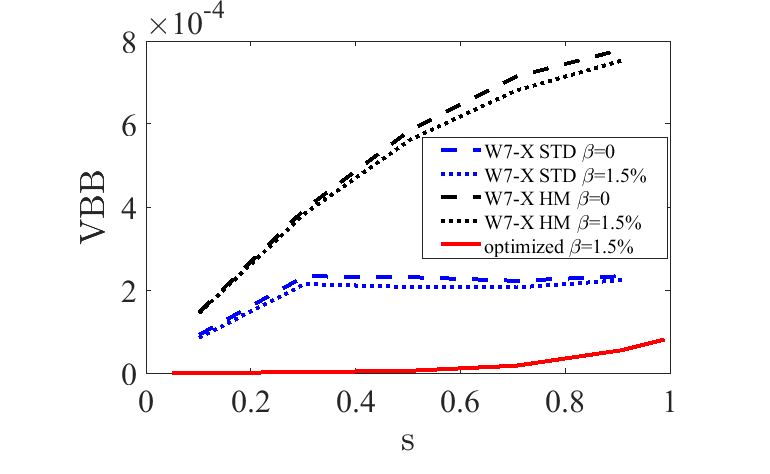}
		\caption{Values of the proxies VBM and VBB versus radial position for the optimized configuration, with   $\beta=1.5\%$ and the reference W7-X configurations.}%$B_0=5~\rm{T}$, $R=1.9~\rm{m}$,
		\label{fig:VarBMinMax}
	\end{figure}
	%*************************************
	%*************************************
	\begin{figure}
		\centering
		\includegraphics[draft=false,  trim=25 67 30 20, clip, width=8.5cm]{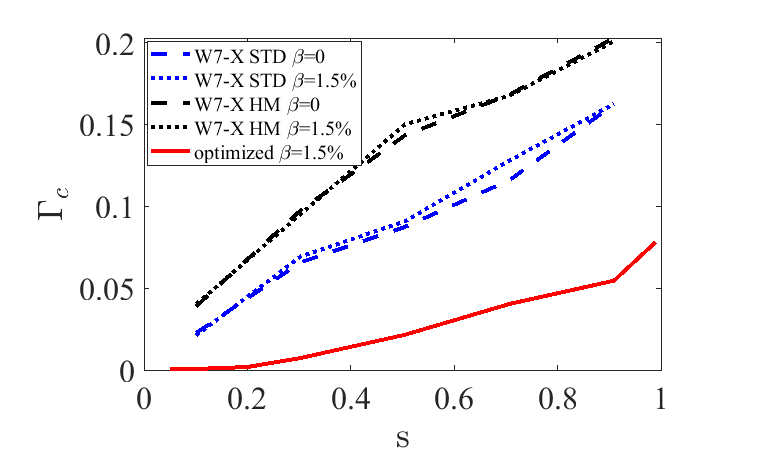}
		\includegraphics[draft=false,  trim=25 20 30 20, clip, width=8.5cm]{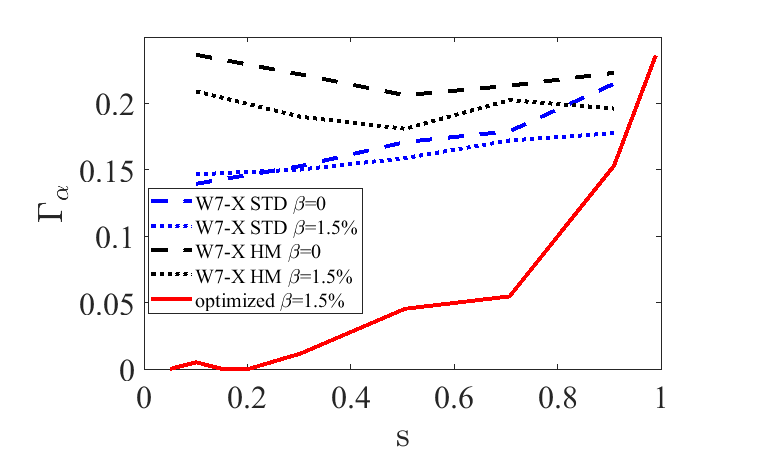}
		\caption{Values of $\Gamma_c$ and $\Gamma_{\alpha}$ proxies versus radial position for the optimized configuration, with  $\beta=1.5\%$ and the reference W7-X configurations..}%$B_0=5~\rm{T}$, $R=1.9~\rm{m}$,
		\label{fig:GammasProxies}
	\end{figure}
	%*************************************

In figure \ref{fig:GammasProxies} we show the values of the proxies $\Gamma_c$ and $\Gamma_{\alpha}$  obtained for the new configuration and compare them with the reference cases. The improvement of these metrics with $\beta$ for the reference W7-X configurations is clear.  
For the values of $\beta$ considered here, the values of the $\Gamma_c$ and $\Gamma_{\alpha}$ proxies are better (smaller) for the standard configuration of W7-X than for the high mirror one. The improvement of these proxies in the optimized configuration with respect to the reference ones is also clear. Note that $\Gamma_{\alpha}$ provides a (simplified) quantitative prediction of the fast ion prompt losses \cite{velasco_model_2021}, then, very small prompt losses of fast ions, below 1\%, can be expected in this configuration for particles born in the core, $s<0.3$. This prediction should be confirmed by more precise calculations, however, and this is what we  do in section \ref{secASCOT}, by means of guiding-center calculations with the code ASCOT. 
 
		%***************************************************************************************
		%***************************************************************************************
		\subsection{Evaluation of fast ion losses with guiding-center calculations}\label{secASCOT}
		%***************************************************************************************
		%***************************************************************************************
 In this section, we address the confinement of energetic ions in the new configuration by means of collisionless guiding-center calculations with ASCOT. 
 In order to do so, we launch energetic ions at specific positions and follow their trajectories until they are lost at the plasma boundary. 
 
 The final goal is making predictions on the confinement of alpha particles in a reactor and assessing whether they would remain in the plasma for a sufficiently long time as to thermalize and give energy to the bulk ions (slowing down time) or they would escape before reaching thermalization. Note that we do not consider collisions in the ASCOT simulations.  Specifically, we follow energetic ions with a normalized ion Larmor radius  comparable to that of alphas in a reactor.
 We use as reference reactor parameters those from \cite{Drevlak2014}, $a=2~\rm{m}$, $A=10$, $B=5~\rm{T}$, $T=14~\rm{keV}$, which give a normalized Larmor radius $\rho^*=\rho_L/a=2.7 \cdot 10^{-2}$ for alphas of energy $E=3.5~ \rm{MeV}$. 
 These fast ions are launched at fixed radial positions with uniform angular distribution over the flux surface and isotropic distribution in velocities. We calculate the fraction of ions lost after {8000 transit times, which is more than the number of toroidal turns (6000)} that alpha particles in a reactor perform in a slowing-down time. The results of these calculations are presented in figures \ref{fig:ResASCOTThreePos} to   \ref{fig:ResASCOTPitch} and discussed in the following paragraphs.
  
  %*************************************
  \begin{figure}
  	\centering
  	\includegraphics[draft=false,  trim=35 67 40 20, clip, width=8.5cm]{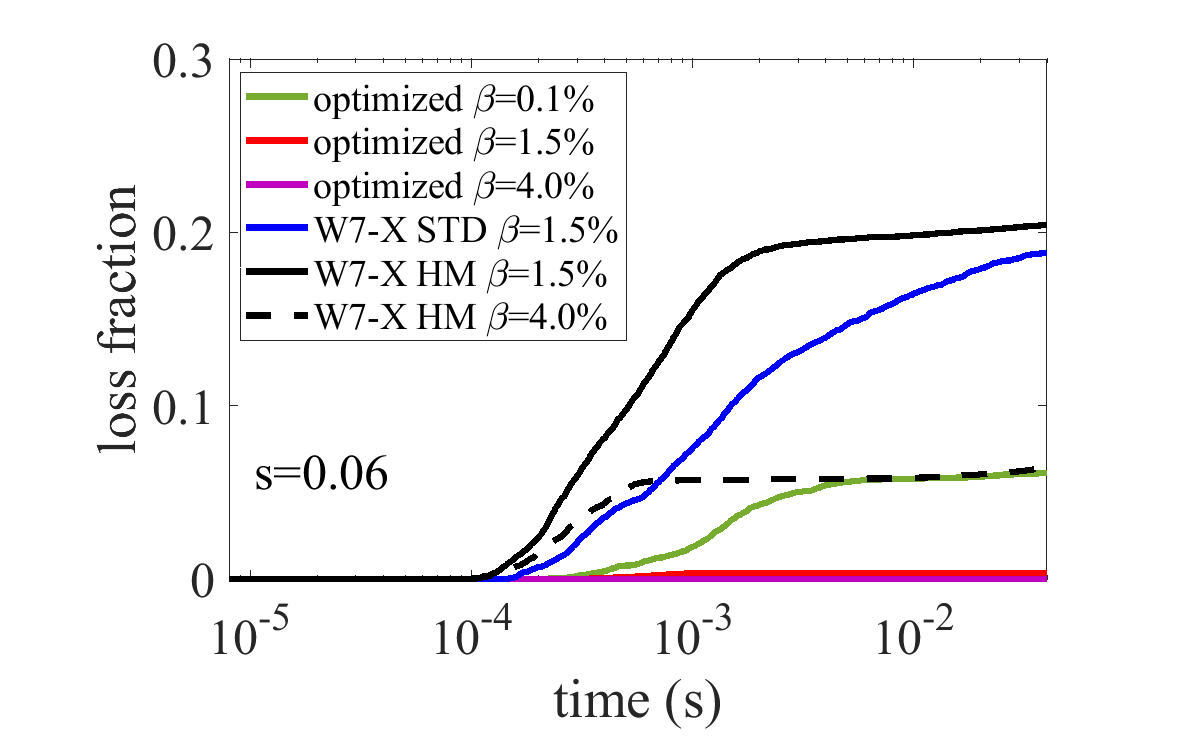}
  	\includegraphics[draft=false,  trim=35 67 40 20, clip, width=8.5cm]{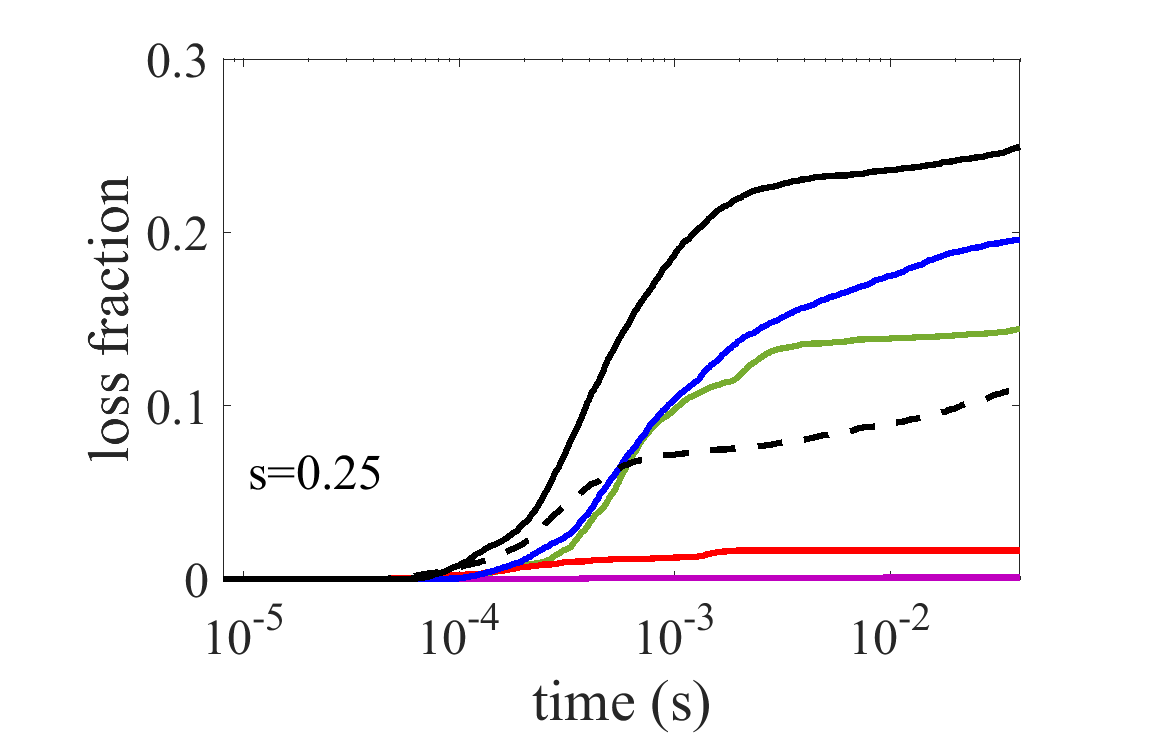}
  	\includegraphics[draft=false,  trim=35 5 40 20, clip, width=8.5cm]{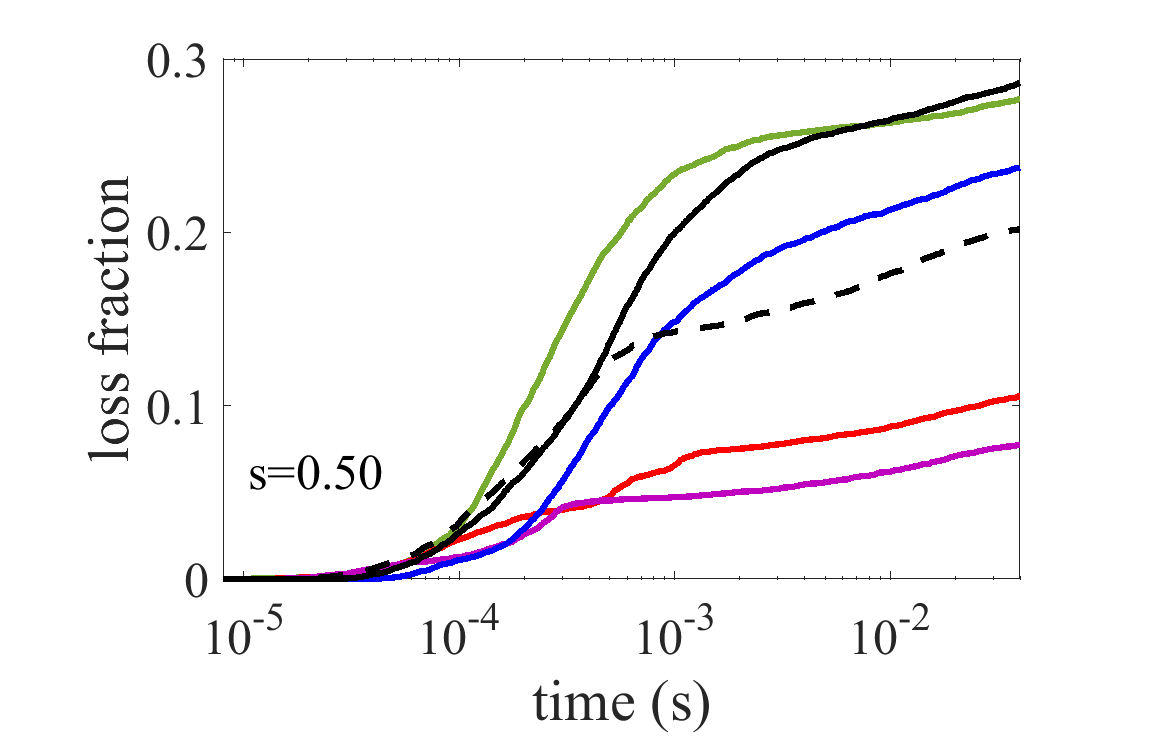}
  	\caption{Fraction of fast ions lost versus time in the optimized configuration, with   $\beta=1.5\%$ and the reference configurations, for ions born at the radial positions $s=0.06$ (top), $s=0.25$ (middle) and $s=0.5$ (bottom).}%$B_0=5~\rm{T}$, $R=1.9~\rm{m}$,
  	\label{fig:ResASCOTThreePos}
  \end{figure}
  %*************************************

 Figure \ref{fig:ResASCOTThreePos} shows the loss fraction versus time for ions launched at the radial positions $s=0.06, 0.25, 0.50$. We show the loss fraction for the optimized configuration at three values of $ \beta$,  0.1\%, 1.5\% and 4\%, and also for the reference W7-X configuration at $\beta=1.5\% $ and  $\beta=4\%$  {(only W7-X HM, as this configuration is expected to perform better at $\beta$=4\% for fast ion confinement)}. Even at the smallest $\beta$ value, 0.1\%, the loss fraction of particles born at $r/a<0.5$ is smaller in the optimized configuration  than that for the reference configuration at $\beta<4\%$. For $\beta=1.5\%$ the losses are much smaller (more than 20 times) than for the reference configurations at same value of $\beta$. {This is a very relevant result, since previous QI configurations have usually required a high value of $\beta$ for confining the fast ions, sometimes showing \textit{increased} losses at low or moderate $\beta$ before the maximum-$J$ property starts to develop, see e.g.~\cite{velasco_model_2021, goodman_constructing_2022}}. At $\beta=4\%$ no ions born at $s=0.06$ are lost after {8000} transit times. The loss fraction increases with the radial position of birth of the fast ions. For ions born at $s=0.25$, at $\beta=0.1\%$ the loss fraction is comparable or smaller than for the reference configuration at $\beta<4\%$, and for $\beta=1.5\%$ the loss fraction is around 10 times smaller than for for reference configuration at $\beta<4\%$. At $\beta<4\%$, the loss fraction is in the range of $10^{-3}$ after {8000 transit} times for ions born at $s=0.25$. In the outermost radial position considered, $s=0.5$, the loss fraction is very similar for $\beta=1.5\%$ or $\beta=4\%$ and  only slightly smaller than that of the reference configurations. 
 We have to point out that most of the alphas in a reactor are expected to be generated in the core region, where the density and temperature profiles have large enough values to make the fusion reaction rate relevant.

  %*************************************
  \begin{figure}
  	\centering
  	\includegraphics[draft=false,  trim=35 68 60 10, clip, width=8.05cm]{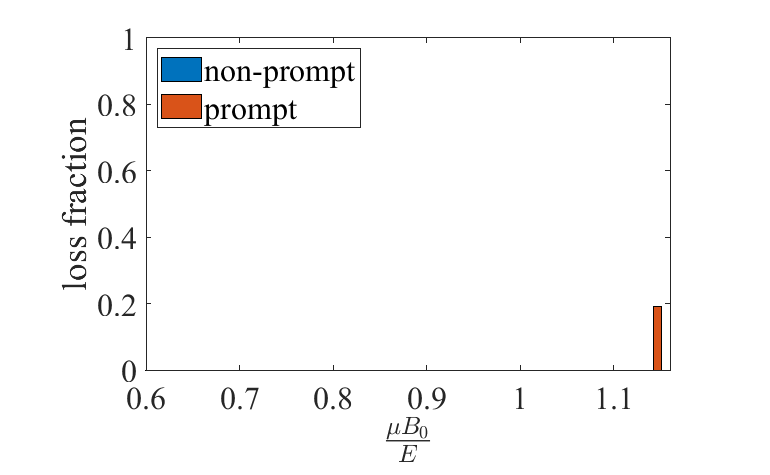}
  	\includegraphics[draft=false,  trim=35 68 60 10, clip, width=8.05cm]{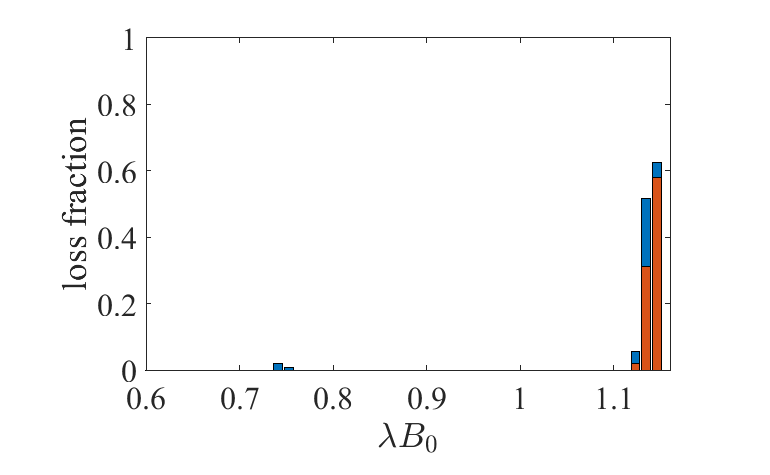}
  	\includegraphics[draft=false,  trim=35 0 60 10, clip, width=8.05cm]{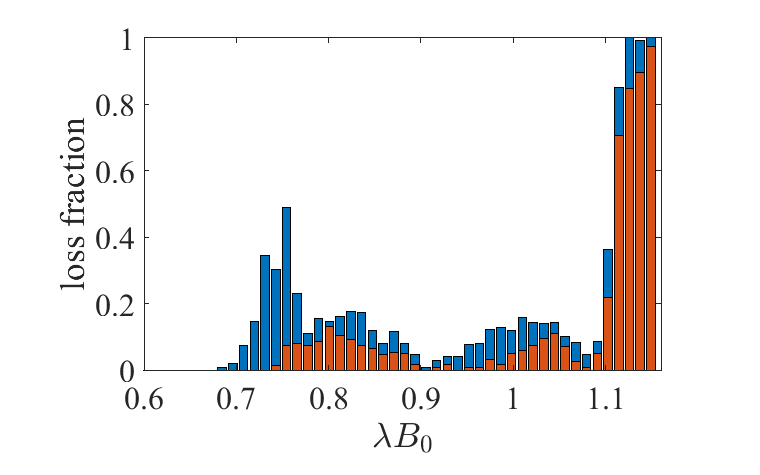}
  	\caption{Fraction of fast ion lost versus {normalized pitch-angle, $\lambda B_0$}, for the optimized configuration, with  $\beta=1.5\%$, and for ions born at three radial positions, $s=0.06$ (top), $s=0.25$ (middle) and $s=0.5$ (bottom).}
  	\label{fig:ResASCOTPitch}
  \end{figure}
  %*************************************
  
  Next, we analyze the distribution of energies of the  fast ions that are lost. 
  In figure \ref{fig:ResASCOTPitch} we show the loss fraction versus $\lambda B_0$ for ions born at the radial positions,  $s=0.06, 0.25, 0.5$, for the optimized configuration at $\beta=1.5\%$. 
  %Here, $\mu$ is the magnetic moment $\mu=mv_{\perp}^2/2B$. 
  In this figure, the losses are separated into two categories: prompt and non-prompt. The former correspond to losses for times $t<10^{-3}~\rm{s}$ and the latter for times  $t>10^{-3}~\rm{s}$. For ions born at $s=0.06$ the losses are very small and correspond almost exclusively to prompt losses of very deeply trapped ions, with large values of {$\lambda B_0\sim 1.15$}. The small fraction of lost ions can be explained by the closeness to omnigeneity for the innermost radial region of this configuration. For ions born at $s=0.25$, the losses slightly increase, and a component of non-prompt losses, with comparable size as that of the prompt fraction, appears. Most of the ions promptly lost are very deeply trapped, which again reinforces the need for a very good alignment of $B$ minima to minimize the losses. Losses of very deeply trapped ions are the most difficult to avoid and require very careful tailoring of the magnetic configuration. A tiny fraction of losses are non-prompt and correspond to barely trapped ions   with {$\lambda B_0\sim 0.75$}. 
  Finally, for the last radial position of birth studied, $s=0.5$, the distribution of energies of the lost ions is qualitatively different, and extends to the region of barely trapped ions. The prompt losses show {a dominant peak around $\lambda B_0\sim 1.15$ in the deeply trapped region, and two smaller peaks centered around $\lambda B_0\sim 0.8$ and $\lambda B_0\sim 1.05$. No significant prompt losses are observed for $\lambda B_0< 0.75$}, while the distribution of non-prompt {extends from the deeply trapped to the barely trapped region, with a dominant peak around $\lambda B_0 \sim 0.75$, close to the trapping-passing limit, at $\lambda B_0 \sim 0.69$. No passing ions are lost.} 
  
  {We note that proxies such as $\Gamma_{\alpha}$, $\Gamma_c$ and VBB target losses caused by the radial drift of fast ions being, on average, directed in the radial direction. These losses take place on a time scale between  $10^{-4}\rm{s}$ and $10^{-3}\rm{s}$ and thus mainly manifest themselves as prompt losses. Our results show that optimizing $\Gamma_{\alpha}$ and $\Gamma_c$ indeed reduce the prompt losses, but also have a positive indirect effect on other types of losses that take place on a longer time scale \cite{paul_energetic_2022}. Some of these losses should be positively affected by a reduction of VBT.}

%*************************************  

	  	%***************************************************************************************
		%***************************************************************************************
		\subsection{Evaluation of the bootstrap current}\label{secBootstrap}
		%***************************************************************************************
		%***************************************************************************************

Having shown that the confinement of energetic ions, which was the central point of the optimization, is very good, we analyze in this section the boostrap current.

 %*************************************
 \begin{figure*}
 	\centering
 	\includegraphics[draft=false,  trim=10 67 60 0, clip, width=8.45cm]{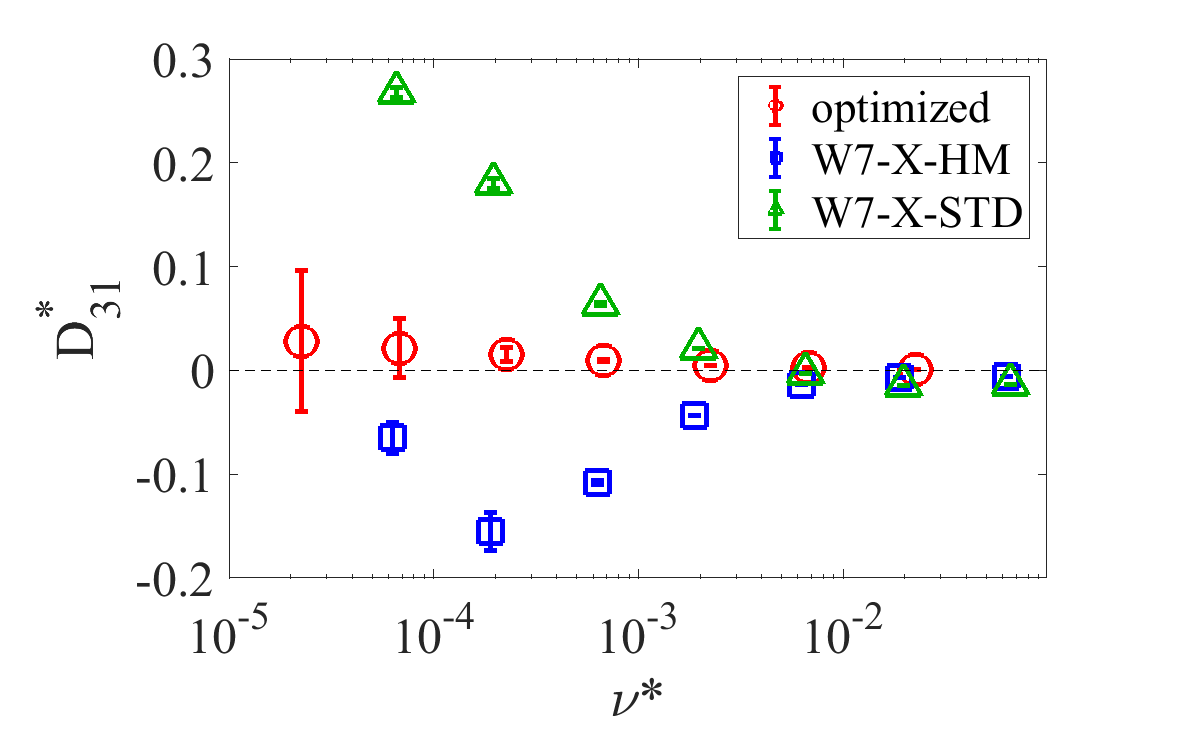}
 	\includegraphics[draft=false,  trim=10 67 60 0, clip, width=8.45cm]{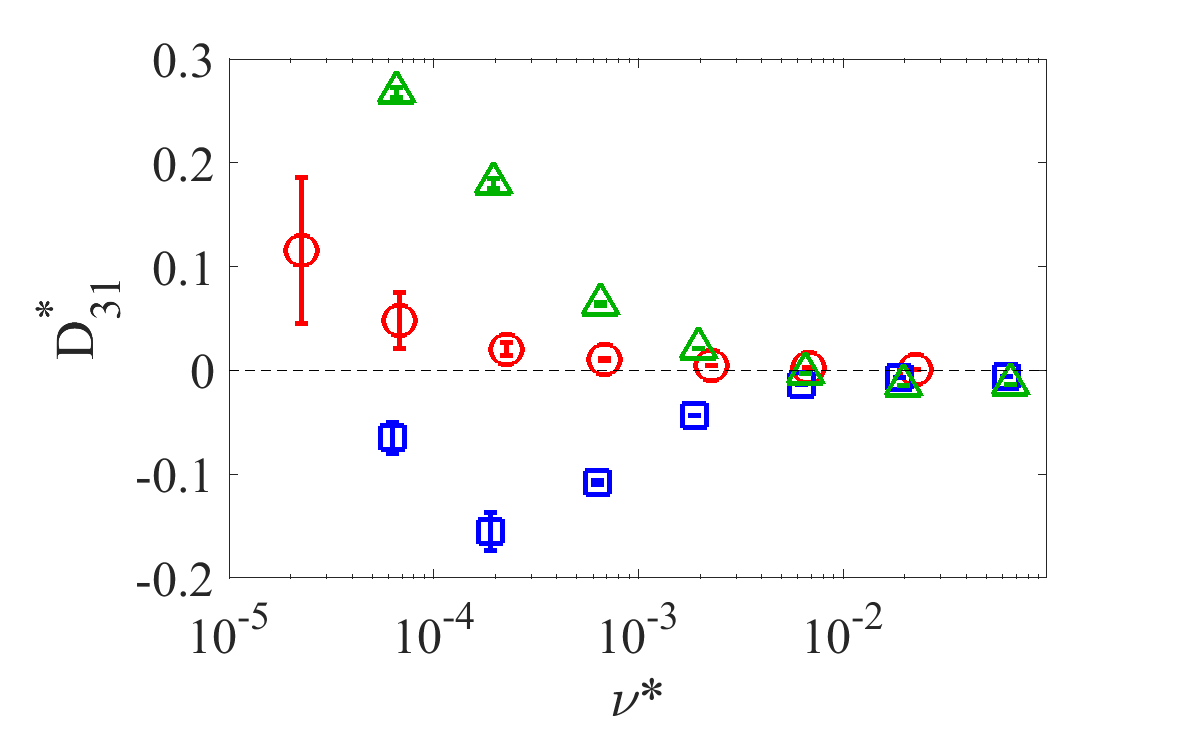}\\
 	\includegraphics[draft=false,  trim=10 67 60 0, clip, width=8.45cm]{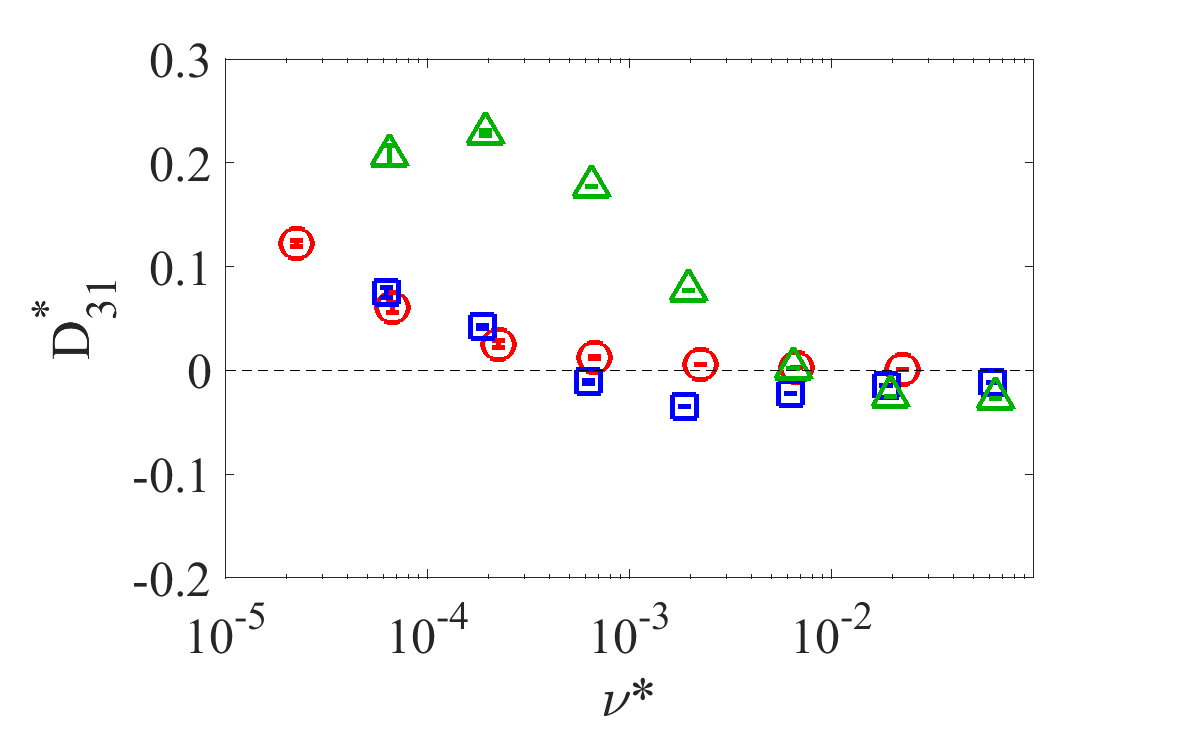}
 	\includegraphics[draft=false,  trim=10 67 60 0, clip, width=8.45cm]{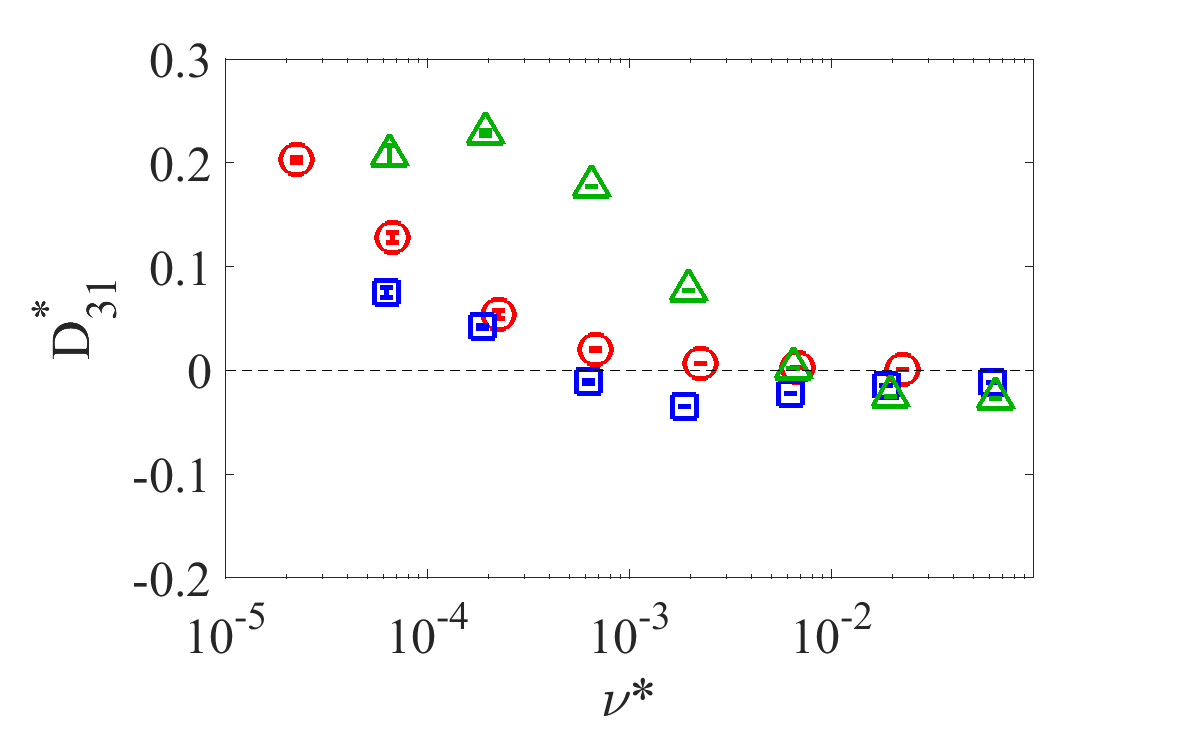}\\
 	\includegraphics[draft=false,  trim=10 5 60 0, clip, width=8.45cm]{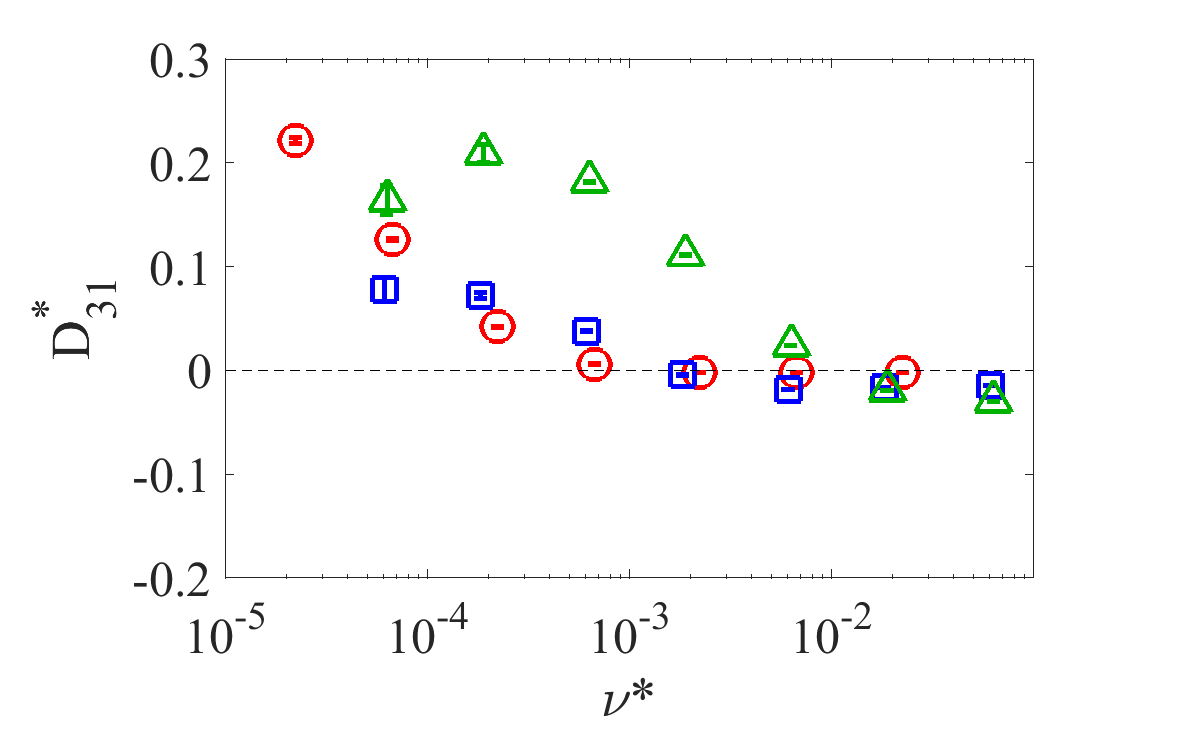}
 	\includegraphics[draft=false,  trim=10 5 60 0, clip, width=8.45cm]{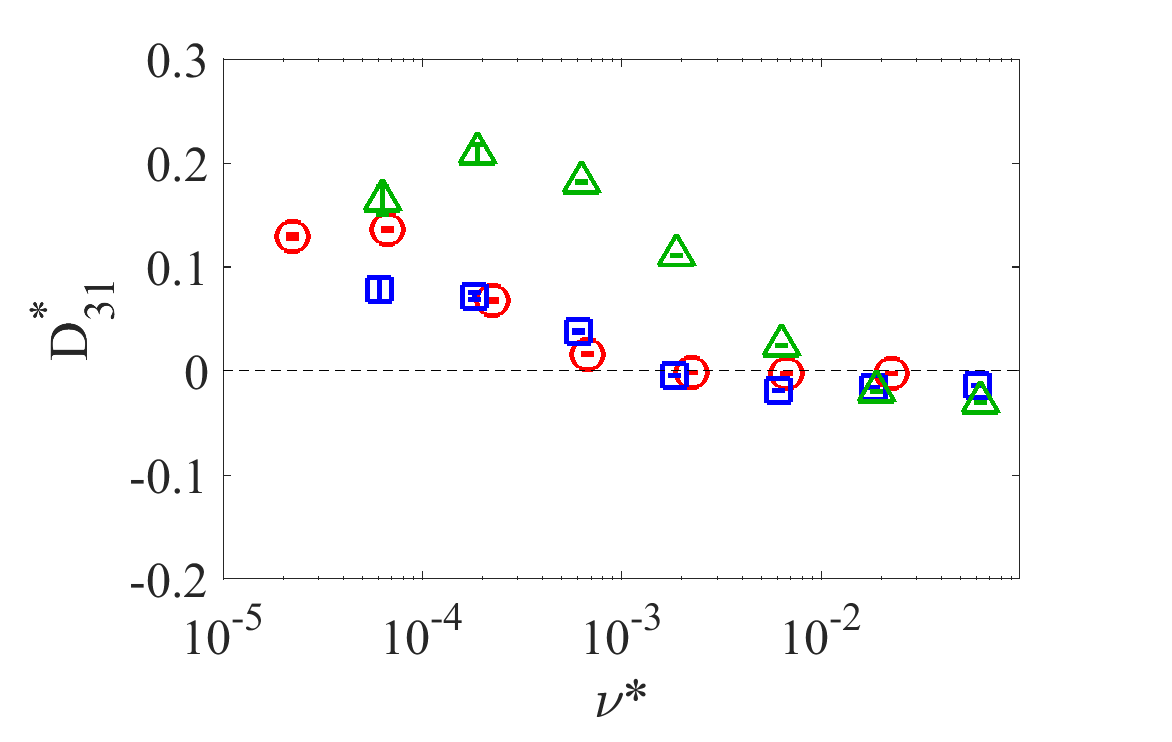}
 	\caption{Values of the $D_{31}^*$ coefficient versus collisionality, for the optimized configuration and the two W7-X reference configurations, for three radial positions $s=0.06$ (top), $s=0.25$ (middle) and $s=0.5$ (bottom). Results for the optimized configuration at two values of $\beta$ are shown, $\beta=1.5\%$ on the left panels and $\beta=4\%$ on the right ones.  	}
 	\label{fig:BootstrapCurr}
 \end{figure*}
 %*************************************%***************************************************************************************
 The configuration dependence of the bootstrap current in a stellarator can be encapsulated in the so called  mono-energetic coefficient $D_{31}$. The dependence on the plasma parameters is then introduced by multiplication by a combination of the plasma gradients and convolution with a Maxwellian distribution function, see e.g. \cite{beidler_benchmarking_2011}. This calculation has to be performed with codes that solve numerically de drift-kinetic equation. Because these calculations need to resolve with good enough accuracy a narrow region of phase-space in the boundary between trapped and passing particles, it requires a substantial computing time and it is not usually included in the optimization loop. Simple formulas such as that in \cite{shaing_bootstrap_1989}  can capture reasonably well the size of the bootstrap current\cite{beidler_benchmarking_2011},  but lack the accuracy to be employed in stellarator optimization.
 A fast and accurate calculation of the bootstrap current for general stellarator configurations is not possible presently. However, it can be demonstrated that for an  omnigenous magnetic field with poloidally closed $B$ contours the bootstrap current is identically zero \cite{subbotin_integrated_2006,Helander2009,LandremanM2012}. 
 
 Due to the lack of an appropriate proxy, we did not include a target for the bootstrap current in the optimization process. However, we could expect that for this configuration that is close to QI, it would be small. In this section, we present the calculations of the $D_{31}$ coefficient performed a posteriori with the code DKES \cite{Hirshman1986DKES} for the optimized configuration.  
 Figure  \ref{fig:BootstrapCurr} shows the values of the normalized\footnote{The transport coefficients are usually normalized to the values for an axisymmetric tokamak in the plateau regime (see \cite{beidler_benchmarking_2011})} $D_{31}^*$ coefficient calculated with DKES versus collisionality for three radial positions $s=0.06, 0.25, 0.5$ and two values of  $\beta=1.5\%, 4\%$. Here, $\nu^*=R\nu/ \iotab v$ is the collisionality,  with $\nu$ the collision frequency and $v$ the thermal velocity. We show results for the high mirror and the standard configurations of W7-X in the same figure \footnote{{The W7-X configurations here used are vacuum configurations}.}. The calculations are done assuming no radial electric field, $E_r$. 
 The region of parameter space corresponding to small $E_r$ is the most relevant one for electrons, which are expected to contribute the most to the bootstrap current in ion root plasmas of low collisionality regimes, see  \cite{beidler_benchmarking_2011}.
 It is shown in the figure that the values of the coefficient  $D_{31}^*$ are comparable to or smaller than those for the high mirror configuration of W7-X, and significantly smaller than those for the standard configuration in the three radial positions considered and for a wide range of collisionality values. {As we discussed earlier, t}he bootstrap current in the high mirror configuration is considered sufficiently small for a reactor, while that for the standard is considered too large. The opposite situation occurs with $\epsilon_{eff}$, which is considered small enough for a reactor in the standard configuration while in the high mirror it is too large \cite{Beidler2021}. The optimized configuration here presented has sufficiently small values of both bootstrap current and effective ripple.
 
  It is remarkable that in the optimized configuration, for the innermost radial position, $s=0.06$, the values of the  $D_{31}^*$  coefficient are  zero, within error bars, for the whole collisionality range studied in the case with $\beta=1.5\%$. At $\beta=4\%$, $D_{31}^*$ is only slightly larger than zero for the low collisionality regimes with $\nu^*\leq10^{-3}$. For collisionalities $\nu^*\geq10^{-3}$, $D_{31}^*$ is zero for the three radial positions, and the two values of $\beta$ considered.  
 The dependence of $D_{31}^*$ with $\beta$ is small, as can be observed by comparing the left and right panels of figure \ref{fig:BootstrapCurr}.
 
 The small values obtained for $D_{31}^*$ {are} a foreseeable consequence of the closeness to quasi-isodynamicity of this configuration, even though no bootstrap current metric was included in the optimization. Based on the results discussed in this section, we can expect that plasmas in this magnetic configuration will show  small values of bootstrap current. 
 
%***************************************************************************************
%***************************************************************************************
%***************************************************************************************
%***************************************************************************************
\section{Feasibility study of magnetic coils}\label{secCoils}
%***************************************************************************************
%***************************************************************************************
%***************************************************************************************
Once we have analyzed in detail the physics properties of the new  configuration, the last question that we address in this work is the conceptual design of coils that could generate it. We follow a two-step approach: first optimize the shape of the magnetic configuration and afterwards try to find coils that produce this configuration. In principle, it is not guaranteed that the magnetic configuration can be reproduced, to a sufficient degree of accuracy,  with realistic coils that are simple enough as to be built. The construction of coils introduces additional engineering constraints that should be properly taken into account. An exhaustive analysis taking into account all engineering criteria required for the actual construction of the coils exceeds the objectives of this work. Here, we perform a basic preliminary assessment of single-filament coils that could generate the magnetic configuration presented in the previous sections. With these simplified coils we calculate the MHD equilibrium with VMEC in free-boundary mode including the coil currents, which we compare with the previously generated equilibrium described in section \ref{secConfigProps}, obtained through a fixed-boundary calculation. We analyze the configuration generated with the coils and evaluate to what extent the field errors degrade the benign  properties of the original one.   We will not address the analysis of the coil-generated configuration with the same level of detail as for the fixed-boundary equilibrium but will concentrate on the confinement of energetic ions, which was the main focus of the optimization, and is expected to be very sensitive to small changes in the magnetic field structure. 
%*************************************
\begin{figure}
	\centering
	\includegraphics[draft=false,  trim=450 172 370 55, clip, width=6.5cm]{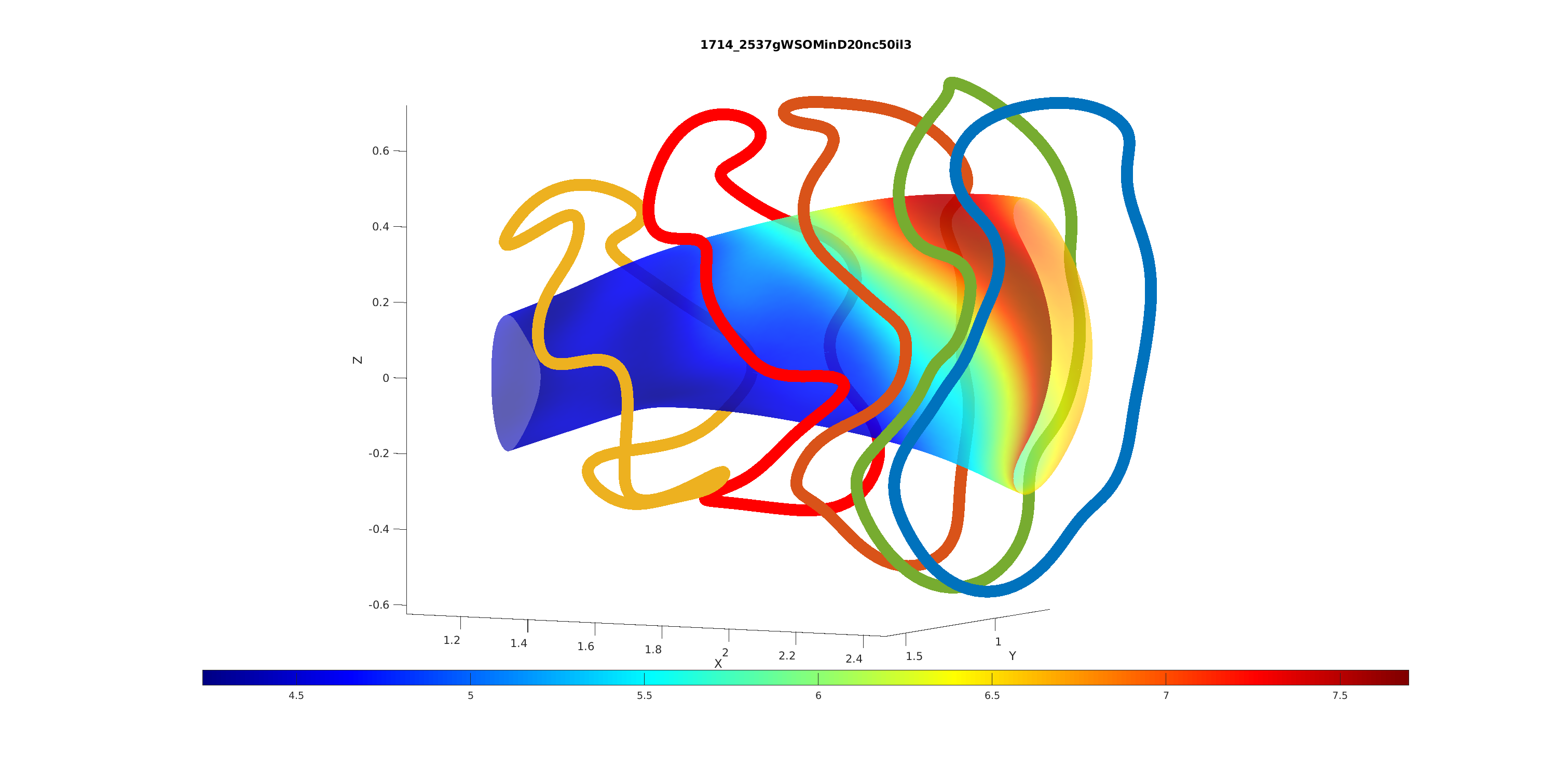}	
	\raisebox{-0.15\height}{
		\includegraphics[draft=false,  trim=40 20 70 40, clip, width=6.5cm]{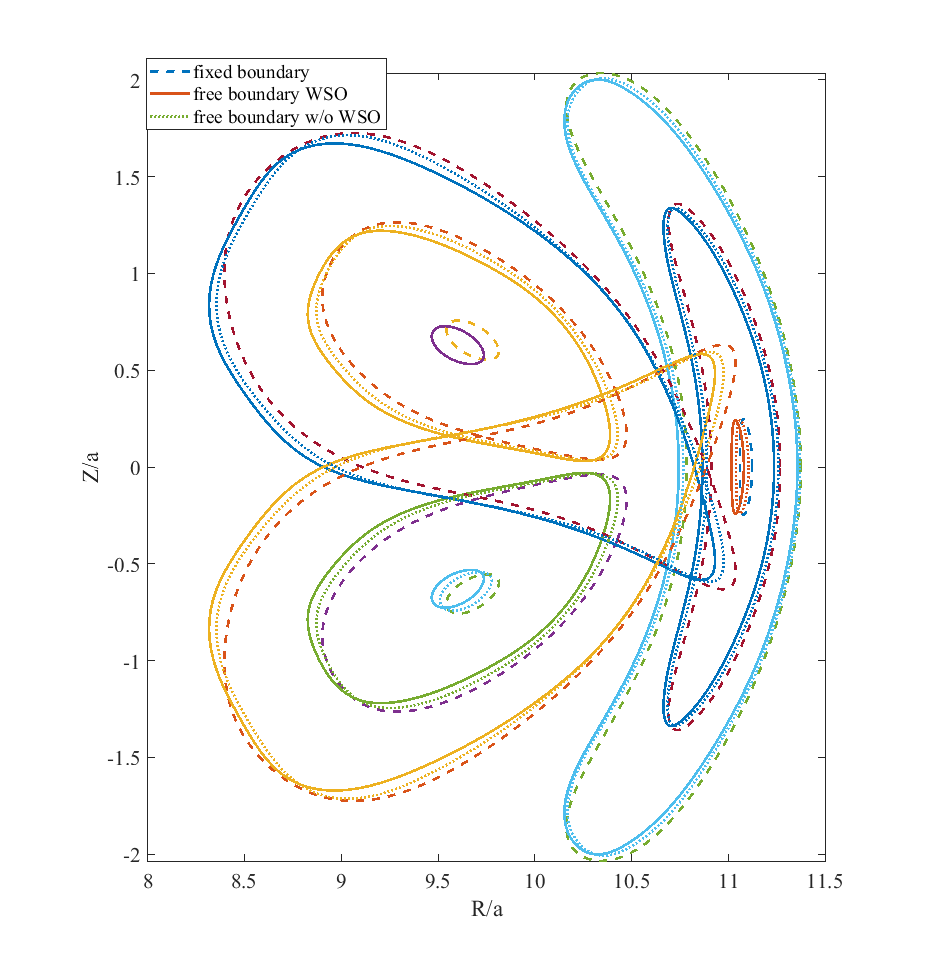}
	}
	\caption{Top: Set of coils designed for the optimized configuration using winding surface optimization and a minimum distance from coils to plasma, $d_{m}=a$. Bottom: Poincaré plots for the fixed-boundary calculation of the original optimized configuration and the free-boundary calculation including the optimized coil set with  $d_{m}=a$. Lines corresponding to a 
		coil set without winding surface optimization and fixed distance $d=~a$ (shown in figure \ref{fig:coilSetFixD20y25}) are also included for comparison. }
	\label{fig:coilSetOptimizedWS}
\end{figure}

For the coil design, we use the code REGCOIL \cite{landreman_improved_2017}, which allows reaching a compromise between coil complexity and magnetic field error in a controlled way. 
In addition to REGCOIL, for this phase of the analysis, we use the codes MAKEGRID \cite{STELLOPT_doecode_12551} and BNORM \cite{merkel_integral_1986}, already included in the STELLOPT suite, for generating the grid required for free-boundary MHD calculation with VMEC and for the calculation of the normal component of the magnetic field over the winding surface (the surface on which the current of the coils lay), respectively.

REGCOIL can assume a winding surface at a constant distance from the plasma or use a predefined (calculated by other means) winding surface. The results from the constant-distance case can be improved by relaxing this  constraint and optimizing the winding surface, which in our case is done using adjoint methods  \cite{paul_adjoint_2018}. For reference, we present the coils obtained for winding surfaces at constant distance from the plasma in the Appendix A while the best coils obtained using the winding surface optimization are presented directly in this section.  

Based on the analysis performed for  fixed distance from plasma to coils (see Appendix A), we concentrate on cases with 10 coils per period and  $d_{m}=0.9~a$, $d_m=a$, and $d_m~1.25~a$, where $d_{m}$ is the minimum distance from the plasma to the winding surface, which is a constraint used in the winding surface optimization, and $a$ is the minor radius. The best results were obtained for $d_{m}=a$.

	%*************************************
	\begin{figure}
		\centering
		\includegraphics[draft=false,  trim=25 68 40 0, clip, width=8.5cm]{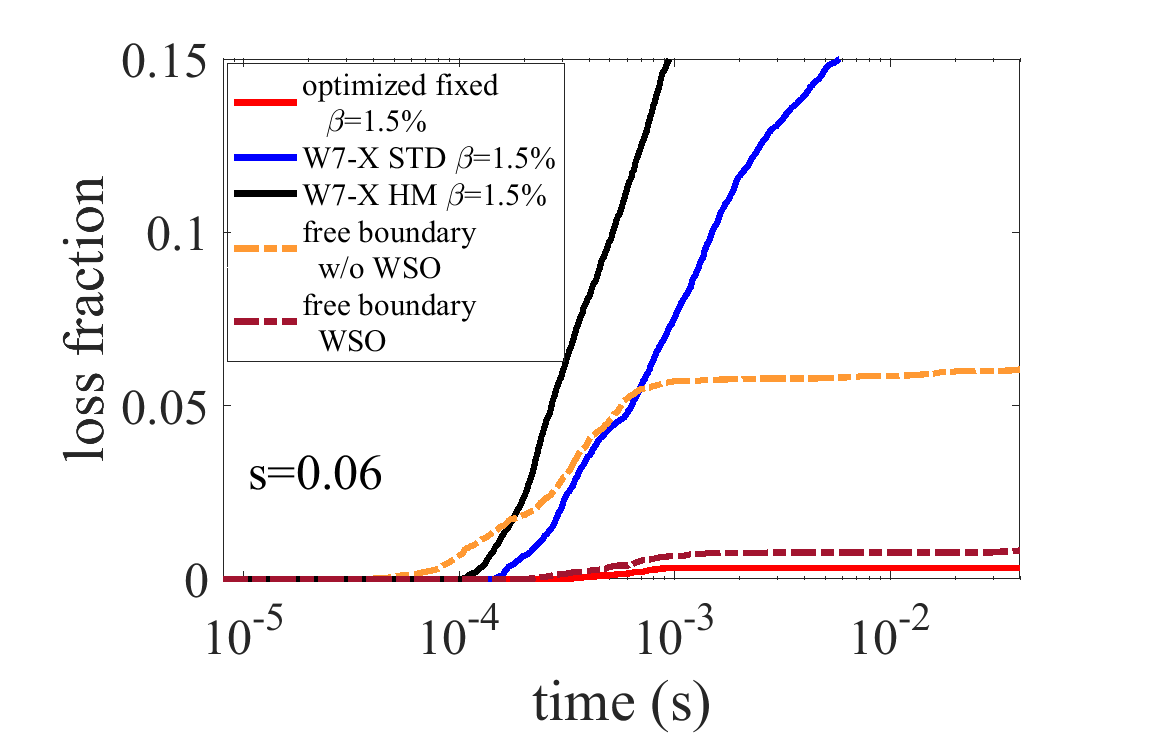}
		\includegraphics[draft=false,  trim=25 0 40 0, clip, width=8.5cm]{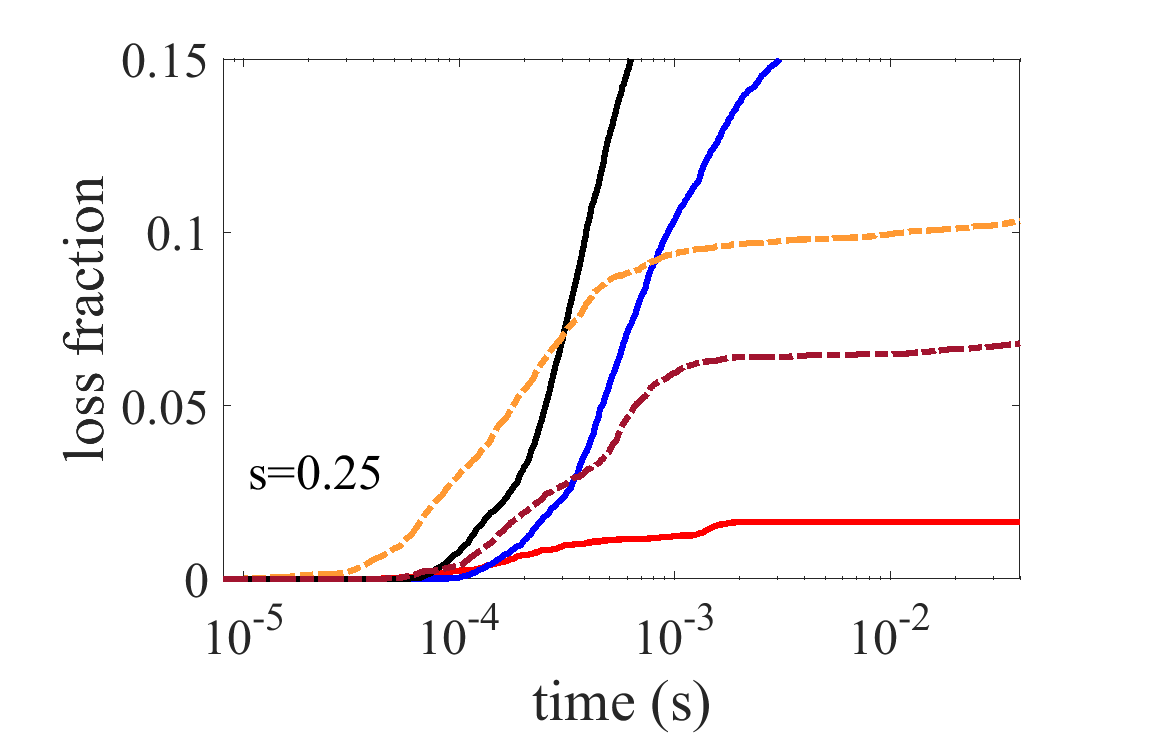}
		\caption{Loss fraction of energetic ions born at $s=0.06$ (top) and $s=0.25$ (bottom) calculated for the optimized (fixed-boundary equilibrium) configuration, the free-boundary configuration obtained with non-optimized coil sets (w/o WSO) and the free-boundary configuration obtained with the optimized coil set (WSO). Calculation for the reference W7-X configurations are also shown for comparison.}
		\label{fig:LossFractionWCoils}
	\end{figure}
	%*************************************

The coil set obtained using a combination of REGCOIL and adjoint methods for a minimum distance of the winding surface to plasma of $d_{m}=a$ is shown in figure \ref{fig:coilSetOptimizedWS}. The coil complexity and minimum distance between coils in this optimized set are comparable to those of the coils obtained with fixed-distance winding surface (see Appendix A), while the configuration obtained with these coils is better, as we will see next. 
Figure \ref{fig:coilSetOptimizedWS}-bottom shows  Poincaré plots at several toroidal sectors for the original fixed-boundary calculation of the optimized configuration from section \ref{secConfigProps}, and the free-boundary calculation including the optimized coil set. 
We also include, for comparison, lines corresponding to a 
 non-optimized coil set with fixed distance $d=a$ (shown in figure \ref{fig:coilSetFixD20y25}).  
 The separations from the fixed-boundary case in both free-boundary calculations are comparable in size. 
 
 %*************************************
 \begin{figure*}
 	\centering
 	\includegraphics[draft=false,  trim=68 60 110 15, clip, width=5.5cm]{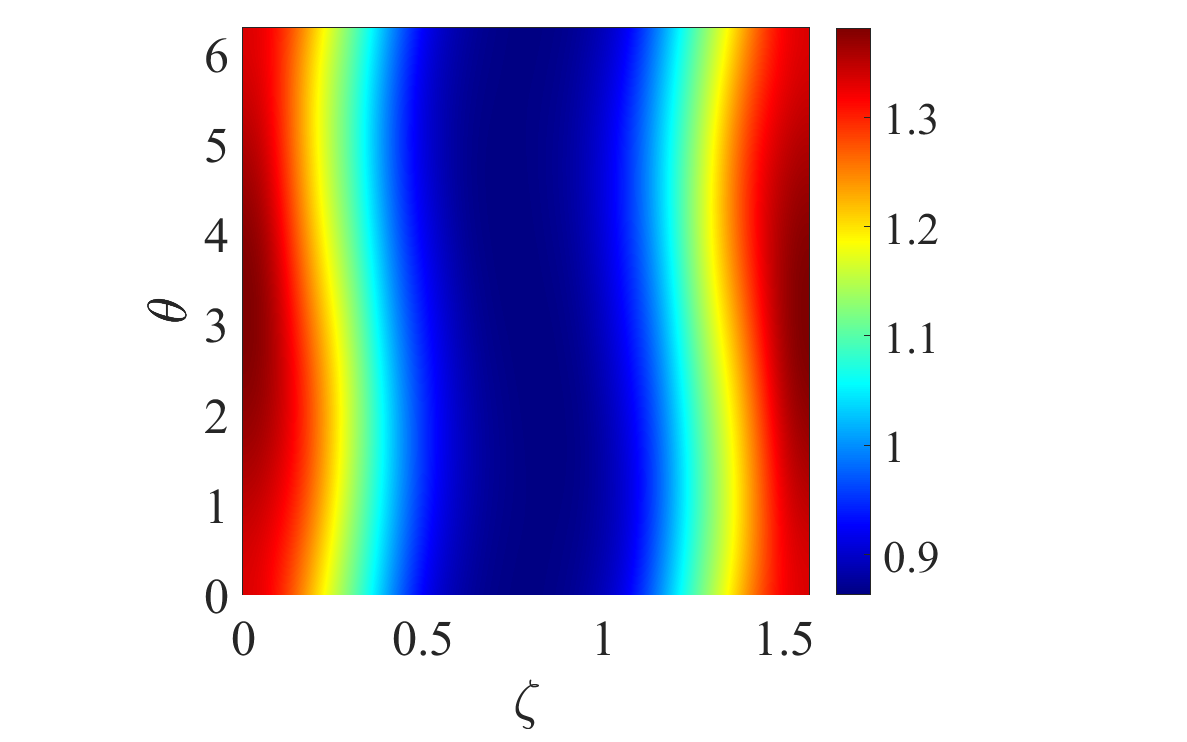}
 	\includegraphics[draft=false,  trim=68 60 110 15, clip, width=5.5cm]{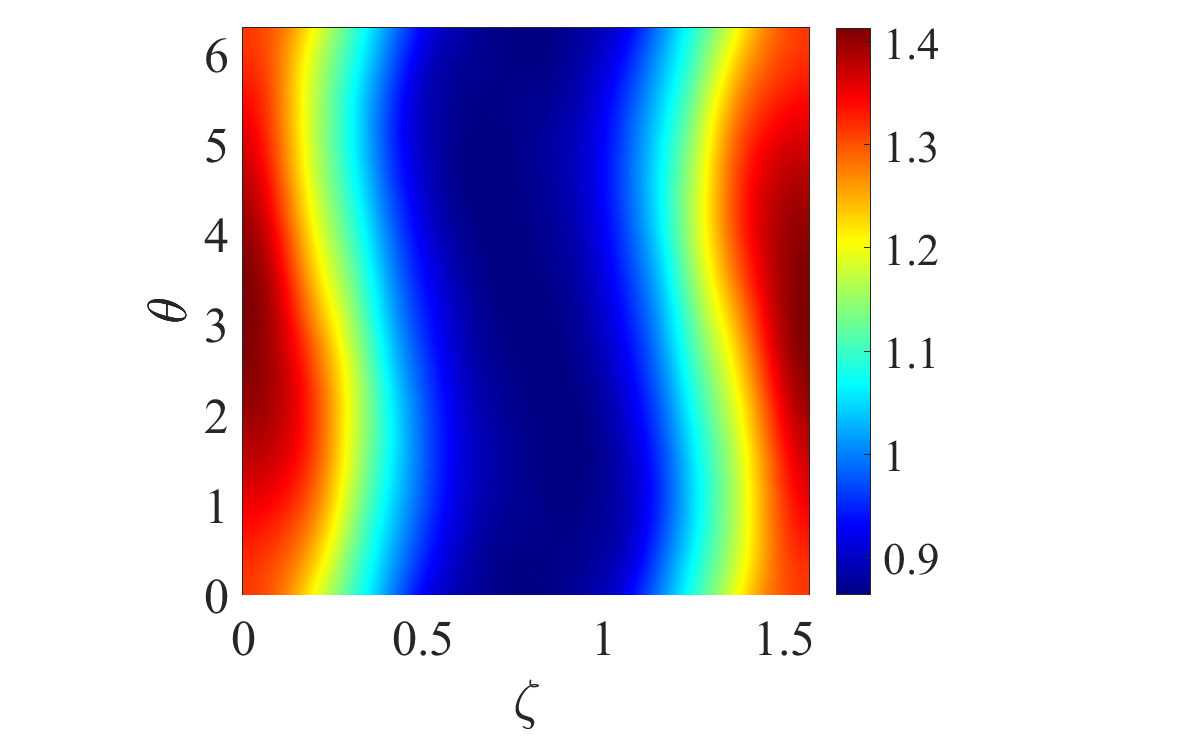}
 	\includegraphics[draft=false,  trim=68 60 110 15, clip, width=5.5cm]{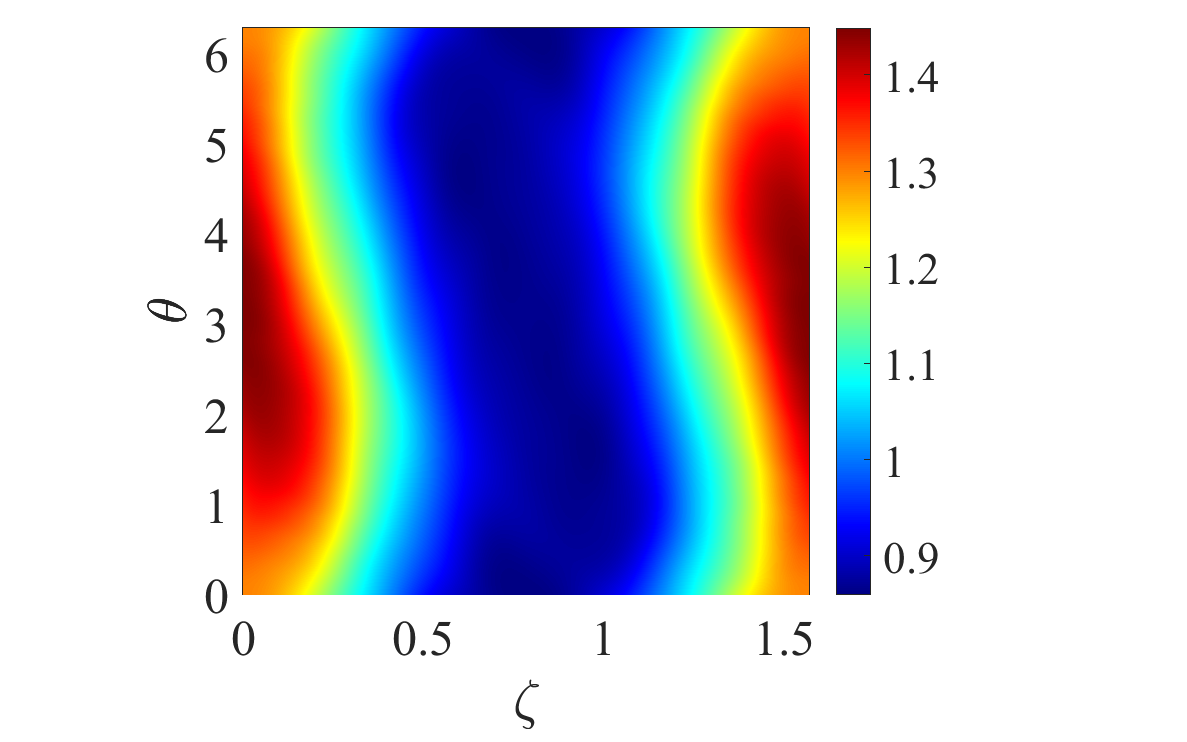}\\
 	\includegraphics[draft=false,  trim=68 60 110 15, clip, width=5.5cm]{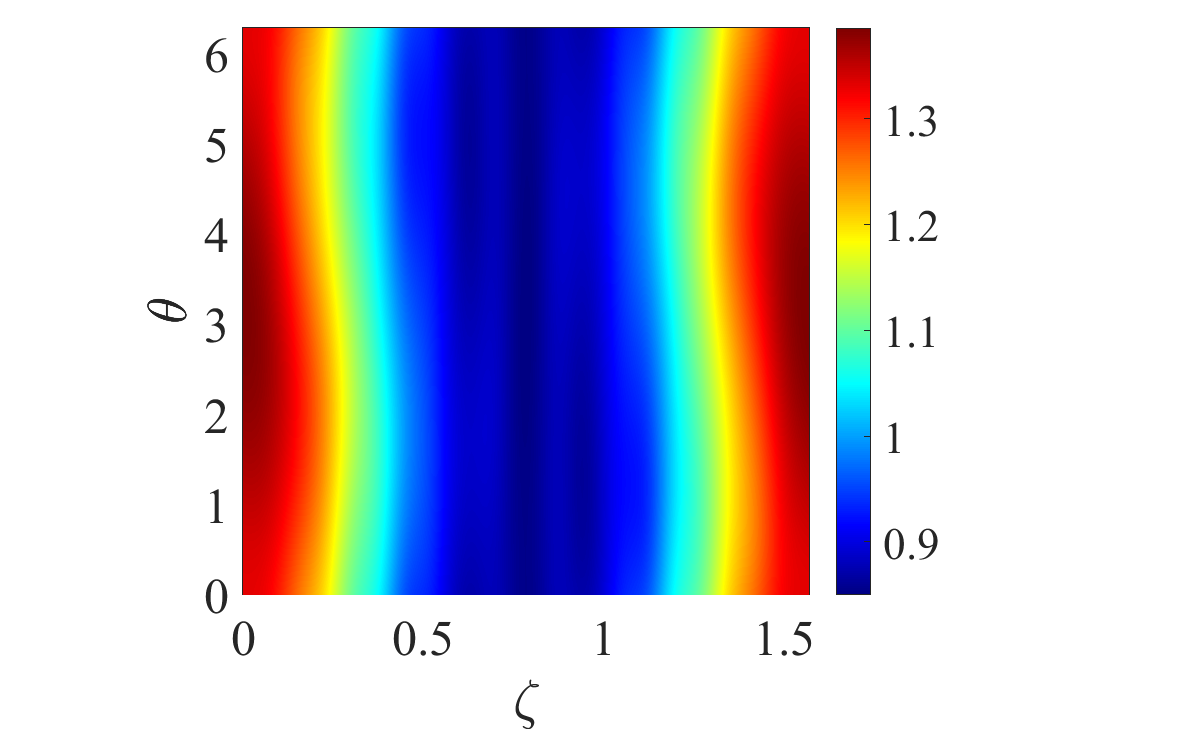}
 	\includegraphics[draft=false,  trim=68 60 110 15, clip, width=5.5cm]{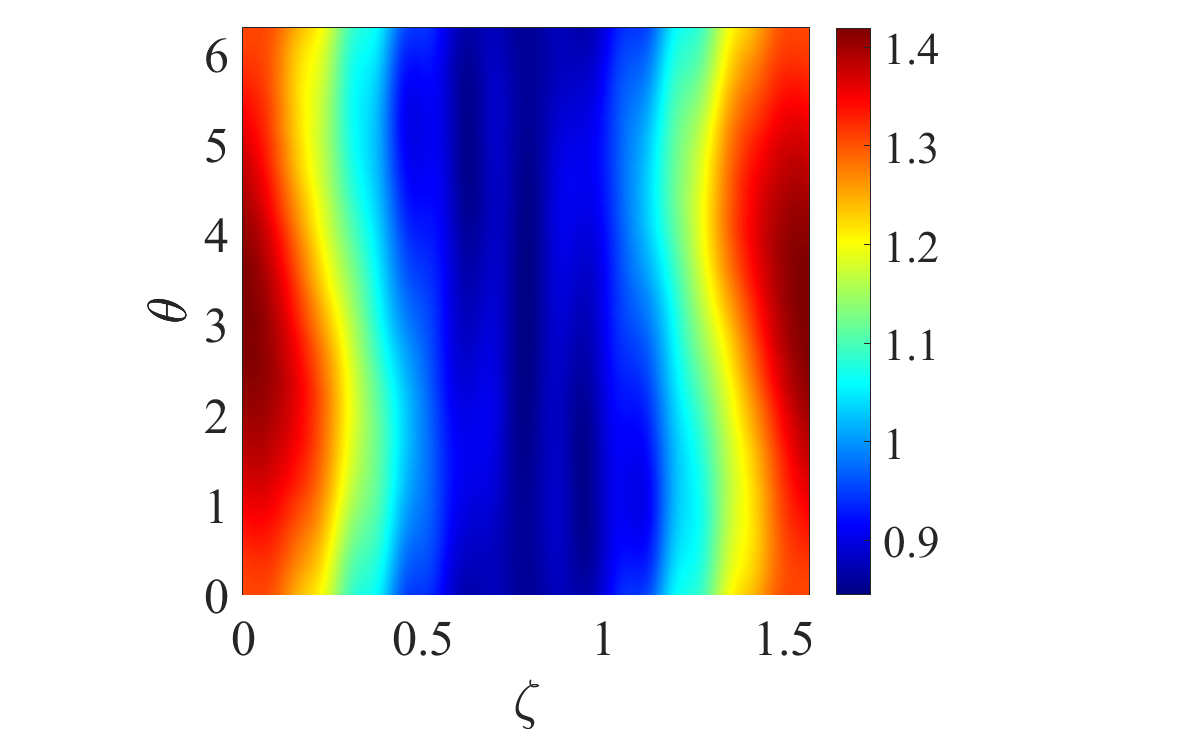}
 	\includegraphics[draft=false,  trim=68 60 110 15, clip, width=5.5cm]{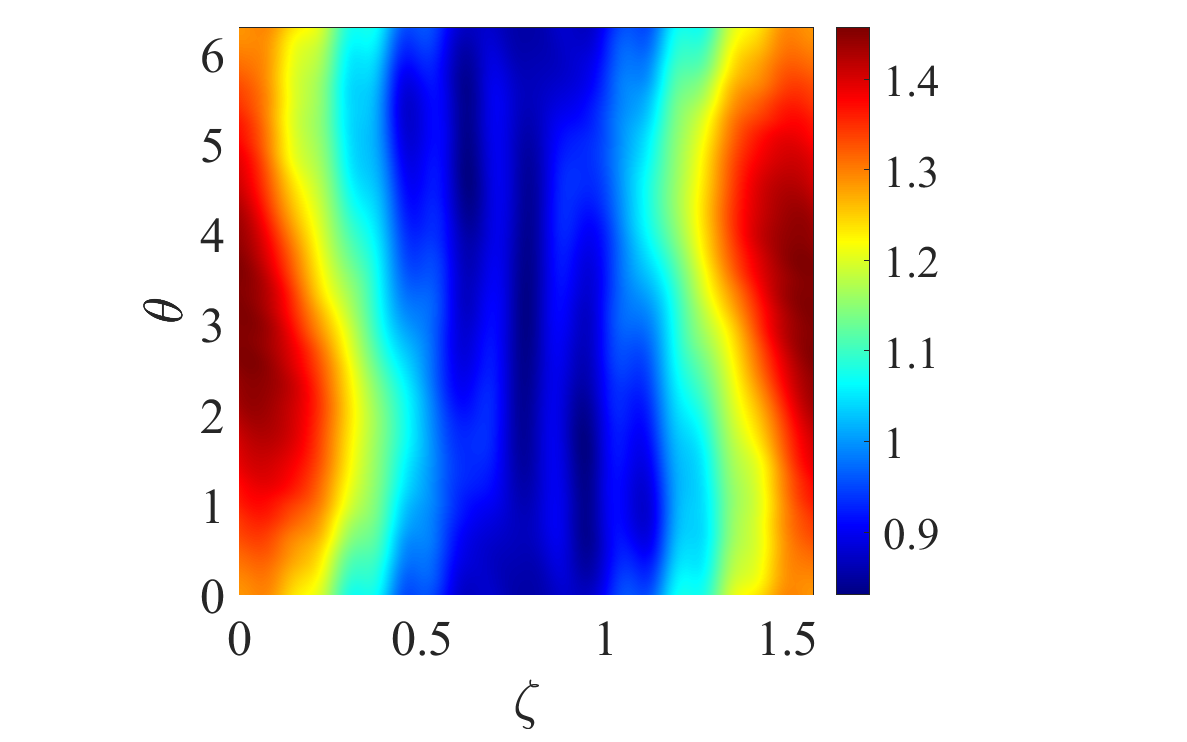}\\
 	\includegraphics[draft=false,  trim=68 5 110 15, clip, width=5.5cm]{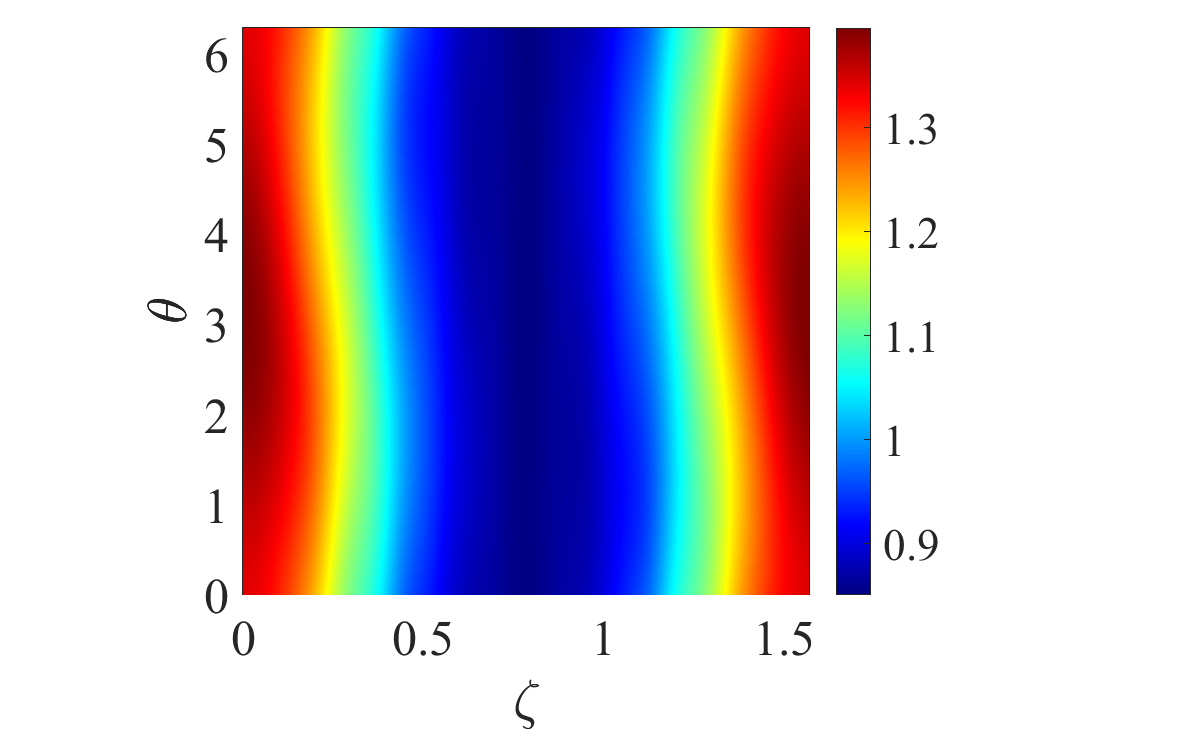}
 	\includegraphics[draft=false,  trim=68 5 110 15, clip, width=5.5cm]{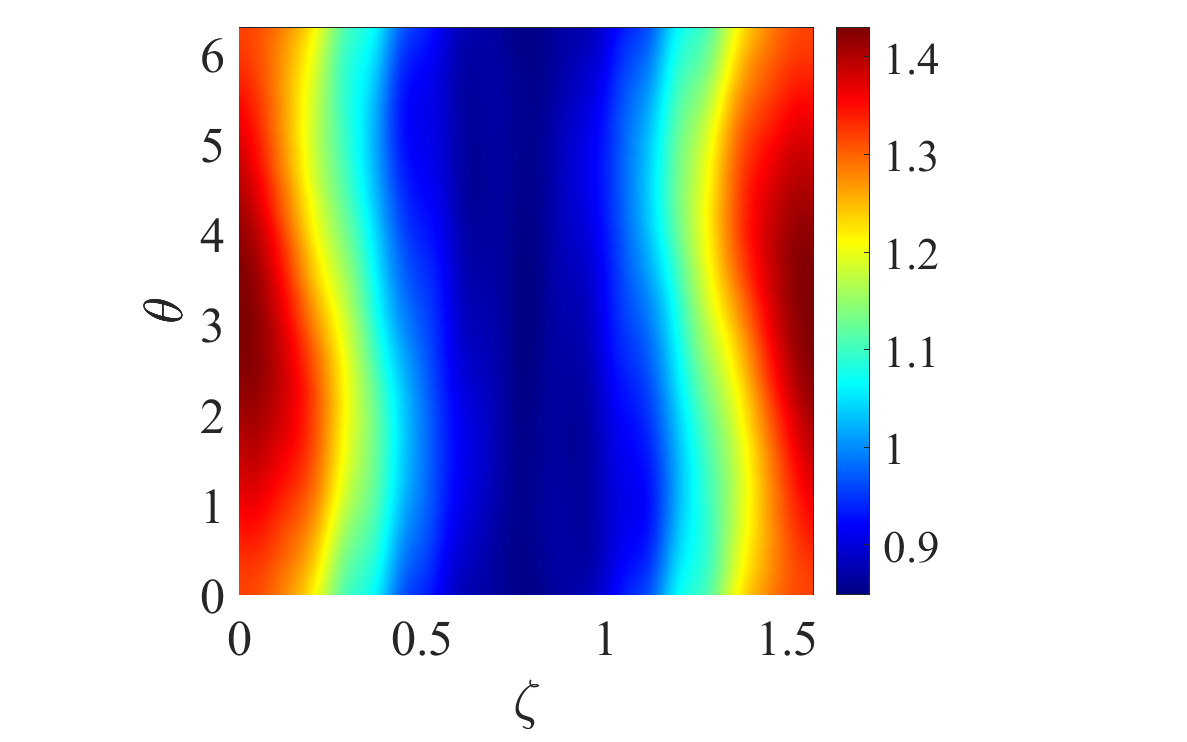}
 	\includegraphics[draft=false,  trim=68 5 110 15, clip, width=5.5cm]{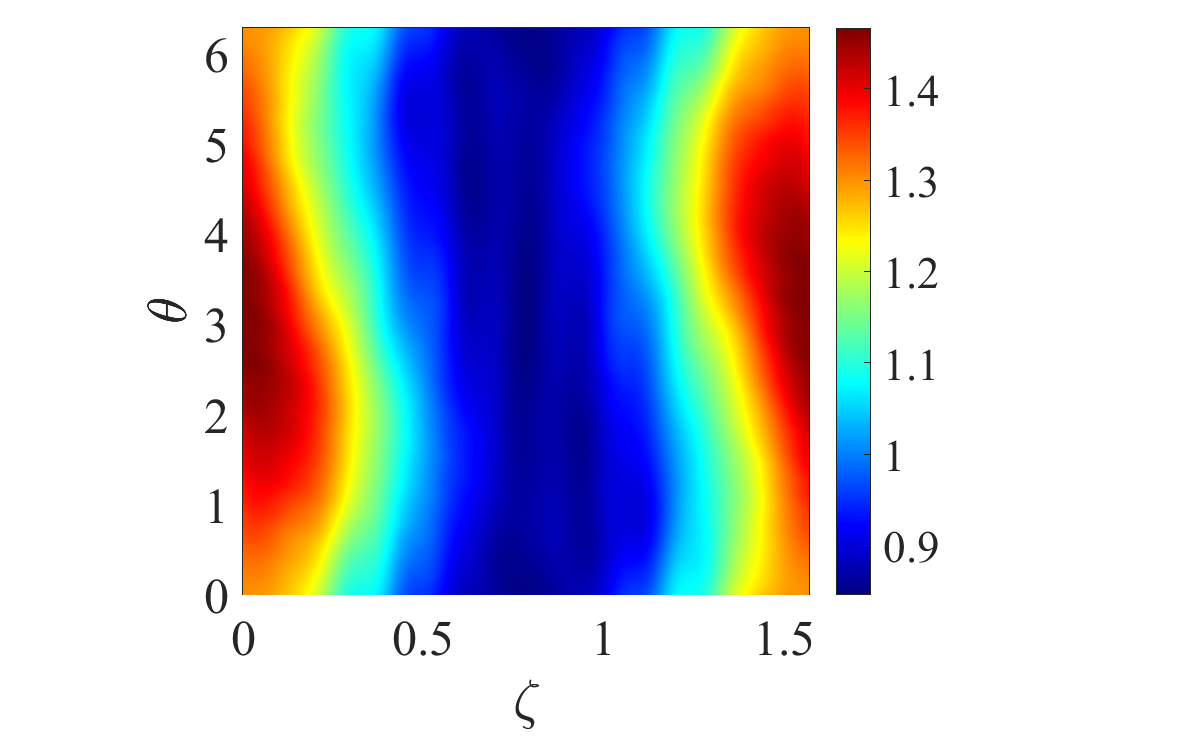}\\
 	\caption{Magnetic field strength (normalized to $B_0$) versus Boozer angles over the flux surface for three radial positions, $r/a=0.25$ (left), $r/a=0.5$ (middle) and $r/a=0.75$ (right), for the original optimized (fixed boundary) configuration (top row), the magnetic configuration obtained including the non-optimized coil set from figure \ref{fig:coilSetFixD20y25}-left (middle row) and the configuration obtained with the winding-surface-optimized coil set from figure \ref{fig:coilSetOptimizedWS} (bottom row).}
 	\label{fig:BMapsCoils}
 \end{figure*}
 %*************************************

For the configuration obtained in a free-boundary calculation with VMEC including the coils shown in figure  \ref{fig:coilSetOptimizedWS} we have evaluated the fast ion losses with guiding-center simulations with ASCOT, following the procedure explained in section \ref{secASCOT}. We focus on the $\beta=1.5\%$ case. 
The results of loss fraction versus time are shown in figure \ref{fig:LossFractionWCoils}. The loss fraction of energetic ions born at $s=0.06$ and $s=0.25$ calculated for the (fixed-boundary) optimized configuration presented in section \ref{secConfigProps} at $\beta=1.5\%$, the free-boundary configuration obtained with the best non-optimized coil set for $d=a$ (shown in figure \ref{fig:coilSetFixD20y25})) and the free-boundary configuration obtained with the optimized coil set from figure \ref{fig:coilSetOptimizedWS} are compared. Results for the reference W7-X configurations are also included.
The benefit of the winding surface optimization for the confinement of energetic ions is clear from the figure. For the innermost position, $s=0.06$, this optimized coil set produces a configuration in which the fast ion losses are significantly smaller than those for the configurations obtained without optimization of the winding surface. Using different levels of regularization does not produce big changes (not shown),  while the  winding surface optimization improves the configuration from the point of view of the confinement of fast ions. The loss fraction for the optimized-coils configuration is comparable to the original optimized configuration (fixed-boundary equilibrium) from section \ref{secConfigProps} obtained without considering coils.

For the radial position $s=0.25$, the configuration including the optimized coils has losses  significantly larger than the original (fixed boundary) configuration, but still much smaller than those obtained for non-optimized coil sets and much smaller than for the reference W7-X configurations. Note that the prompt losses, for times $t< 10^{-3}~\rm{s}$, for the configuration with non-optimized coils are larger than those for the reference configurations, while the optimized coil set allows keeping the prompt losses at levels comparable to the reference configurations and much smaller for longer times. 

Note that the optimized coil set, even producing significant field errors, as shown in figure \ref{fig:coilSetOptimizedWS}, generates a configuration with very good confinement of energetic ions. This positive result can be in part understood by looking at the magnetic  field maps. In figure \ref{fig:BMapsCoils} we show the magnetic field strength versus Boozer angles for three magnetic flux surfaces at $s=0.25, 0.5, 0.75$ for the original configuration from section \ref{secConfigProps}, the best configuration obtained with a non-optimized coil set from figure \ref{fig:coilSetFixD20y25} and for a configuration obtained with optimized coils (Figure \ref{fig:coilSetOptimizedWS}). 
It is clear from the figure that the coil set obtained with the winding surface optimization has a reduced coil ripple as compared to the coil set without optimization, which strongly impacts the confinement of energetic ions. The coil ripple significantly increases with radius.

 Hence, using REGCOIL and the winding surface optimization with adjoint methods, we have obtained a set of simplified coils that allow generating the magnetic configuration with enough  accuracy to keep the confinement of energetic ions at levels comparable to those of the original configuration in the plasma core.

%***************************************************************************************
%***************************************************************************************
\section{Summary and conclusions}\label{secSumandConc}
%***************************************************************************************

In this work, we present a four-field period quasi-isodynamic configuration of moderate aspect ratio optimized for fast-ion confinement at low plasma $\beta$. The configuration has been obtained using the stellarator optimization suite STELLOPT, which has been extended by including new metrics related to the confinement of energetic ions and the proximity to quasi-isodinamicity. 
The well-known proxy $\Gamma_c$ and the newly defined model,  $\Gamma_{\alpha}$ \cite{velasco_model_2021} have been included in STELLOPT and shown to be of great help in the optimization of the confinement of energetic ions. We have also defined several metrics related with the proximity to quasi-isodynamicity, the variance of the $B$ extrema, maxima and minima, and the width of the central magnetic field valley. A measure of the radial derivative of $B$ was also defined as a new target in STELLOPT.

With these tools and following a multi-stage optimization strategy, we have found a new configuration with four periods and aspect ratio $A=9.94$ that exhibits very good confinement of energetic ions at low $\beta$.  It has low magnetic shear and a rotational transform profile with $\iotab<1$, allowing a 4/4 island divertor. The magnetic field structure is close to omnigenous and has poloidally closed $B$ contours, particularly in the innermost region. 
This configuration has a significant magnetic well, which ensures ideal MHD stability and the ballooning stability has been confirmed up to values of $\beta=5\%$ by means of calculations with the code COBRA. It also shows very good confinement of thermal particles, with an effective ripple below 0.5\% in the plasma core. 
The good confinement of energetic ions predicted by the proxies $\Gamma_c$ and $\Gamma_{\alpha}$ has been confirmed with guiding-center calculations carried out with ASCOT. 
The fast ion confinement is particularly good in the plasma core, and it is good even for values of $\beta$ as low as $\beta=0.1\%$, improving up to a very good level at $\beta=1.5\%$. For reactor scale plasmas, with $\beta=4-5\%$, the confinement of energetic ions is outstanding.

Furthermore, a small bootstrap current is expected for plasmas in this configuration as
deduced from calculations of the $D_{31}^*$ coefficient carried out with DKES. These results agree with the theoretical expectation that the bootstrap current is close to zero for configurations sufficiently close to quasi-isodynamicity.

A preliminary study of the feasibility of magnetic coils for this configuration has been carried out. For this study, the code REGCOIL has been used in combination with the optimization of the winding surface using adjoint methods \cite{paul_adjoint_2018}. A set of coils has been designed that allows keeping the good confinement of energetic ions at the same level as the configuration without coils in the plasma core and at acceptable levels up to middle radius. {In future work, we plan to design magnetic coils taking into account additional criteria. In addition, an effort to search for quasi-isodynamic configurations with different periodicities that are optimized for fast ion confinement is underway. }

{The results presented in this paper represent an important advance in design of QI stellarators, as they alleviate one of the major issues in the road towards a helias reactor: good fast ion confinement is obtained at low $\beta$ and is found to be consistent with MHD stability and other neoclassical criteria. Note that good fast-ion confinement is required at low $\beta$ even if the operational design point of the reactor corresponds to a high-$\beta$ value. Otherwise}, {large fast-ion losses might happen during the initial phases of a discharge, damaging the plasma facing components of the reactor. The reason why the configuration presented in this paper exhibits good confinement of fast ions even at very low $\beta$ is that it (very approximately) satisfies the maximum-$J$ property for values of $\beta$ that are much lower than configurations reported in the literature so far ({some} previous QI configurations tend to be minimum-J at low $\beta$, and approach the maximum-$J$ property only for high $\beta$ values~\cite{faustin_fast_2016,velasco_model_2021,subbotin_integrated_2006}).} {Moreover, this} has another} {foreseeable and very relevant} {consequence: {the 
	reduction} of TEM 
	 turbulence~{already at low $\beta$, which  would also help reaching and maintaining the operation point of a stellarator reactor based on this configuration}. We will discuss this in detail in a future publication devoted to the analysis of turbulent transport in the optimized configuration described in this paper.}

%***************************************************************************************
\section*{Acknowledgments}
%***************************************************************************************

The authors thank M. Landreman and E. Paul for their help with REGCOIL and the adjoint methods for the design of magnetic coils, to S. Lazerson and C. Zhu for their help with STELLOPT and to all the contributors to this optimization suite. We thank A. Alonso, D. Carralero{, J. M. García Regaña and G. Godino for continuous discussions} and C. Beidler, M. Drevlak, and A. Bader for their useful comments}. 
This work has been carried out within the framework of the EUROfusion Consortium,
funded by the European Union via the Euratom Research and Training Programme (Grant Agreement No 101052200-EUROfusion). Views and opinions expressed are however those of the authors only and do not necessarily reflect those of the European Union or the European Commission. Neither the European Union nor the European Commission can be held responsible for them. 
This work has been partially funded by the Ministerio de Ciencia, Innovaci\'on y Universidades of Spain  under projects PGC2018-095307-B-I00 and PID2021-123175NB-I00.

%***************************************************************************************
\section*{Apendix A. Coil  designs with fixed distance from the plasma}\label{apendice}
%***************************************************************************************
		
In this appendix, we present several coil designs for the optimized configuration from section \ref{secConfigProps}, obtained using REGCOIL with the constraint of a fixed distance between the coil surface and the plasma, which are comparable in complexity to the coil set shown in figure \ref{fig:coilSetOptimizedWS} but less accurate than that reproducing the original configuration. To illustrate the benefit of performing an optimization of the winding surface, we also compare the optimized winding surface and the winding surface at constant distance from the plasma boundary.

	%*************************************
	\begin{figure}
		\centering
		\includegraphics[draft=false,  trim=400 175 300 55, clip, width=7.5cm]{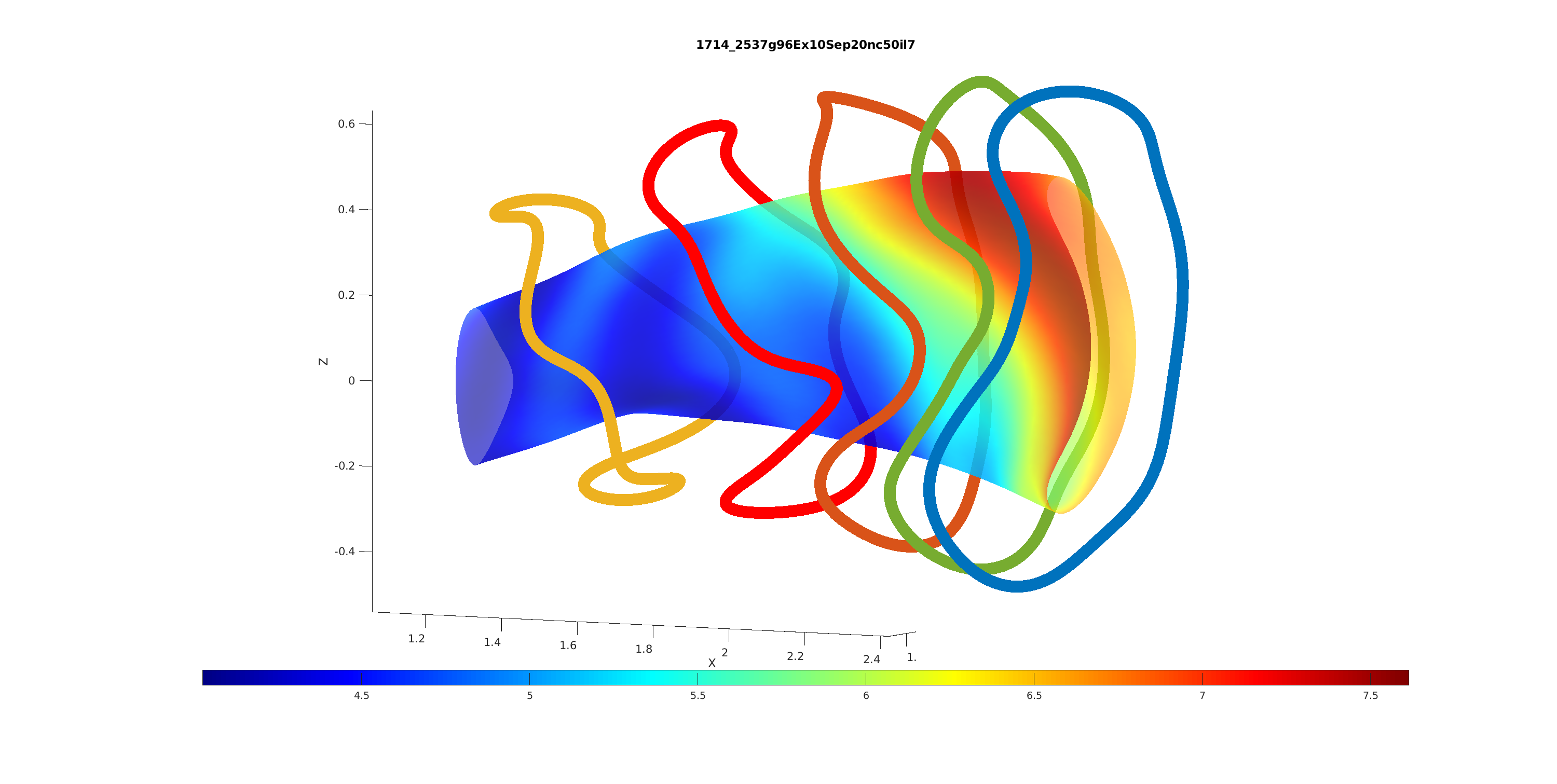}\\
		\includegraphics[draft=false,  trim=400 175 300 55, clip, width=7.5cm]{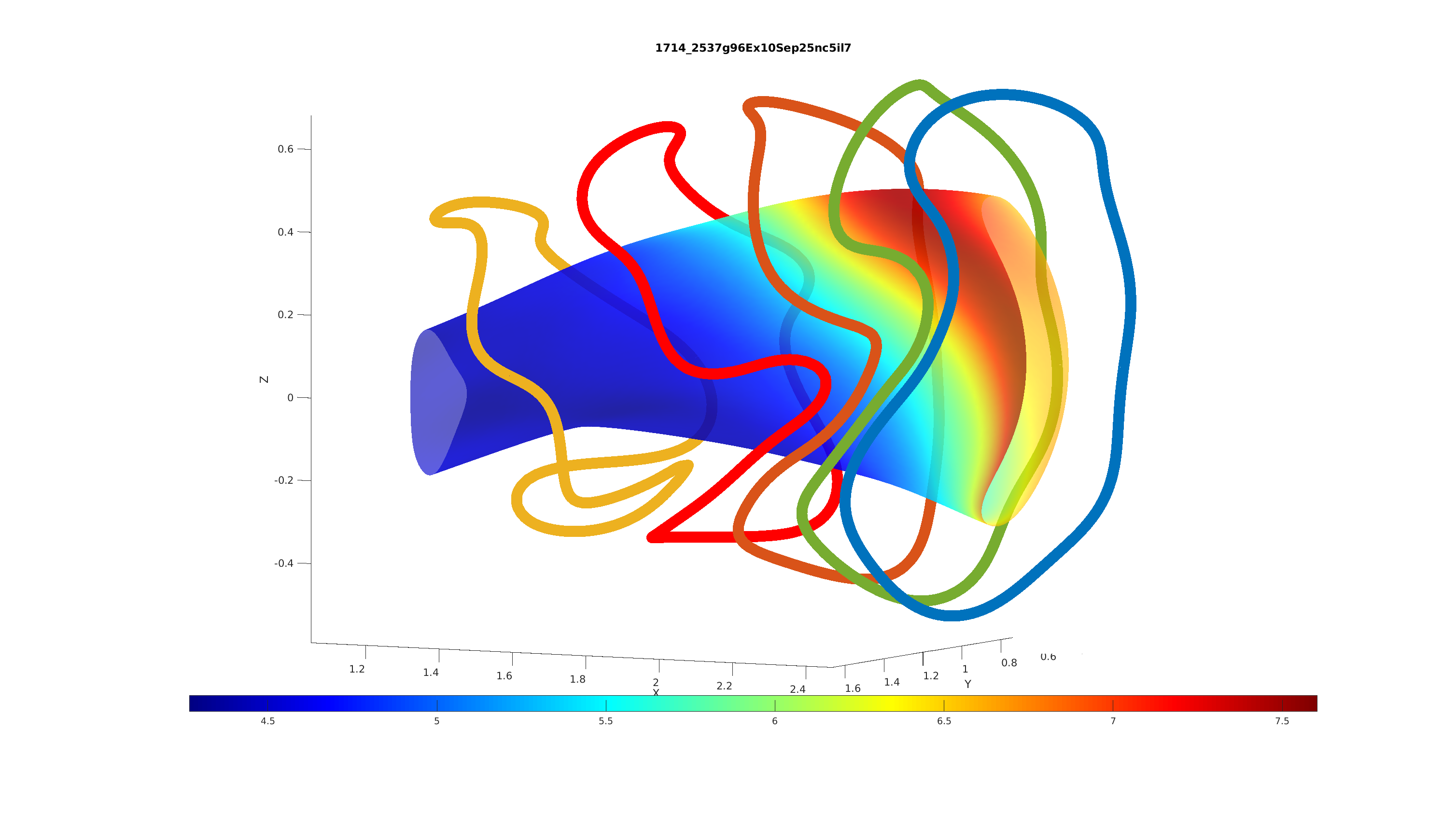}
		\caption{Coil sets designed for the optimized configuration at two values of the (constant) distance coils-plasma, $d=a$ (top) and $d=1.25~a$ (bottom).}
		\label{fig:coilSetFixD20y25}
	\end{figure}
	%*************************************

In a first stage, designs with 8, 10, and 12 coils per period were explored  and 10 was selected as the optimal number, because it balances the coil-to-coil distance, coil complexity, and field errors. With fewer coils the magnetic configuration is poorly reproduced while using more coils the minimum distance between them becomes too small.
We studied the dependence of the resulting configurations with the coil-plasma distance for $d=0.75~a$ to $d=2~ a$,   with a the minor radius, %d=15, 16, 18, 20, 22, 25, 30, 35,~\rm{and}~ 40 ~\rm{cm}$, 
finding that for $d<0.9~a$ %$d<18~\rm{cm}$ 
the ripple introduced by the coils is too large, which makes it very difficult to reproduce the configuration with acceptable fidelity. For $d>1.1~a$ %$d>22~\rm{cm}$ 
the coil ripple significantly decreases, but reproducing the magnetic field accurately requires too complicated coils and the minimum coil-to-coil distance reduces significantly. Two examples of coils obtained for fixed distances $d=a$ and $d=1.25~a$, are shown in figure \ref{fig:coilSetFixD20y25}, which correspond to the best trade-off between coil complexity and field errors for these plasma-to-coil distances obtained by tuning the regularization parameter \cite{landreman_improved_2017}. The results obtained with these two coil sets for the proxies $\Gamma_{\alpha}$ and $\Gamma_c$ were very similar, while the increased complexity of the coils for the case of $d=1.25~a$ as compared to $d=a$ is clear in the figure. The coils in the latter case have larger curvatures and smaller minimum coil-to-coil distances.

For coil-to-plasma distances $d=0.9~a$, $d=a$, and $d=1.25~a$ we performed detailed calculations of fast ion confinement with ASCOT. In all cases, the free-boundary calculation of the MHD equilibria obtained with VMEC including the coils resulted in a magnetic configuration in which the confinement of energetic ions is significantly poorer than that for the fixed-boundary calculation without coils, although still better than for the reference W7-X configurations (see figure \ref{fig:LossFractionWCoils}  for an example). All the cases analyzed with reasonable values of the regularization parameter produced magnetic configurations in which the confinement of energetic ions was very similar. These results suggested the need of an optimization of the winding surface.

%*************************************
\begin{figure*}
	\centering
	\includegraphics[draft=false,  trim=120 20 120 0, clip, width=6.5cm]{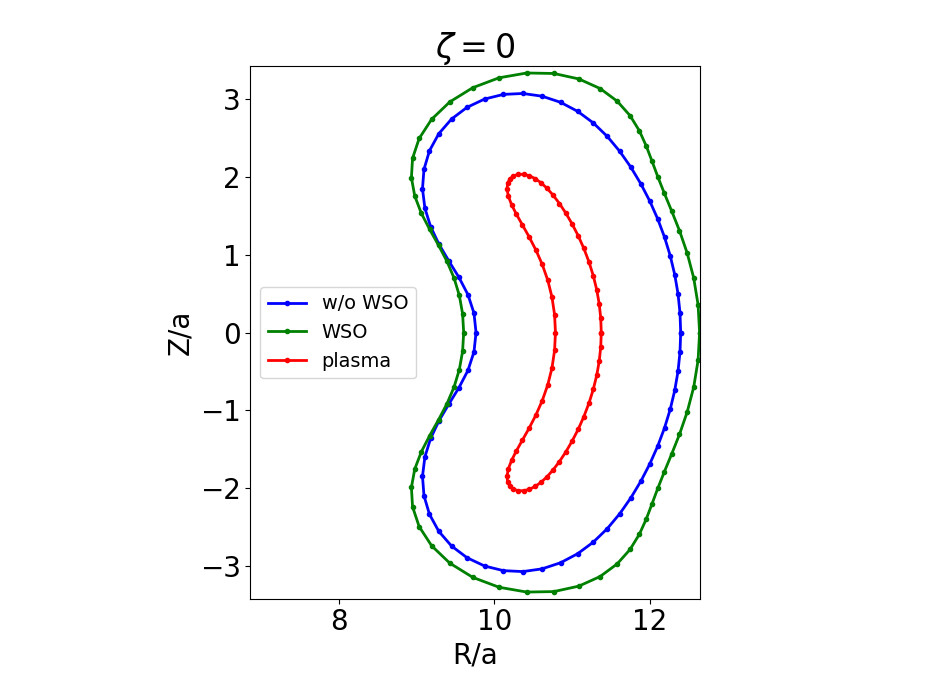}
	\includegraphics[draft=false,  trim=120 20 120 0, clip, width=6.5cm]{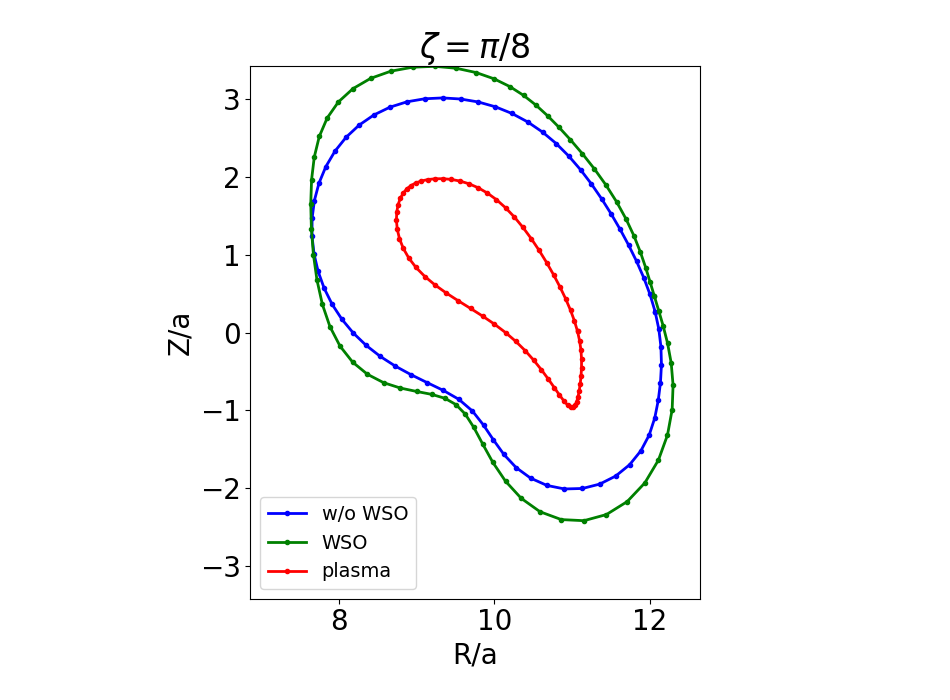}\\
	\includegraphics[draft=false,  trim=120 20 120 0, clip, width=6.5cm]{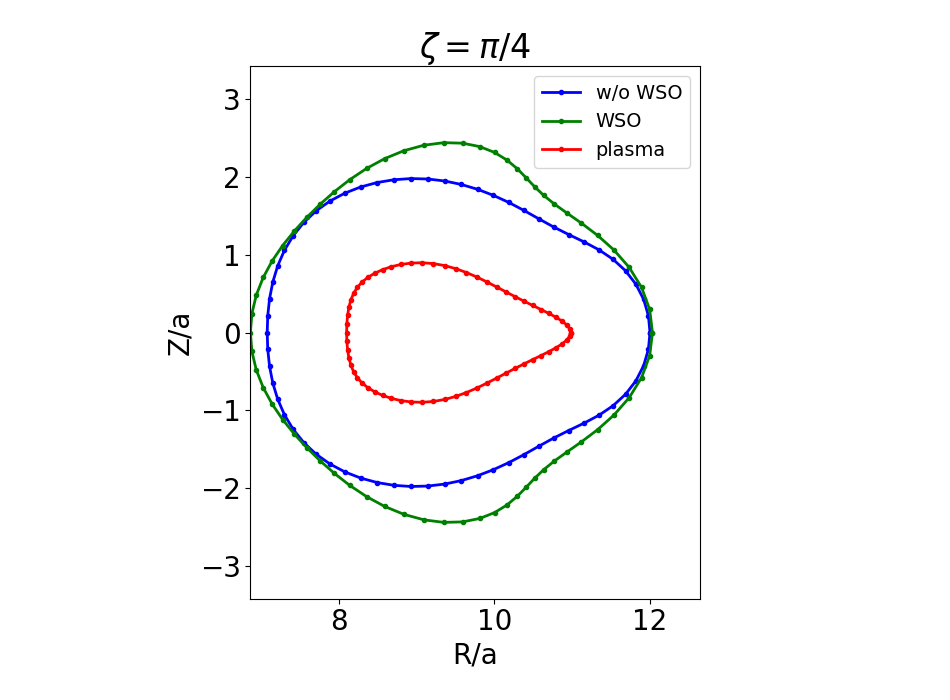}
	\includegraphics[draft=false,  trim=120 20 120 0, clip, width=6.5cm]{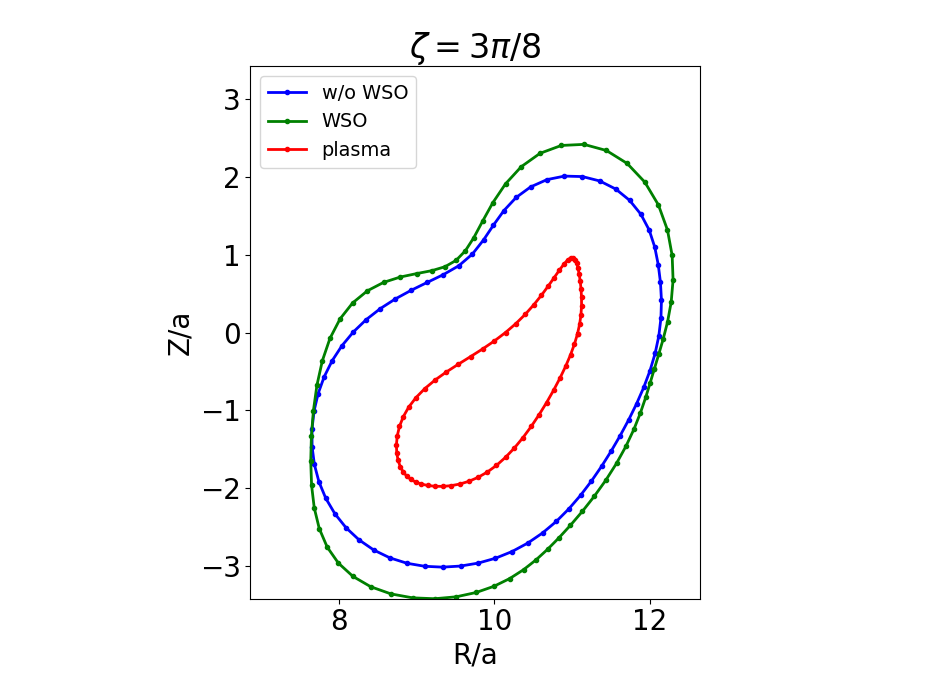}
	\caption{Poincare plots showing the projection, at several toroidal angles, of the winding surface for the case with constant  distance coils-plasma with $d=a$ (blue) and for a case with optimized winding surface with minimum distance from coils-to plasma $d_{m}=a$ (green). The plasma is shown in red.}
	\label{fig:WindingSurfaces20cm}
\end{figure*}
%*************************************
{
As described in section \ref{secCoils} we have carried out an optimization of the winding surface using adjoint methods \cite{paul_adjoint_2018}.  This way, the constraint of the constant distance from plasma to the winding surface is relaxed and a new optimized winding surface is computed. A constraint in the minimum distance from the coil surface to the plasma is set.
The   plots in figure \ref{fig:WindingSurfaces20cm} show the projection of the winding surface for the coil set with constant  distance from coils  to plasma, $d=a$, shown in figure \ref{fig:coilSetFixD20y25},  and for the coil set with an optimized winding surface from section \ref{secCoils} in which the minimum distance from coils to plasma is $d_{m}=a$. The separation of the optimized winding surface from the constant-distance surface is clear in the figure. This optimization of the winding surface translates into a significant improvement of the fast ions confinement (see section \ref{secCoils}).}

%
%%***************************************************************************************
\section*{References}
%%***************************************************************************************

\bibliography{Sanchez_ISHW22_JPP}

\providecommand{\newblock}{}
\begin{thebibliography}{10}
\expandafter\ifx\csname url\endcsname\relax
  \def\url#1{{\tt #1}}\fi
\expandafter\ifx\csname urlprefix\endcsname\relax\def\urlprefix{URL }\fi
\providecommand{\eprint}[2][]{\url{#2}}
% Bibliography created with iopart-num v2.1
% /biblio/bibtex/contrib/iopart-num

\bibitem{hall_three-dimensional_1975}
Hall L~S and McNamara B 1975 {\em Physics of Fluids\/} {\bf 18} 552
  \urlprefix\url{https://aip.scitation.org/doi/10.1063/1.861189}

\bibitem{cary_omnigenity_1997}
Cary J~R and Shasharina S~G 1997 {\em Physics of Plasmas\/} {\bf 4} 3323--3333
  \urlprefix\url{http://aip.scitation.org/doi/10.1063/1.872473}

\bibitem{LandremanM2012}
Landreman M and Catto P~J 2012 {\em Phys. Plasmas\/} {\bf 056103} 1--16

\bibitem{parra_less_2015}
Parra F~I, Calvo I, Helander P and Landreman M 2015 {\em Nuclear Fusion\/} {\bf
  55} 033005

\bibitem{rosenbluth_lowfrequency_1968}
Rosenbluth M~N 1968 {\em The Physics of Fluids\/} {\bf 11} 869--872
  \urlprefix\url{https://aip.scitation.org/doi/10.1063/1.1692009}

\bibitem{Helander2013}
Helander P, Proll J~H~E and Plunk G~G 2013 {\em Physics of Plasmas\/} {\bf 20}
  1--9

\bibitem{Nuhrenberg1988}
Nührenberg J and Zille R 1988 {\em Physics Letters A\/} {\bf 129} 113--117

\bibitem{Boozer1995}
Boozer A~H 1995 {\em Plasma Physics and Controlled Fusion\/}  A103

\bibitem{zarnstorff_physics_2001}
Zarnstorff M~C, Berry L~A, Brooks A, Fredrickson E, Fu G~Y, Hirshman S, Hudson
  S, Ku L~P, Lazarus E, Mikkelsen D, Monticello D, Neilson G~H, Pomphrey N,
  Reiman A, Spong D, Strickler D, Boozer A, Cooper W~A, Goldston R, Hatcher R,
  Isaev M, Kessel C, Lewandowski J, Lyon J~F, Merkel P, Mynick H, Nelson B~E,
  Nuehrenberg C, Redi M, Reiersen W, Rutherford P, Sanchez R, Schmidt J and
  White R~B 2001 {\em Plasma Physics and Controlled Fusion\/} {\bf 43}
  A237--A249 \urlprefix\url{https://doi.org/10.1088/0741-3335/43/12a/318}

\bibitem{anderson_helically_1995}
Anderson F~S~B, Almagri A~F, Anderson D~T, Matthews P~G, Talmadge J~N and
  Shohet J~L 1995 {\em Fusion Technology\/} {\bf 27} 273--277
  \urlprefix\url{https://doi.org/10.13182/FST95-A11947086}

\bibitem{spong_physics_2001}
Spong D, Hirshman S, Berry L, Lyon J, Fowler R, Strickler D, Cole M, Nelson B,
  Williamson D, Ware A, Alban D, Sánchez R, Fu G, Monticello D, Miner W and
  Valanju P 2001 {\em Nuclear Fusion\/} {\bf 41} 711--716

\bibitem{landreman_magnetic_2022}
Landreman M and Paul E 2022 {\em Physical Review Letters\/} {\bf 128} 035001

\bibitem{nemov_evaluation_1999}
Nemov V~V, Kasilov S~V, Kernbichler W and Heyn M~F 1999 {\em Physics of
  Plasmas\/} {\bf 6} 4622--4632
  \urlprefix\url{http://aip.scitation.org/doi/10.1063/1.873749}

\bibitem{landreman_optimization_2022}
Landreman M, Buller S and Drevlak M 2022 {\em Physics of Plasmas\/} {\bf 29}
  082501 \urlprefix\url{https://aip.scitation.org/doi/10.1063/5.0098166}

\bibitem{subbotin_integrated_2006}
Subbotin A~A, Mikhailov M~I, Shafranov V~D, Isaev M~Y, Nührenberg C,
  Nührenberg J, Zille R, Nemov V~V, Kasilov S~V, Kalyuzhnyj V~N and Cooper W~A
  2006 {\em Nuclear Fusion\/} {\bf 46} 921--927
  \urlprefix\url{https://doi.org/10.1088/0029-5515/46/11/006}

\bibitem{Helander2009}
Helander P and Nuhrenberg J 2009 {\em Plasma Physics and Controlled Fusion\/}
  {\bf 51}

\bibitem{Wobig1999}
Wobig H 1999 {\em Plasma Physics and Controlled Fusion\/} {\bf 41} A159

\bibitem{Grieger1992}
Grieger G~e 1992 {\em Fusion Technol.\/} {\bf 21} 1767–1778

\bibitem{Beidler2021}
Beidler C~D, Smith H~M, Alonso A, Andreeva T, Baldzuhn J, Beurskens M~N~A,
  Borchardt M, Bozhenkov S~A, Brunner K~J, Damm H, Drevlak M, Ford O~P, Fuchert
  G, Geiger J, Helander P, Hergenhahn U, Hirsch M, Höfel U, Kazakov Y~O,
  Kleiber R, Krychowiak M, Kwak S, Langenberg A, Laqua H~P, Neuner U, Pablant
  N~A, Pasch E, Pavone A, Pedersen T~S, Rahbarnia K, Schilling J, Scott E~R,
  Stange T, Svensson J, Thomsen H, Turkin Y, Warmer F, Wolf R~C and Zhang D
  2021 {\em Nature\/} {\bf 596} 221--226
  \urlprefix\url{https://www.nature.com/articles/s41586-021-03687-w}

\bibitem{Alonso2022}
Alonso J~A, Calvo I, Carralero D, Velasco J~L, García-Regaña J~M, Palermo I
  and Rapisarda D 2022 {\em Nuclear Fusion\/} {\bf 62} 036024
  \urlprefix\url{https://doi.org/10.1088/1741-4326/ac49ac}

\bibitem{faustin_fast_2016}
Faustin J, Cooper W, Graves J, Pfefferlé D and Geiger J 2016 {\em Nuclear
  Fusion\/} {\bf 56} 092006
  \urlprefix\url{https://iopscience.iop.org/article/10.1088/0029-5515/56/9/092006}

\bibitem{velasco_model_2021}
Velasco J, Calvo I, Mulas S, Sanchez E, Parra F, Cappa A and {the W7-X Team}
  2021 {\em Nuclear Fusion\/} {\bf 61} 116059
  \urlprefix\url{https://iopscience.iop.org/article/10.1088/1741-4326/ac2994}

\bibitem{goodman_constructing_2022}
Goodman A, Mata K~C, Henneberg S~A, Jorge R, Landreman M, Plunk G, Smith H,
  Mackenbach R and Helander P 2022 Constructing precisely quasi-isodynamic
  magnetic fields arXiv:2211.09829 [physics]
  \urlprefix\url{http://arxiv.org/abs/2211.09829}

\bibitem{jorge_single-field-period_2022}
Jorge R, Plunk G~G, Drevlak M, Landreman M, Lobsien J~F, Mata K~C and Helander
  P 2022 {\em Journal of Plasma Physics\/} {\bf 88} 175880504
  \urlprefix\url{https://www.cambridge.org/core/journals/journal-of-plasma-physics/article/singlefieldperiod-quasiisodynamic-stellarator/9B2A5FDCCD7774E4F91BE45E75FDC6B0}

\bibitem{mata_direct_2022}
Mata K~C, Plunk G~G and Jorge R 2022 {\em Journal of Plasma Physics\/} {\bf 88}
  905880503 publisher: Cambridge University Press

\bibitem{bozhenkov_high-performance_2020}
Bozhenkov S~A, Kazakov Y, Ford O, Beurskens M~N~A, Alcuson J~A, Alonso J~A,
  Baldzuhn J, Brandt C, Brunner K~J, Damm H, Fuchert G, Geiger J, Grulke O,
  Hirsch M, Höfel U, Huang Z, Knauer J~P, Krychowiak M, Langenberg A, Laqua
  H~P, Lazerson S~A, Marushchenko N~B, Moseev D, Otte M, Pablant N~A, Pasch E,
  Pavone A, Proll J, Rahbarnia K, Scott E~R, Smith H~M, Stange T, Stechow A~v,
  Thomsen H, Turkin Y, Wurden G~A, Xanthopoulos P, Zhang D and Wolf R~C 2020
  {\em Nuclear Fusion\/}
  \urlprefix\url{http://iopscience.iop.org/10.1088/1741-4326/ab7867}

\bibitem{Yamada2005}
Yamada H, Harris J, Dinklage a, Ascasibar E, Sano F, Okamura S, Talmadge J,
  Stroth U, Kus a, Murakami S, Yokoyama M, Beidler C, Tribaldos V, Watanabe K
  and Suzuki Y 2005 {\em Nuclear Fusion\/} {\bf 45} 1684--1693

\bibitem{warmer_w7-x_2016}
Warmer F, Beidler C~D, Dinklage A and and R~W 2016 {\em Plasma Physics and
  Controlled Fusion\/} {\bf 58} 074006
  \urlprefix\url{https://doi.org/10.1088/0741-3335/58/7/074006}

\bibitem{lion_deterministic_2022}
Lion J, Warmer F and Wang H 2022 {\em Nuclear Fusion\/} {\bf 62} 076040
  \urlprefix\url{https://doi.org/10.1088/1741-4326/ac6a67}

\bibitem{lion_general_2021}
Lion J, Warmer F, Wang H, Beidler C~D, Muldrew S~I and Wolf R~C 2021 {\em
  Nuclear Fusion\/} {\bf 61} 126021
  \urlprefix\url{https://doi.org/10.1088/1741-4326/ac2dbf}

\bibitem{beidler_helias_2001}
Beidler C, Harmeyer E, Herrnegger F, Igitkhanov Y, Kendl A, Kisslinger J,
  Kolesnichenko Y, Lutsenko V, Nührenberg C, Sidorenko I, Strumberger E, Wobig
  H and Yakovenko Y 2001 {\em Nuclear Fusion\/} {\bf 41} 1759--1766
  \urlprefix\url{http://stacks.iop.org/0029-5515/41/i=12/a=303?key=crossref.2418e6b2f2563dcab2d7c900edabb55c}

\bibitem{STELLOPT_doecode_12551}
Lazerson S, Schmitt J, Zhu C, Breslau J and STELLOPT~Developers A 2020 Stellopt
  [Computer Software] \url{https://doi.org/10.11578/dc.20180627.6}
  \urlprefix\url{https://doi.org/10.11578/dc.20180627.6}

\bibitem{nemov_b_2005}
Nemov V~V, Kasilov S~V, Kernbichler W and Leitold G~O 2005 {\em Physics of
  Plasmas\/} {\bf 12} 112507
  \urlprefix\url{http://aip.scitation.org/doi/10.1063/1.2131848}

\bibitem{nemov_poloidal_2008}
Nemov V~V, Kasilov S~V, Kernbichler W and Leitold G~O 2008 {\em Physics of
  Plasmas\/} {\bf 15} 052501
  \urlprefix\url{http://aip.scitation.org/doi/10.1063/1.2912456}

\bibitem{velasco_knosos_2020}
Velasco J~L, Calvo I, Parra F~I and García-Regaña J~M 2020 {\em Journal of
  Computational Physics\/}  109512
  \urlprefix\url{http://arxiv.org/abs/1908.11615}

\bibitem{sanchez_cobra_2000}
Sanchez R, Hirshman S~P, Whitson J~C and Ware A~S 2000 {\em Journal of
  Computational Physics\/} {\bf 161} 576--588
  \urlprefix\url{https://www.sciencedirect.com/science/article/pii/S0021999100965148}

\bibitem{VMECFBTutorial}
VMEC 2022 Vmec free boundary tutorial
  https://princetonuniversity.github.io/STELLOPT/VMEC%20Free%20Boundar%%20Run

\bibitem{merkel_integral_1986}
Merkel P 1986 {\em Journal of Computational Physics\/} {\bf 66} 83--98
  \urlprefix\url{https://www.sciencedirect.com/science/article/pii/0021999186900550}

\bibitem{hirvijoki_ascot_2014}
Hirvijoki E, Asunta O, Koskela T, Kurki-Suonio T, Miettunen J, Sipilä S,
  Snicker A and Äkäslompolo S 2014 {\em Computer Physics Communications\/}
  {\bf 185} 1310--1321
  \urlprefix\url{https://www.sciencedirect.com/science/article/pii/S0010465514000277}

\bibitem{Hirshman1983}
Hirshman S~P and Whitson J~C 1983 {\em Physics of Fluids\/} {\bf 26} 3553--3568

\bibitem{greene_brief_1998}
Greene J~M 1998 {\em Comments Plasma Phys. Controlled Fusion\/}  389

\bibitem{bader_advancing_2020}
Bader A, Faber B~J, Schmitt J~C, Anderson D~T, Drevlak M, Duff J~M, Frerichs H,
  Hegna C~C, Kruger T~G, Landreman M, McKinney I~J, Singh L, Schroeder J~M,
  Terry P~W and Ware A~S 2020 {\em Journal of Plasma Physics\/} {\bf 86}
  \urlprefix\url{https://www.cambridge.org/core/product/identifier/S0022377820000963/type/journal_article}

\bibitem{ciematBranch}
STELLOPT/CIEMAT 2022 Stellopt (see ciemat branch)
  https://github.com/PrincetonUniversity/STELLOPT/ tree/CIEMAT
  \urlprefix\url{https://github.com/PrincetonUniversity/STELLOPT/ tree/CIEMAT}

\bibitem{drevlak_optimisation_2019}
Drevlak M, Beidler C~D, Geiger J, Helander P and Turkin Y 2019 {\em Nuclear
  Fusion\/} {\bf 59} 016010
  \urlprefix\url{https://doi.org/10.1088%2F1741-4326%2Faaed50}

\bibitem{landreman_calculation_2021}
Landreman M and Zhu C 2021 {\em Plasma Physics and Controlled Fusion\/} {\bf
  63} 035001 \urlprefix\url{https://doi.org/10.1088/1361-6587/abd13d}

\bibitem{henneberg_representing_2021}
Henneberg S~A, Helander P and Drevlak M 2021 {\em Journal of Plasma Physics\/}
  {\bf 87} 905870503

\bibitem{hudson_differentiating_2018}
Hudson S~R, Zhu C, Pfefferlé D and Gunderson L 2018 {\em Physics Letters A\/}
  {\bf 382} 2732--2737
  \urlprefix\url{https://www.sciencedirect.com/science/article/pii/S037596011830759X}

\bibitem{murakami_neoclassical_2002}
Murakami S, Wakasa A, Maa berg H, Beidler C, Yamada H, Watanabe K and Group
  L~E 2002 {\em Nuclear Fusion\/} {\bf 42} L19--L22
  \urlprefix\url{https://iopscience.iop.org/article/10.1088/0029-5515/42/11/101}

\bibitem{Freidberg1987}
{J Freidberg} and Freidberg J 1987 {\em Ideal {Magnetohydrodynamics}\/}
  \urlprefix\url{file:////media/DATOS/BIBLIOTECA/libros/PlasmaPhysics/Freidberg-Magnetohydrodinamics.djvu}

\bibitem{Dinklage2013}
Dinklage a, Yokoyama M, Tanaka K, Velasco J, López-Bruna D, Beidler C, Satake
  S, Ascasíbar E, Arévalo J, Baldzuhn J, Feng Y, Gates D, Geiger J, Ida K,
  Isaev M, Jakubowski M, López-Fraguas A, Maaßberg H, Miyazawa J, Morisaki T,
  Murakami S, Pablant N, Kobayashi S, Seki R, Suzuki C, Suzuki Y, Turkin Y,
  Wakasa A, Wolf R, Yamada H and Yoshinuma M 2013 {\em Nuclear Fusion\/} {\bf
  53} 063022
  \urlprefix\url{http://stacks.iop.org/0029-5515/53/i=6/a=063022?key=crossref.7e15326693f98ac81904c8bd2c8a01e7}

\bibitem{Drevlak2014}
Drevlak M, Geiger J, Helander P and Turkin Y 2014 {\em Nuclear Fusion\/} {\bf
  54} 073002
  \urlprefix\url{http://stacks.iop.org/0029-5515/54/i=7/a=073002?key=crossref.3d7d752a9376634b7391fc8c2cb31aa2}

\bibitem{paul_energetic_2022}
Paul E~J, Bhattacharjee A, Landreman M, Alex D, Velasco J~L and Nies R 2022
  {\em Nuclear Fusion\/} {\bf 62} 126054 publisher: IOP Publishing
  \urlprefix\url{https://dx.doi.org/10.1088/1741-4326/ac9b07}

\bibitem{beidler_benchmarking_2011}
Beidler C, Allmaier K, Isaev M, Kasilov S, Kernbichler W, Leitold G, Maaßberg
  H, Mikkelsen D, Murakami S, Schmidt M, Spong D, Tribaldos V and Wakasa a 2011
  {\em Nuclear Fusion\/} {\bf 51} 076001

\bibitem{shaing_bootstrap_1989}
Shaing K~C, Crume E~C, Tolliver J~S, Hirshman S~P and van Rij W~I 1989 {\em
  Physics of Fluids B: Plasma Physics\/} {\bf 1} 148--152
  \urlprefix\url{https://aip.scitation.org/doi/10.1063/1.859093}

\bibitem{Hirshman1986DKES}
Hirshman S~P, Shaing K~C, van Rij W~I, Beasley C~O and Crume E~C 1986 {\em
  Physics of Fluids\/} {\bf 29} 2951--2959
  \urlprefix\url{http://scitation.aip.org/content/aip/journal/pof1/29/9/10.1063/1.865495}

\bibitem{landreman_improved_2017}
Landreman M 2017 {\em Nuclear Fusion\/} {\bf 57} 046003
  \urlprefix\url{http://stacks.iop.org/0029-5515/57/i=4/a=046003?key=crossref.d965f662ac2b69d6691056725320fc00}

\bibitem{paul_adjoint_2018}
Paul E, Landreman M, Bader A and Dorland W 2018 {\em Nuclear Fusion\/} {\bf 58}
  076015
  \urlprefix\url{https://iopscience.iop.org/article/10.1088/1741-4326/aac1c7}

\end{thebibliography}

\end{document}